\documentclass[11pt]{article}
\usepackage[margin=1in]{geometry}

\usepackage[authoryear]{natbib}

\usepackage{thm-restate}
 \usepackage{booktabs} %
\usepackage{tabularx}
\usepackage{amsmath, xr, graphicx, amsthm, amssymb}
\usepackage{thmtools}

\usepackage{float,subcaption,placeins}

\newtheorem{lemma}{Lemma}

\renewcommand{\proof}{\noindent{\textbf{Proof. }}}

 \newcommand{\Halmos}{$\square$}

\newcommand\bbR{\ensuremath{\mathbb{R}}} %

\renewcommand{\b}[1]{\ensuremath{\overline{#1}}}

\begin{document}
\title{Designing Informative Rating Systems: Evidence from an Online Labor Market}

\newcommand\blfootnote[1]{%
	\begingroup
	\renewcommand\thefootnote{}\footnote{#1}%
	\addtocounter{footnote}{-1}%
	\endgroup
}

\author{
	Nikhil Garg\\
	Stanford University\\
	\texttt{nkgarg@stanford.edu} \\
	\and
	Ramesh Johari\\
	Stanford University\\
	\texttt{rjohari@stanford.edu} \\
}

\maketitle

\begin{abstract}{%
\makeatletter{}%

Platforms critically rely on rating systems to learn the quality of market participants. In practice, however, these ratings are often highly inflated, and therefore not very informative.  In this paper, we first investigate whether the platform can obtain less inflated, more informative ratings by altering the {\em meaning} and {\em relative importance} of the \textit{levels} in the rating system.  Second, we seek a principled approach for the platform to make these choices in the design of the rating system.

First, we analyze the results of a randomized controlled trial on an online labor market in which an additional question was added to the feedback form. Between treatment conditions, we vary the question phrasing and answer choices; in particular, the treatment conditions include several {\em positive-skewed verbal rating scales} with descriptive phrases or adjectives providing specific interpretation for each rating level. The online labor market test reveals that current inflationary norms can in fact be countered by re-anchoring the meaning of the levels of the rating system. In particular, the positive-skewed verbal rating scales yield rating distributions that significantly reduce rating inflation and are much more informative about seller quality. 

Second, we develop a model-based framework to compare and select among rating system designs, and apply this framework to the data obtained from the online labor market test. Our simulations demonstrate that our model-based framework for scale design and optimization can identify the most informative rating system and substantially improve the quality of information obtained over baseline designs.

Overall, our study illustrates that rating systems that are informative in practice \textit{can} be designed, and demonstrates \textit{how} to design them in a principled manner.

		\blfootnote{This work benefited from substantial implementation and experimental efforts led by Cary Luu. We also thank Michael Bernstein, Hayden Brown, Ashish Goel, John Horton, Shane Kinder, and participants of the Market Design workshop at EC'18. This work was funded in part by the Stanford Cyber Initiative, the National Science Foundation Graduate Research Fellowship under grant DGE-114747, and National Science Foundation grant 1544548. }
	}
\end{abstract}%
\makeatletter{}%
\section{Introduction}

{\em Rating systems} are an integral part of modern online markets. Marketplaces for products (Amazon and eBay), ridesharing (Lyft and Uber), housing (Airbnb), and freelancing all employ rating systems to vet platform participants. Buyers rely on ratings to choose which products to buy and how much to pay, and platforms use ratings to identify both poor and great performers, and in ranking search results.  Ratings are consequential: a high score typically directly translates to more visibility and sales.  Indeed, without effective mechanisms to collect feedback after matches, online markets would be ``flying blind'' in reducing search frictions between buyers and sellers.

Despite their central importance, extensive prior work suggests the standard rating systems of many online platforms are {\em not sufficiently informative}, i.e., ratings do not sufficiently discriminate between high and low quality sellers.  A major causal factor in this lack of informativeness is {\em rating inflation}, where most participants predominantly receive high ratings.   Heavily skewed rating distributions lead to systems in which noise dominates, and as a result buyers are challenged to extract meaningful signal from available rating scores.

Several empirical studies have established the prevalence of rating inflation.  On eBay, more than 90\% of sellers studied between 2011 and 2014 had a rating of at least 98\% positive, and more transactions result in a dispute than in negative feedback~\citep{nosko_limits_2015}. On the online freelancing platform oDesk, average ratings rose by one star over seven years~\citep{filippas2018reputation}.  On Uber, an average rating of $4.6$ out of $5$ stars puts a driver at risk of deactivation~\citep{cook_ubers_2015}.  On Airbnb, almost $95\%$ of hosts have an average rating of $4.5$-$5$ out of $5$ stars~\citep{zervas_first_2015}. On Amazon, ratings tend to be bimodal with a big peak near the most positive score and then a (much) smaller one near the most negative one~\citep{hu_overcoming_2009}. Numerous other works report similar findings; \citet{tadelis_reputation_2016} provides a thorough review of the literature.

The empirical literature concludes that inflated ratings are less informative about quality differences among participants.  For example, \cite{filippas2018reputation} notes that the increase in average ratings at oDesk could not be explained solely by higher seller performance, indicating that rating informativeness dropped over time {as ratings inflated}.   As a consequence {of inflation}, negative ratings carry outsized influence, because they are so rare; for example,~\citet{cabral_dynamics_2010} find that on eBay a seller's \textit{first} negative feedback reduces her weekly sales growth rate from $5\%$ down to $-8\%$.

In this paper we investigate whether platforms can improve the quality of information obtained by changing the design of the {\em rating scale} that they employ.  In particular, we ask: by carefully choosing both the meaning and importance of different answer choices in a rating scale, can platforms elicit higher quality information from their raters, i.e., such that the platform recovers the true relative qualities of sellers with fewer ratings?  Our main contributions are as follows.

{\bf Reducing rating inflation via {positive-skewed verbal scales}.}  First, we establish evidence that a careful choice of the rating scale can in fact strongly reduce rating inflation.  In particular, we analyze a test in the live rating system of a large online labor market.  In this test, {the platform} asks buyers to choose from a list of phrases (e.g., {\em Best Freelancer I've Hired}) or adjectives (e.g., {\em Fantastic!}).  Our results show that platforms can effectively combat rating inflation by using positive-skewed verbal scales: the rating distribution obtained from such scales is substantially more dispersed than under the ``standard'' star rating scale.  Most starkly, in our experiment, $80.6\%$ of freelancers received the best possible numeric (i.e., star) rating, but less than $35.8\%$ were rated with the highest-ranked verbal phrase across non-numeric treatment cells.  We further provide evidence that inflation {\em over time} can be countered: ratings on our additional question did not inflate over the test time period, in contrast to an inflation of about $0.3$ points (on a five star scale) over a similar time period after the introduction of a new numeric rating system on the same platform, cf.~\cite{filippas2018reputation}.  Our findings suggest that in platforms today, the \textit{norm} is that any \textit{acceptable} experience is given the top numeric rating, with the rest of the scale reserved for various degrees of \textit{unacceptable} experiences.

{\bf {Positive-skewed verbal scales} yields more informative ratings.}  Second, we establish evidence that the {verbal scales we tested yield} {\em more informative ratings}.  In particular, we show ratings given with the positive-skewed verbal adjective scales are more predictive of whether a freelancer will be re-hired by the client in the near-future: clients are up to $30.8\%$ more likely to rehire the freelancer during the test period after giving them a top rating from a positive-skewed scale than after giving them the top numeric score. In addition, for each freelancer we estimate their quality through the experimental data itself (carefully handling endogeneity concerns) and then produce an estimated joint distribution of freelancer quality and the ratings they receive with a given ratings scale. The distributions qualitatively reveal that positive-skewed verbal scales are much more informative about freelancer quality than are numeric scales.

{\bf A principled approach to comparing rating system designs.}  Third, we provide a principled approach to comparison of different rating system designs.  In particular, we develop a metric on the joint distribution of seller quality and resulting ratings that directly reflects the typical goal of a rating system: to learn about sellers {\em as quickly as possible}.  We develop a mathematical framework where the performance of a rating system is measured through the {\em large deviations} rate of convergence of the seller ranking via observed score to the true underlying seller quality ranking.
This rate  is the exponent in the exponential decay of the Kendall's $\tau$ distance between the estimated and true seller rankings over time.

We develop a stylized model for rating system design within which we calculate these convergence rates. We define a fictitious ``marketplace'' in which sellers accumulate ratings over time, with match rates proportional to their quality. Buyers rate sellers using a multi-level {\em rating scale}, i.e., buyers are asked to answer a multiple choice question (e.g., 1-5 stars, or a set of adjectives describing the interaction) when rating the seller. The platform can choose amongst several rating scale options that differ in their levels (e.g., adjectives); these scale options induce different buyer rating behavior. The platform can also set the scores to assign to these adjectives (e.g., the seller might receive a ``5'' if the buyer selects the best adjective, and a ``3.7'' if they select the second best adjective). Within this marketplace, different design choices (scale choices and scores) differ in the rates of convergence to the true quality ranking they induce, with higher rates reflecting better designs.

We show that given behavioral data of how buyers have rated sellers under various rating system designs {in our test}, this framework can be effectively employed to compare and select {among} the designs.  In particular, we apply this framework to the data from our online labor market test. This process reveals the quantitative gains in convergence rate obtained by verbal rating scales over the naive numeric rating system.  Interestingly, our framework also reveals that the first order effect on the rate of convergence comes from the choice of verbal descriptions on the scale; optimizing the choice of scores yields a lower order improvement in performance.\vspace{0.1in}

Taken together, our results suggest that platforms have much to gain by optimizing the {\em meaning} of the levels in their rating systems, and in particular using positive-skewed verbal rating scales instead of numeric scales. %
Our managerial insight is that ratings on online platforms are not doomed to be highly inflated; rating behavior is responsive to how the system is designed, and \textit{good} rating behavior can be both quantified and obtained through a structured design methodology. Our entire approach of experimenting with various rating scales and then choosing amongst them in a principled manner also provides a framework for doing the same in other ratings contexts, including where other behavioral challenges (such as bias or deflation) may be present. For example, in Section~\ref{sec:mturkrepeatanalysis} in the Appendix, we repeat our experiment, design approach, and analysis in a synthetic rating setting on Amazon Mechanical Turk where we have access to expert ratings on item quality.

The remainder of the paper is organized as follows.  Section~\ref{sec:related} contains related work. In Section~\ref{sec:labor}, we describe the labor market test, with results presented in Section~\ref{sec:laborresults}. In Section~\ref{sec:adjectives} we describe a model and approach to evaluating and designing a multi-level rating scale. Finally, in Section~\ref{sec:laborapplyresults}, we apply our design approach to the data from our labor market experiment. The Appendix contains additional information and robustness analyses for our labor market test, a second application of our design approach via a synthetic experiment on Mechanical Turk, and proofs.

\section{Related literature}
\label{sec:related}

Challenges in designing effective online rating systems are well-documented. To help explain the empirical inflation findings discussed above, one branch of the literature focuses on how ratings are given after bad experiences, and in particular conditions under which buyers either don't leave a review at all or leave a {\em positive} review.  On Airbnb, for example, \citet{fradkin_determinants_2017} find that inducing more reviews resulted in more negative reviews, suggesting that those with negative experiences are less likely to normally submit a review.  Though historically this inflation has been thought of as a \textit{strategic} response to potential retaliation, recent evidence indicates that \textit{social} pressure also plays a role. For example, sellers incentivize reviews (of any kind) by offering discounts, potentially creating an implicit social obligation for reviewers to reciprocate with a positive review~\citep{li_money_2014,cabral_dollar_2014}. Such effects, along with outright fraud and sellers asking for higher ratings, contribute to rating inflation.

\subsection{Platform measures to counter or encourage inflation} Platforms are aware of the inflation problem and have invested in fixing it. Most existing solutions try to decrease retaliatory pressure from sellers or to encourage more buyers to submit reviews. In 2007, eBay implemented one-sided feedback (i.e., only buyers rating sellers), with anonymous ratings presented only in aggregate; the platform later eliminated negative buyer ratings altogether~\citep{bolton_engineering_2013}. Through a test with private feedback, oDesk reports that such feedback predicts both future private \textit{and} public feedback better than does public feedback, and there is evidence that buyers utilize private ratings more than they do public ratings~\citep{filippas2018reputation}.\footnote{``Public ratings'' are shown publicly, non-anonymized, e.g., ``\textit{A} rated \textit{B} 5 stars.'' ``Private ratings'' are either shown as a summary statistic, e.g., ``B averages 4.6 stars'', or not shown at all and used only internally by the platform.} Other work has attempted to align buyer incentives with providing informative reviews~\citep{gaikwad_boomerang:_2016}, but the approach has not yet been widely adopted. Despite such fixes, the problem of inflation largely remains on online platforms,
consistent with the hypothesis that norms have shifted so that even average experiences are given the top numeric value.

This literature suggests that many initially effective ideas may not have a first
order effect in increasing the informativeness of ratings, especially in the long-term: rating behavior on online platforms is not static. \cite{filippas2018reputation} show that inflation happens over time: on the same online labor market as in our test, average public ratings over a span of nine years went from below 4 stars to about 4.8 stars. This view is consistent with the ``disequilibrium'' view of rating system design described by \cite{nosko_limits_2015}.

\subsection{Survey design and rating inflation in other contexts}

Rating inflation and the question of rating system design are also prevalent in other contexts. For example, grade inflation in education is an oft-recognized issue~\citep{johnson2006grade}.  {Proposed solutions include} include forcing educators to deflate grades (either by assigning quotas to each grade or by eliciting rankings) or standardizing grades after the fact~\citep{lackey2006grade,blum2017nine}. Similar methods are used to evaluate employees~\citep{shaout2014performance} and athletes. In baseball, for example, scouts rate athletes on a numeric scale that spans from 20 to 80~\citep{gines2017tastes}; however, scouts differ in how they evaluate talent or otherwise have heterogeneous biases, and teams may use sophisticated systems to calibrate the information provided by each scout~\citep{reiter_astroball:_2018}. On online platforms, in contrast, it may not be desirable to impose ratings quotas on buyers or feasible to assess the rating ability of individual buyers ({though these are interesting avenues for future work}).

An alternate approach to counter grade inflation is adding and labeling rating levels (e.g., plus-minus grading, or providing suggested mappings from relative ranking to grade) in order to \textit{behaviorally} induce more dispersed grade distributions from educators~\citep{lackey2006grade,blum2017nine}. This solution is similar to the well-studied idea of using labels for scales in survey responses, in which the specific design of rating scales -- including the specific words, number of words, and their positive-negative balance -- is known to affect responses \citep{krosnick_survey_1999,parasuraman_marketing_2006,klockars_influence_1988,hicks_choosing_2000}. In such solutions, the raters are not forced or even explicitly asked to answer in a certain manner; rather, the question and answer choices are presented in a way such that raters naturally behave as the survey designer wishes them to.

{Our behavioral results are consistent with this latter literature, {\em despite} the presence of incentive issues as discussed above: scale design can have a first order effect on the quality of responses in real rating systems.}  %
{Although this finding aligns with the survey design literature}, as discussed above our study is preceded by a long line of rating systems literature in which substantive changes (making ratings private, trying to prevent retaliation, or UI changes) do not in practice lead to {sufficiently} informative rating systems. Given the potential costs for giving negative ratings posited by previous work, it is not clear \textit{a priori} that any change will induce raters to do so; {our work provides one path forward}.

Beyond this behavioral insight, we provide a theoretical framework that a survey designer in any context can use to pinpoint the most informative design for their setting in a principled manner. For example, in the Appendix we apply our approach to a setting more similar to standard survey design and crowd-sourcing, and it yields a non-trivial rating system design that outperforms others.

\subsection{Theoretical analyses of ratings} Recent literature has attempted to explain rating behavior, and inflation in particular, through a variety of models~\citep{immorlica_emergence_2010, cabral_dynamics_2010, filippas2018reputation,fradkin_determinants_2017}.  Much of this work seeks to understand how buyer incentives may result in an equilibrium in which they provide with dishonest ratings, or how sellers may be incentivized to accumulate high ratings and then give low effort.  For example, \cite{filippas2018reputation} posit that high ratings are unavoidable when sellers are affected by negative ratings, as buyers are incentivized to incorrectly give positive ratings even after negative experiences. %

Several recent works also study the speed of learning in rating systems and other similar contexts~\citep{che_optimal_2015,johari_matching_2016,ifrach_bayesian_2017,acemoglu_fast_2017,papanastasiou2017crowdsourcing}. In these works, the platform influences \textit{which matches occur} through its design, and this affects the learning rates. In contrast, we take the matches as given and show how the platform can meaningfully design \textit{what it learns from each match}. Finally, in a related paper, we consider the optimal design of \textit{binary} rating systems, for which far more theoretical structure exists~\citep{garg_binary_2018}.

\makeatletter{}%
\section{Online labor market experiment description}
\label{sec:labor}
Our work focuses on whether we can improve the design of the feedback systems used in online platforms.
{As we have noted}, the literature suggests that despite substantial effort across a variety of platforms, rating behavior has not changed for the better over time: average ratings on platforms tend to be extremely high or ``inflated'' (see discussion in Section \ref{sec:related}).  {A significant consequence of this inflation is that current} ratings systems and their resulting distribution of ratings do not provide information that can effectively and efficiently differentiate high {quality participants} from low quality participants.

In this section, we {propose} a simple but under-explored innovation in the design of a rating system: using \textit{positive-skewed} verbal phrases in the rating scale.  We study the effect of such a change through the results of a randomized controlled trial on the rating system of a large online labor market.  In this test, new ratings questions were introduced in a feedback form clients submit upon finishing a job with a freelancer.

The section is structured as follows.  In Section~\ref{sec:motivhypo}, we further discuss our motivation and hypotheses. In Section~\ref{sec:empiricalcontext}, we briefly describe the online labor market. Section~\ref{sec:methodsubsec} contains our method and the treatment conditions.  {We discuss the results in Section~\ref{sec:laborresults}; as we show there, our results demonstrate that our proposed design changes successfully curb rating inflation and lead to substantially more informative ratings.}

\subsection{Motivation and hypothesis}
\label{sec:motivhypo}
We aim to design rating scales for online platforms that lead to more informative ratings. Motivated in part by the emergence of the rating norms discussed in the introduction, where 5 stars is routinely considered ``average,'' we are interested in evaluating the effectiveness of changes that can counter this norm:  in particular, we consider rating scales where the answer choices are \textit{positive-skewed}, with \textit{specific} descriptions attached to each rating.

Our hypothesis is that {\em such positive-skewed scales lead to less ``inflated'' ratings than standard, numeric rating scales, and as a result, produce more informative ratings}. (By ``inflated'',  we mean ratings where a large majority of the rating distribution is on the highest rating score).

This hypothesis is motivated by the idea that raters feel a cost if they are dishonest in their ratings, and that {\em this cost is an increasing function in how dishonest she perceives herself to be}.  Crucially, this quantity would vary both with experience quality and the rating system design. With standard numeric rating systems and today's norms, a rater arguably does not consider herself dishonest for rating mediocre experiences $5/5$, because that is what 5 stars has come to mean.  On the other hand, suppose a platform provides explicit guidance on what ratings mean (e.g., ``5 stars means best experience you've had''); raters would thus face a higher cost of dishonesty for giving low quality sellers a high ratings. This hypothesis is consistent with the {\em self-concept maintenance} literature, where people are understood to be more likely to be dishonest when they can convince themselves that they are acting honestly~\citep{mazar_dishonesty_2008}. Models with such costs have been considered in, e.g., \cite{filippas2018reputation} and \cite{fradkin_determinants_2017}.

Finally, note that while our test design allows us to measure the effects of other design changes, we did not hypothesize any other specific effects (direction or magnitude) {\em a priori}, except for the benefits of positive-skewed rating scales. These alternate design changes let us compare the relative benefits of possible solutions to the rating inflation problem.

\subsection{Empirical context}
\label{sec:empiricalcontext}
The test ran on a large, online labor market. In this market, clients seek the services of freelancers across a variety of categories (e.g., software development, graphic design, and translation). Clients may choose to contract with a freelancer for a job based on work history, prior ratings, the freelancer's proposal, and potentially an initial conversation. A client-freelancer pair may work on multiple jobs together during their time on the platform.

At the end of each job, the client is asked to fill out a feedback form in which they rate the freelancer's work through a series of multiple choice and free-form questions. This labor market has both private and public ratings, and private ratings are aggregated and made available to potential future clients as part of a freelancer's public score. Both private and public ratings are high on the platform: even the average private feedback score is over $8.5/10$. See~\citet{filippas2018reputation}, which analyzes ratings over time on the same labor market, for an in-depth description of the status-quo rating system and its performance.

\subsection{Method}
\label{sec:methodsubsec}
We now describe our test method. The authors were involved in test design and analysis of anonymized data, but not implementation or deployment.

The test added a question to the feedback form given to clients after they close a job. This question appeared with the current private rating questions and was marked optional. All clients were still asked the existing private rating questions, including rating the freelancer on a numeric $0-10$ scale. The answer choices were displayed vertically after the question.

The test ran over a 90 day period in Summer 2018, with a pilot in January 2018 over 5 days. We report the set-up and results of the long test; pilot results are nearly identical. %

\subsubsection{Treatment conditions}
\label{sec:treatmentconditions}
There were six treatment conditions that included an additional question on the feedback form. The question phrasing and answer choices differed between the treatment conditions. See Table~\ref{tab:labortreatments} for a detailed list of the treatment conditions. There were four different types of answer choices: (1) comparing against a client's expectations (\textit{Expectation}); (2) descriptive adjectives (\textit{Adjectives}); (3) comparing against the average freelancer the client has hired, as well as two variants (\textit{Average; Average, not affect score; Average, randomized}); and (4) a numeric scale with no descriptions attached to the ratings (\textit{Numeric}).

The non-numeric treatments describe possible ways to design multiple choice rating systems that add more specificity to the rating scale. The choices themselves are skewed toward the positive end: each scale has two ``negative'' choices, one ``neutral'' choice, and 3 ``positive'' choices, in increasing levels of effusiveness. This imbalance was chosen so that (a) clients could give ``positive'' feedback to most freelancers while still allowing the platform to disambiguate the very best from others, and (b) to emphasize that the best ratings should be reserved for the very best freelancers.

The \textit{Numeric} treatment, giving freelancers the option of giving $0-5$ stars, helps disambiguate between novelty effects of introducing new questions and the idiosyncratic effects of the question itself. As in the other treatments, this question is asked \textit{in addition to} the the existing rating questions on the site, which include a $0-10$ overall rating question.
Furthermore, the question phrasing is identical in the \textit{Adjectives} and \textit{Numeric} treatments; only the answer choices differ. This design thus teases out the different effects of the type of question itself and the answer choices.

{We include two additional variants as follows:
(a) a variant with additional text emphasizing that the answer will not impact the freelancer's publicly displayed rating (\textit{Average, not affect score}), and (b) a variant where we randomize the order of the answer choices (\textit{Average, randomized}).} The first variant tests the additional informational gain from clients knowing for certain that a low rating will not affect the freelancer.  The second variant helps assess the propensity of clients to not read all the answer choices before responding.

In addition to the six treatments, a \textit{Control} condition was included, in which no additional question is asked (replicating the status quo feedback form).

\begin{table}[]
\small
{
	\begin{tabularx}{\linewidth}{>{\hsize=.35\hsize}X>{\hsize=1.0\hsize}X>{\hsize=1.67\hsize}X}
		\hline
		\textbf{Treatment}            & \textbf{Additional Question}                                                                             & \textbf{Answer choices}                                                                                                                                         \\ \hline
		\textit{Expectation}              & How did this freelancer compare to your expectations?                                                    & Much worse than I expected, Worse than I expected, About what I expected, Better than I expected, Far better than I expected, Beyond what I could have expected \\
\hline
		\textit{Adjectives}                & How would you rate this freelancer overall?                                                              & Terrible, Mediocre, Good, Great, Phenomenal, Best possible freelancer!                                                                                          \\
\hline
		\textit{Average}                   & How does this freelancer compare to others you have hired?                                               & Worst Freelancer I've Hired, Below Average, Average, Above Average, Well Above Average, Best Freelancer I've Hired                                              \\
\hline
		\textit{Average, not affect score} & How does this freelancer compare to others you have hired? (This will not impact the freelancer's score) & Same as Average group                                                                                                                                           \\
\hline
		\textit{Average, randomized}       & How does this freelancer compare to others you have hired?                                               & Same as Average group, but in random order                                                                                                            \\
\hline
		\textit{Numeric}                   & How would you rate this freelancer overall?                                                              & 0, 1, 2, 3, 4, 5                                                                                                                                                \\ \hline
	\end{tabularx}
}
	\caption{Treatments groups for labor market test \label{tab:labortreatments}}
\end{table}

 \subsubsection{Allocation to treatment groups} \label{sec:allocation} Allocation was done at the client level when they first closed a job and landed on the feedback form after the start of the test.  Clients who had closed less than two jobs in the past were excluded, as several of the treatment conditions ask clients to compare the freelancer to past experiences. Each treatment condition was allocated 15$\%$ of the clients, and the remaining 10$\%$ of clients were allocated to \textit{Control}.  After being allocated to a treatment group, a given client was assigned the same treatment for the duration of the test and was thus shown the same additional question for any further jobs she may have closed.  (During the pilot in January, 2018, $40\%$ of clients were allocated to \textit{Control} and $10\%$ to each treatment condition.)

 Due to a bug in the allocation code during the test, $1,086$ out of the $66,755$ clients who submitted feedback were assigned to different treatment conditions on different closed jobs. We disregard all such clients in our analysis to eliminate the possibility of contamination between treatment cells. To confirm experimental validity, we show in the Appendix that otherwise the randomization was effective: the distribution of clients in different cells are essentially identical on all observed covariates.  This bug does bias the client population in our data in one way, however: clients who closed more jobs in the test period were more likely to experience the bug, and thus to be incorrectly assigned to multiple treatment cells.  As a consequence, the client population on which we carry out our analysis skews {slightly} away from the highest volume clients on the platform.

 \subsubsection{Number of responses and data preprocessing}

 $75,592$ unique clients landed on the feedback page, and $66,755$ clients submitted feedback for at least one job. We remove the  clients mistakenly assigned to multiple treatment cells (the bug described above), as well as seven clients who were correctly assigned but who closed more than 200 jobs during the test period. Table~\ref{tab:counts} contains, for each treatment cell, the numbers of clients assigned, clients who submitted a job, and clients and jobs in our dataset after the pre-processing.

\begin{table}[]
	\centering
	\small
	\begin{tabular}{l|ccccc|c}
		& \textbf{Assigned} &  \multicolumn{2}{c}{\textbf{Submissions}} & \multicolumn{2}{c}{\textbf{Analyzed}} & \\
		\textbf{Condition}  & {Clients} & {Clients} &   {Jobs}  & {Clients} & {Jobs} & Mean treatment response  \\ \hline
		Control              & 7576 &  7179  &23554 & 6880             & 21850     &  -  \\
		Expectation          &11271 &  10073 &28880  & 9718             & 27156     & 3.34    \\
		Adjectives           &11101 &  9966  &28413 & 9616             & 26370     & 3.65   \\
		Average              &11375 & 10135  & 28372 & 9807             & 26605     &3.76     \\
		Average, not affect score &11500 & 10295& 28882 & 9944             & 27536   &3.78      \\
		Average, randomized     &11466 &  10258&28663 & 9895             & 26978     &3.46    \\
		Numeric                 &11303 & 10120&32153 & 9802             & 27677    &4.59
	\end{tabular}
\caption{Number of clients and jobs in each cell, and mean treatment response}
\label{tab:counts}

\end{table}

\section{Labor market test results}
\label{sec:laborresults}

{In this section we provide results that demonstrate that the positive-skewed verbal scales reduced inflation and produced more informative ratings.} In Section~\ref{sec:marginaldist} we show the verbal rating scales result in \textit{deflated} ratings compared to the numeric scale; we report simple marginal distributions of the rating choices made by clients in different treatment cells, both overall and across time throughout the experiment. Then in Section~\ref{sec:jointdistributions} we show that such verbal scales are more \textit{informative} than the numeric scale: the rating choices made by clients in the verbal treatment cells better correspond to exogenous signals of a given freelancers quality, {in two ways.  First, we show that the verbal scales are more predictive of whether clients tend to rehire a freelancer.  Second, we show that the verbal scales yield ratings that are more strongly correlated with the average rating a freelancer receives by distinct clients across treatment cells; this latter quantity serves as an empirical proxy for freelancer ``quality''.} The resulting {approximate} joint distributions of freelancer quality and ratings received motivate our model in the next section, where we develop a measure to compare rating scales by the joint distributions they induce. %

\begin{figure}
	\centering
	\begin{subfigure}[b]{.5\textwidth}
	\includegraphics[width=1.07\linewidth]{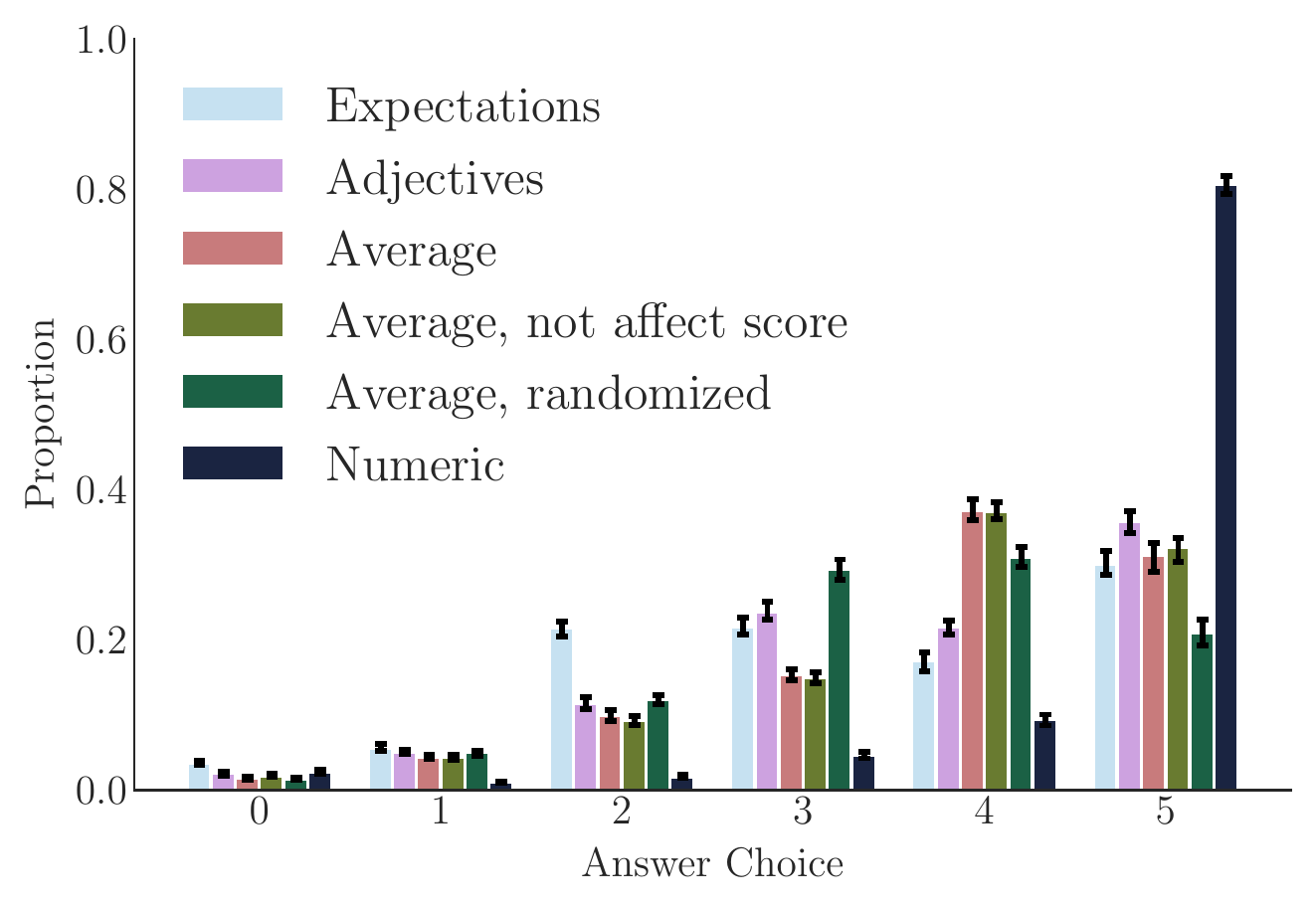}\vfill
	\caption{Rating distributions over entire time period}
	\label{fig:laborfeedbackdist}
	\end{subfigure}\hfill
	\begin{subfigure}[b]{.5\textwidth}
	\includegraphics[width=1.04\linewidth]{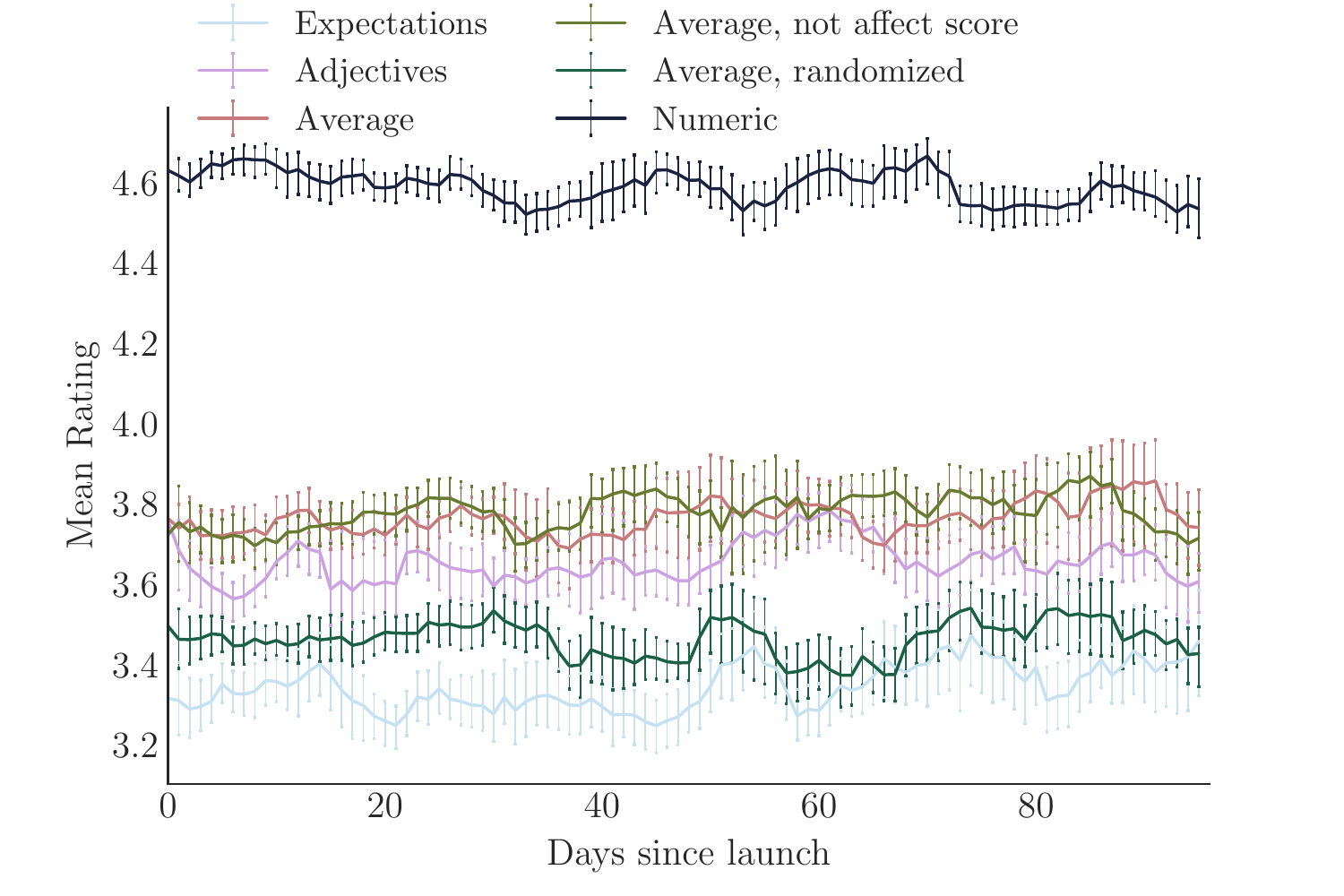}\vfill
	\caption{Mean ratings over time in a 7 day sliding window}
	\label{fig:over_time_plot}
\end{subfigure}\hfill

\caption{Labor market test marginal rating distributions. Error bars are  $95\%$ boot-strapped confidence intervals, where the bootstrapped sampling is done at the client level. }
\end{figure}

\subsection{Verbal rating scales counter inflation}
\label{sec:marginaldist}
We start our analysis of the results by looking at the marginal rating distributions in each treatment, i.e., how many freelancers received each possible rating in each treatment cell. These distributions provide evidence that the non-numeric scales provide more dispersed and deflated ratings. Furthermore, we find that the verbal rating scales are resistant to inflation throughout the course of the experiment; {this surprising finding stands in contrast to prior work on rating inflation over time.}

\subsubsection{{Snapshot analysis of ratings}}
Figure~\ref{fig:laborfeedbackdist} shows the marginal rating distributions for each treatment group, and Table~\ref{tab:counts} contains the mean treatment response in each group, for the entire experiment period. There is a large and significant difference between the rating distribution from the numeric scale and each of the other treatment groups. Each treatment cell is different from each of the others at $p<10^{-100}$ using the Kolmogorov-Smirnov two-sample test, except for the \textit{Average} and \textit{Average, not affect score} cells, where $p>0.1$. While the \textit{Numeric} treatment ratings follow the J-curve pattern usually seen in ratings, the other treatments are far more evenly distributed as desired. Most starkly, $80.6\%$ of ratings on the \textit{Numeric} scale are $5/5$, while at most $35.8\%$ of responses on any other scale received the highest possible rating.

The substantial effect size of the difference between the \textit{Numeric} condition and the other treatments confirms our hypothesis that specific and positive-skewed scales  are an effective way to counter inflation: the answer choices presented to the rater are a first-order determinant of rating behavior. The other changes (emphasizing that the freelancer would not be affected, and randomizing the choices) have comparatively small effects.

Additional analyses are in the Appendix.  In particular, our results there demonstrate that the findings reported in this section remain essentially identical even if we use other approaches to the analysis; for example, if we sample only one job per client, if we include all valid clients (i.e., including those with more than 200 jobs submitted), or if we even include the invalid incorrectly allocated clients.

\subsubsection{{Temporal analysis of ratings}}

The above analysis provides a \textit{snapshot} view of what happens when a new question is added to the rating form. Some of the rating dispersion may be a novelty effect that decreases over time. As \citet{filippas2018reputation} emphasize, a substantial component of rating inflation in online platforms happens over time, on the order of months or even years. Here, we analyze whether ratings on the new questions inflated in the time period of the test.

We find that the rating scales do not inflate substantially. Figure~\ref{fig:over_time_plot} shows the average rating per treatment group over the 90 days after the launch of the test, in a sliding window of 7 days. There is no discernible inflation over time. It is instructive to compare the (lack of) inflationary trend to the inflation after the launch of a new numeric scale on the same platform in 2007, as reported by~\citet{filippas2018reputation}: average ratings inflated from about 3.8 stars to about 4.1 stars in the first three months after the system launched. (Note that introducing a new \textit{Numeric} question in 2018 yields immediately inflated responses, suggesting that current platform users have been conditioned to the norm of inflated ratings.)

One concern with drawing conclusions from the preceding analysis over time is that there may not be enough clients who actually submit multiple jobs during the test period, and so novelty effects may still predominate when looking at overall averages. To study this concern, we analyze the ratings given by the clients who submitted at least 10 ratings each. We then run a regression for treatment response, with a covariate indicating how many previous jobs the client had submitted during the test period. Appendix Section~\ref{onlinesuppsec:inflationovertime} has the associated table and discussion. For such high-volume clients, inflation exists but is slow: ratings may be inflated by a full point after a client has given 100 ratings.

Positively, this finding suggests that as long as new clients continue to enter the platform, ratings should remain deflated over a long time horizon.  Indeed, given that existing norms are strongly biased towards inflationary ratings (as evidenced by clients' responses to the \textit{Numeric} question), it is quite valuable to see no evidence of inflation in the verbal treatment groups within a three month period.  Of course, in principle it remains possible that over a timescale much longer than that of this test, norms would shift again towards inflated ratings.  A longer-term longitudinal analysis of this type of inflationary behavior remains an important direction for future work in this area, though of course data collection over such a long time horizon is a significant obstacle.

\subsection{Verbal rating scales {yield more informative ratings}}
\label{sec:jointdistributions}

The analysis above establishes that buyers behave substantially \textit{differently} with non-numeric rating scales than they do with the numeric scale, and in particular that such scales produce deflated ratings.  In this section, we establish that this change is \textit{beneficial} to the platform in terms of learning about freelancers: that higher ``quality'' freelancers indeed receive better ratings on average with the verbal scales, where ``quality'' is exogenously defined based on signals other than ratings on the given rating scale of interest.

To do this analysis, however, one needs such an exogenous signal on latent freelancer quality; such a strong signal is precisely what is missing on many online platforms with inflated rating systems. In fact, this lack of a signal, especially for new participants on the platform, is the primary motivation for our work. We provide two approaches to overcome this gap and show that indeed the verbal rating scales substantially provide more information to the platform. Our second approach, in particular, provides estimates for the joint distribution of freelancer quality and the ratings they receive in each scale.

\subsubsection{Predicting freelancer rehires}

\begin{figure}
	\centering

		\includegraphics[width=.7\linewidth]{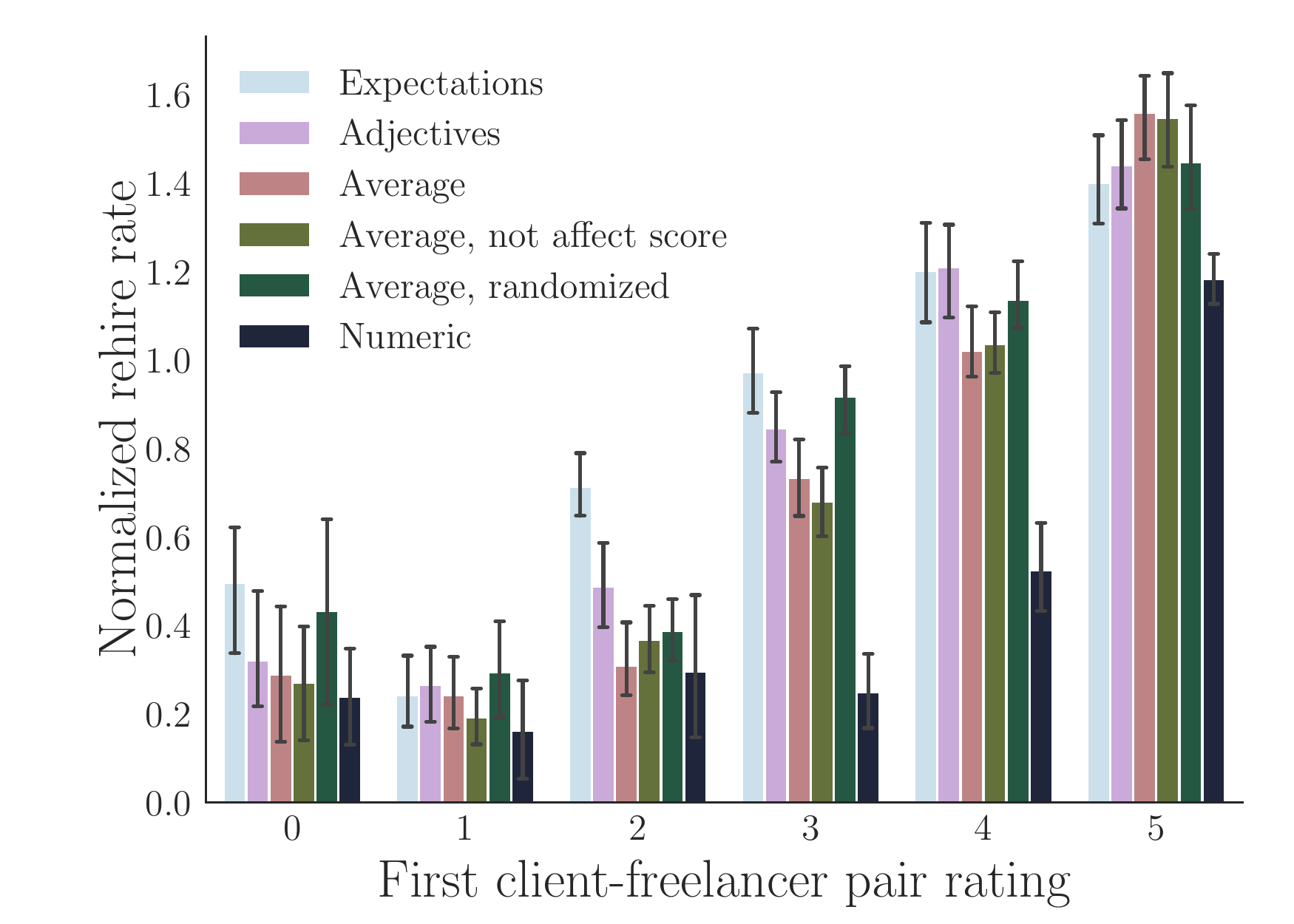}\vfill
		\caption{Likelihood that a client will rehire a freelancer during the time period of the test, given just the first rating the client gives that freelancer during the test period. Values are normalized by the overall mean rehire rate. Confidence intervals are 95$\%$ intervals with bootstrapped sampling done at the client level.}
		\label{fig:rehirerates}
\end{figure}

First, we observe that on this labor market, clients often rehire the same freelancers for jobs in the future. Consistent with the literature, we assume that a client with a more positive experience with a freelancer is more likely to return to the platform and rehire the freelancer~\citep{nosko_limits_2015}. We thus analyze whether the verbal rating scores provide more predictive power on whether a freelancer will be rehired. (This measure is not perfect, as there are others reasons that a rehire may not occur, including that the freelancer does not wish to work with the client. However, the ratings in our test are private, and so could not have directly {exerted this influenced}).

For each client-freelancer pair that completed a job during the experiment period, we consider the rating given by the client to the freelancer on the \textit{first} such completed job. Across condition cells (besides \textit{Control}), there are $125,386$ such first jobs, with $58,787$ unique clients and $110,798$ unique freelancers. We then observe whether the client-freelancer pair completed another contract during the test time period.

The results imply that {\em the verbal rating scales are substantially more informative than the numeric scale} -- even a \textit{single} rating provides more predictive power to the platform. Figure~\ref{fig:rehirerates} shows, for each condition, the likelihood that a freelancer  given a certain rating is to be eventually rehired by the same client, normalized by the overall mean rehire rate. Clients who gave a freelancer anything but a ``5'' on the numeric scale almost never rehired the freelancer. With the positive-skewed verbal scales, by contrast, there is a smoother decline of rehire rate, giving the platform finer-grained insight on whether a freelancer is likely to be rehired. Furthermore, top verbal rating scores better identify truly exceptional freelancers: for example, freelancers are $31.8\%$ more likely to be rehired after receiving the top rating in the \textit{Average} treatment than they are after receiving the top numeric score ($1.56$x and $1.18$x higher than the average rehire rate, respectively). Clients are providing more information when asked to rate freelancers on the verbal scale.

Note, however, that rehiring data during the test period does not provide enough information to construct reliable quality estimates for individual freelancers: a given freelancer typically only matches with a few unique clients and the rehire decision is itself noisy. In the next sub-section, we construct such {freelancer-level} quality estimates by looking at freelancer ratings across cells.

\subsubsection{Correlation with estimated freelancer quality}
\label{sec:fulljointdist}
\begin{figure}
	\centering
	\begin{subfigure}[b]{.5\textwidth}
		\includegraphics[width=\linewidth]{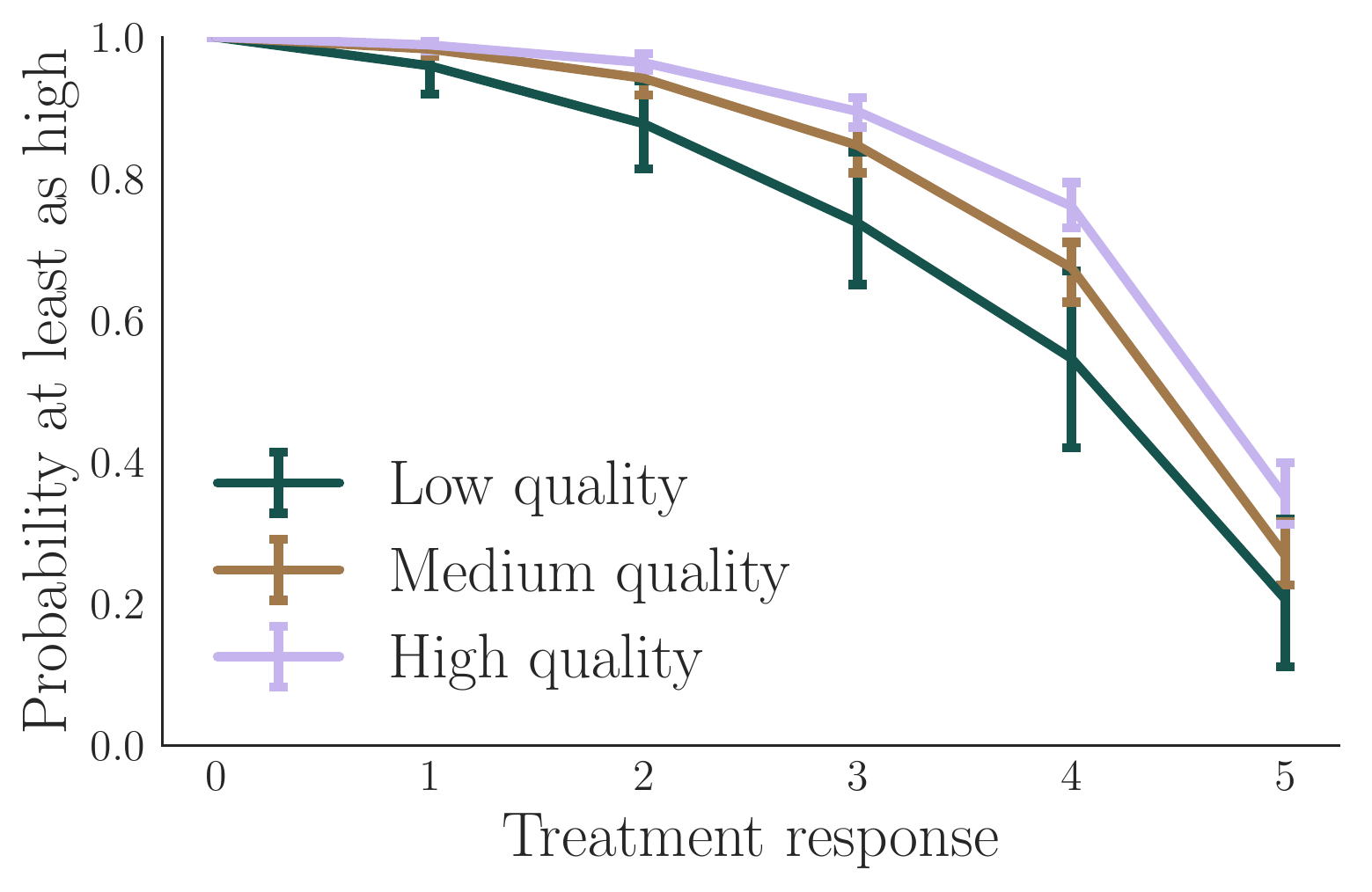}\vfill
		\caption{\textit{Average} treatment}
		\label{fig:labor_joint_treatment3}
	\end{subfigure}\hfill
	\begin{subfigure}[b]{.5\textwidth}
		\includegraphics[width=\linewidth]{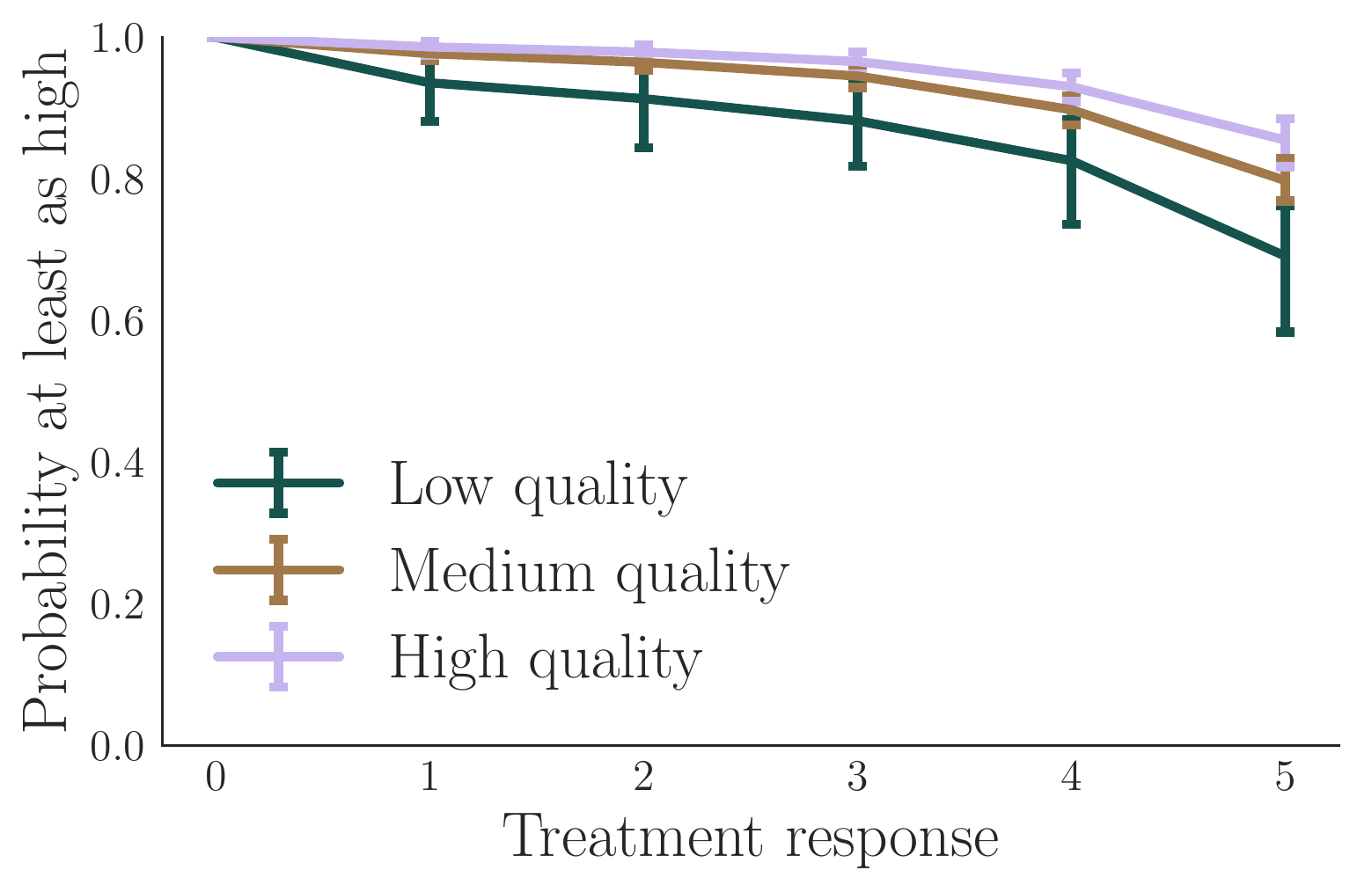}\vfill
		\caption{\textit{Numeric} treatment}
		\label{fig:labor_joint_treatment6}
	\end{subfigure}\hfill
		\caption{Joint distributions of freelancer quality vs. ratings in the \textit{Average} and \textit{Numeric} treatment cells, respectively. Low, Medium, and High quality sellers refer to those with other cell average ratings in $[0, 2), [2.5, 3.5)$ and $[4.5, 5]$, respectively. The $Y$ axis is the probability that a freelancer of a given quality receives a rating at least as high as the $X$ axis. Confidence intervals are $95\%$ intervals with bootstrapped sampling done at the client level.}
	\label{fig:jointdistributions_labor}
\end{figure}
{In this section, we estimate each freelancer's quality and use this estimate to construct a joint distribution of estimated freelancer quality and ratings under a given scale.  We then use this joint distribution to compare different designs.}

Recall that in our experiment design, a given \textit{client} is only in a single treatment cell throughout the test period. On the other hand, a given \textit{freelancer} may complete jobs with and receive ratings from clients across treatment cells. {Thus the ratings any individual freelancer} receives in different treatment cells are independent.

We can leverage this independence to construct approximate joint distributions of freelancer quality and ratings in each treatment cell as follows. For a given treatment cell, we consider all freelancers who received at least three ratings in the \textit{other} treatment cells, {and we estimate a freelancer's quality via} a simple average of these ratings. For each given treatment cell, these {estimates of quality} are exogenous with respect to the ratings received in that cell. {For each treatment cell, we then construct a joint distribution over freelancers of the rating received in that cell, and the estimated quality of that freelancer.}

We note that {given the amount of data we had available}, our estimates of these joint distributions are noisy.  The freelancer quality estimates are only from about three ratings across the various treatment cells, and responses in the cells themselves differ in meaning. In practice, a platform with access to historical performance data across a longer time-period, especially for long-lived sellers, may be able to construct more reliable estimates.

Figure~\ref{fig:jointdistributions_labor} includes two such joint distributions, for the \textit{Average} and \textit{Numeric} treatments, respectively. The Appendix contains the same joint distribution for the other treatment cells;  we also show another way to group freelancers by their average ratings, and similar patterns emerge. In all treatment cells, higher quality freelancers receive better ratings, though to varying degrees.

In the \textit{Numeric} cell, most freelancers receive high ratings independent of quality, and it may be difficult to distinguish high and medium quality freelancers. In contrast, in the \textit{Average} cell there is a larger gap between freelancers of different quality, and qualitatively one expects that this gap is beneficial in terms of learning freelancer quality.  {In this sense, the \textit{Average} cell is providing ratings that are more informative than the \textit{Numeric} cell.}

\subsection{Discussion}
\label{sec:expdiscussion}
These results suggest that there are countervailing forces to ratings inflation that can induce ratings to be more dispersed than in existing systems, by shifting how buyers interpret the scale: a platform can find large improvements over standard rating systems by explicitly defining what each rating means and positive-skewing such descriptions.  In particular, though ratings still tend positive in absolute terms in our verbal scales (over $80\%$ of freelancers receive \textit{Above Average} or better), clients seem hesitant to give most freelancers the best possible score when such a score is interpreted as truly exceptional. This deflationary effect has positive information implications for the platforms. For example, freelancers who receive such a rating are more likely to be rehired by the platform.

Furthermore, this large effect is first order and dwarfs other sources of rating variation on the labor market. For example, in the Appendix we show that rating heterogeneity across market segments is small, on the order of $0.1$ differences in means. Similarly, the treatment with randomized answer choices reveals that clients tend to pick the first choice presented more than others, but again the effect is second order.

{We conclude by noting that our qualitative assessment of the joint distribution of estimated quality and ratings in Section \ref{sec:fulljointdist} is somewhat {\em ad hoc}.  Motivated by this work, in the next section we develop a {\em quantitative} approach to capture the performance gain of verbal rating scales, based on the joint distribution of estimated quality and observed ratings.  In particular, we compare rating system designs in terms of the {\em speed} at which they allow the platform to correctly rank the freelancers.}

\makeatletter{}%
\section{A framework to compare rating scales}
\label{sec:adjectives}

The preceding section establishes, through a variety of metrics, that a platform can improve the information obtained through the rating system through careful choice of the descriptions for each level of a multi-level rating scale.  This finding naturally prompts the question: is there a principled way to compare rating scale designs to find the one that is ``best'' for the platform? We now develop a framework to do so.

 In particular, we take the perspective that the platform's objective is to ensure that the ranking of sellers based on their aggregate rating score converges to the true ranking at the fastest rate possible in the number of ratings received.  We develop a stylized model to formalize this notion and use it to develop an approach to compare and optimize rating systems. The stylized model we consider has the following key elements.  We assume that {\em buyers} enter per time period and match with long-lived {\em sellers}, potentially at varying rates according to the seller's quality. After the match, the buyer rates the seller; the rating behavior depends on the rating scale (answer choices, e.g., the adjectives or other answer phrasings in Table~\ref{tab:labortreatments}). The platform's design levers are the answer choices making up the rating scale, and the \textit{scores} it attaches to those adjectives.  We leverage this stylized model to propose an approach to maximize the rate of convergence (in a large deviations sense) of the estimated ranking based on sellers' aggregate scores, to the true underlying ranking based on sellers' qualities. We apply this methodology to our labor market data (presented in Section~\ref{sec:laborapplyresults}), and to a synthetic dataset collected through Amazon Mechanical Turk (presented in the Appendix).

\subsection{Model}
\label{sec:modeladjectives}

Our model is constructed to emphasize the platform's learning rate of participants through its rating system.  It is deliberately stylized so that we can derive a relatively straightforward method to compare and optimize rating scales.  The key components are as follows.%

{\bf Time}.  Time is discrete: $k = 0, 1, 2, \ldots$.

{\bf Sellers}.  The system consists of a unit mass of sellers, each associated with a quality \textit{type} $\theta$, which is (initially) unknown to the platform. We assume $\theta$ is drawn independently and uniformly at random from a finite and totally ordered set $\Theta$, with $|\Theta| = M$. We use $\theta_i$ to denote the $i$th element of $\Theta$ within this order, for $0\leq i < M$.

In addition, each seller has an {\em aggregate score}, described further below; we let $x_k(\theta)$ denote the aggregate reputation score of the seller of type $\theta$ at time $k$.

{\bf Rating accumulation}.  Sellers accumulate ratings over time by matching with buyers. At each time step, each seller matches with at most a single buyer. We make one key assumption that drives the accumulation of ratings: in particular, that sellers of higher quality are more likely to be matched. %
We consider an analysis that is asymptotic in the number of ratings received by sellers and so we model this visibility benefit by assuming that {\em sellers of higher quality accumulate ratings at a faster rate}. In particular, we assume the existence of a nondecreasing {\em match function} $g(\theta)$, where a seller of type $\theta$ receives $n_k(\theta) = \lfloor k g(\theta) \rfloor$ matches, and thus ratings, up to time $k$.

Our approach to modeling rating accumulation is stylized in at least two important ways.  First, the matching function is artificial: in general, sellers are more likely to match when they have a higher {\em observed} aggregate score, and there may be other heterogeneity as well. Second, we suppose all sellers have the same age: at time $k$, all sellers have had $k$ opportunities to match with buyers.  In reality, of course, sellers have different ages on a marketplace. {These choices allow us to develop a clean approach to optimizing the learning rate; we discuss the consequences further in our empirical investigation in Section \ref{sec:laborapplyresults}.}

{\bf Ratings}.  How are sellers rated? After each match, the seller receives a rating in the form of a multiple choice question answered by the buyer. The platform makes two decisions at the beginning when designing this question.  First, the platform chooses a rating scale $Y$, composed of an ordered set of answer choices $y\in Y$ from which the buyer will choose.%

Second, whenever a seller receives a rating $y \in Y$, the platform gives the seller a {\em score} $\phi(y) \in [0,1]$ depending only on the rating received. The score represents the relative positivity assigned to a rating $y$: high scores positively affect the seller's aggregate score (as we formally describe below).  Platforms often %
use equally spaced scores when translating rater's choices to an aggregate score {(e.g., the choice ``5 stars'' translates to a numeric $5$ when averaging, the choice ``4 stars'' translates to a numeric $4$ when averaging, etc.)}, but we allow the possibility that this choice should also be optimized.

At each rating opportunity (i.e., match made), the seller receives a rating from the set $Y$, and we assume that this rating depends only on the true quality of the seller. In particular, we presume that, given scale $Y$, the probability a seller of type $\theta$ receives a rating $y$ is $\rho(\theta, y | Y)$, with corresponding cumulative mass function $R(\theta,y| Y)$ reflecting the probability a seller of type $\theta$ receives a rating $y$ or higher. In other words, the scale $Y$ induces a joint distribution between the underlying seller quality and the rating choices buyers make. We make the natural assumptions that $R(\theta,y| Y)$ is strictly increasing in $\theta$ and strictly decreasing with $y$.

Let $y_0(\theta), y_1(\theta), y_2(\theta), \ldots$ be the sequence of ratings received by the seller of type $\theta$.  The {\em aggregate score} up to time $k$ of this seller is the average score from ratings received:
\begin{equation}
\label{eq:aggscore}
 x_k(\theta) =  \frac{1}{n_k(\theta)} \sum_{\ell = 0}^{n_k(\theta)} \phi(y_\ell(\theta)).
\end{equation}
(We presume $x_0(\theta) = 0$ for all $\theta$.)  Since $\phi(y) \in [0,1]$ for all $y$, the score $x_k$ also lies in $[0,1]$.

This rating behavior is also a strong assumption. In particular, it does not capture heterogeneity across raters (the types of sellers a buyer matches with may correlate with the buyer's rating behavior in general).  Including such heterogeneity is a direction for future work, and we discuss it further in the conclusion Section~\ref{sec:conclusion}; indeed, empirical identification of such heterogeneity presents an interesting practical challenge.

{\bf System state}.  We represent the state of the system defined above by a joint distribution $\mu_k(\Theta, X)$, which gives the mass of sellers of type $\theta \in \Theta$ with aggregate score $x_k(\theta) \in X$ at time $k$. Throughout our model presentation, we describe the system model as one emerging from interactions between individual buyers and sellers. However, we assume a unit mass of sellers (and some mass of buyers), and so all such descriptions should be viewed as illuminating the evolution of a joint distribution $\mu_k (\Theta, X)$ of the types of sellers on the platform and their current scores. To formally describe the evolution of $\mu_k$, let $E_k = \{ \theta : n_k(\theta) = n_{k-1}(\theta) + 1 \}$.  These are the sellers who receive an additional rating at time $k$; for all $\theta \in E_k^c$, $n_k(\theta) = n_{k-1}(\theta)$.  Next, for each $x, x' \in [0,1]$, define $\chi_k(x, x',\theta | Y, {\phi})$ as:
\[
{\chi_k(x, x',\theta | Y, \phi) =} \{ y : n_k(\theta)x - n_{k-1}(\theta)x' = \phi(y) \}.
\]
The set $\chi$ describes the rating(s) a seller of type $\theta$ at time $k$ with aggregate score $x'$ can receive to transition to aggregate score $x$.  We then have:
\[ \mu_{k+1}(\Theta, X) = \int_{E_k} \int_0^1 \int_X \sum_{y \in \chi_k(\theta, x, x' | Y, \phi)} \rho(\theta, y | Y)  dx \mu_k(d\theta, dx') + \int_{E_k^c} \int_X \mu_k(d\theta, dx'). \]
It is straightforward but tedious to check that the preceding dynamics are well defined, given our primitives.

{\bf Platform objective}.  We assume that the platform wants the ranking of sellers by observed aggregate score to reflect the underlying true quality ranking as closely as possible.

Formally, given $\theta_1 > \theta_2$, define $P_k(\theta_1, \theta_2)$ as follows:
\begin{equation}
\label{eq:Pk}
P_k(\theta_1, \theta_2) = \mu_k(x_k(\theta_1) > x_k(\theta_2)|\theta_1, \theta_2) - \mu_k(x_k(\theta_1) < x_k(\theta_2)|\theta_1, \theta_2).
\end{equation}
This expression captures the ``errors'' made by the ranking according to observed score.  In particular, when $\theta_1 > \theta_2$ but $x_k(\theta_1) < x_k(\theta_2)$, the aggregate score ranking swaps the ordering of sellers $\theta_1$ and $\theta_2$.  Thus, a good rating system has large $P_k(\theta_1, \theta_2)$.

We consider the problem of maximizing the following objective, a scaled version of {\em Kendall's $\tau$ rank correlation} between the estimated ranking of sellers and the true ranking:
\begin{equation}
\label{eq:objective}
W_k = \frac{2}{M(M-1)}\sum_{\theta_1 > \theta_2 \in \Theta} P_k(\theta_1, \theta_2)
\end{equation}
The coefficient ensures that ${W}_k$ remains bounded even as $M$ increases. This objective depends on the model primitives $R$ (rater behavior) and $g$ (matching rates), as well as the platform's decisions $Y$ (levels) and $\phi$ (score).

We note that, in this model, the goal of the rating system is to accurately rank sellers by quality. Another approach may be to directly optimize for total platform revenue or aggregate welfare. This approach would require primarily optimizing which matches occur, a focus of many other works. We optimize information gained \textit{per} match, for which finding the true ranking of sellers is a reasonable objective.  One observation in support of this choice is that the ``deliverable'' for the ratings team in an online platform company is typically an accurate rating that can be an input to models used by other teams throughout the organization. Further, in a model where matching rates are exogenously determined by quality, we conjecture that optimizing other objectives (accuracy and revenue) would produce qualitatively similar results.

{\bf Learning $R(\theta, y | Y)$}. We note that our approach to quantify the learning rate, which we develop in the next subsection, requires the platform to learn the rating joint distribution $R(\theta, y | Y)$ for various potential rating scales $Y$. In our analysis of the labor market experiment, we provide one approach to do so --- using the data collected during the experiment itself to estimate seller qualities $\theta$. In practice, a platform with access to more historical data may rely on estimates of $\theta$ for a group of ``known'' sellers; e.g., these may be long-lived sellers on the platform. The platform can then test new rating scales, and use the resulting data to estimate $R$.

\subsection{Quantifying design performance via convergence rate}
\label{sec:optimizingadjectives}

As noted above, the platform has two design choices it makes: the set of rating levels $Y$, and the score function $\phi$.  We now consider an approximate approach to maximization of the objective $W_k$, by appropriate choice of $Y$ and $\phi$.

No single choice of $Y$ and $\phi$ can simultaneously optimize $W_k$ for all $k$: some designs may be effective in separating the best sellers from the worst quickly, but then never separate all sellers. Further, as long as $\phi(y)$ is strictly increasing, then because $R(\theta, y)$ is strictly increasing in $\theta$, we have, for all $\theta_1\neq \theta_2$, and all choices of $Y$ and $\phi$:
$\lim_{k \to \infty} P_k(\theta_1, \theta_2) = 1$. Using the bounded convergence theorem we conclude that $\lim_{k \to \infty} W_k = 1 $, independent of the design choice $Y$ and $\phi$.  Thus {\em any} design asymptotically  -- with enough ratings -- recovers the true ranking of sellers.

For these reasons, we focus on maximization of the {\em rate} at which $W_k$ converges; we call the design $(Y, \phi)$ that maximizes this rate \textit{optimal}.  We use a {\em large deviations} approach to study the rate of convergence  \citep{dembo_large_2010}, following other works that adopt this approach~\citep{glynn_large_2004,garg_binary_2018}.

We {have} the following result.

\begin{restatable}{theorem}{lemhatW}
	\label{lem:hatW}
	\begin{equation}
	\label{eq:rW}
	r \triangleq - \lim_{k \to \infty} \frac{1}{k} \log (1- {W}_k) = \min_{0 \leq i < M} \inf_{a \in \bbR} \left \{ g(\theta_{i+1}) I(a|\theta_{i+1}) + g(\theta_i) I(a|\theta_i)\right\}
	\end{equation}
	where
	$I(a|\theta) = \sup_z \{ za - \Lambda(z|\theta) \}$, and $\Lambda(z|\theta) = \log \sum_{y \in Y} {\rho(\theta, y | Y)} \exp( z \phi(y))$ is the log moment generating function of a single rating given to seller of type $\theta$.
\end{restatable}

The proof follows from standard results in large deviations analysis and is in the Appendix.

 The expression in \eqref{eq:rW} is called the {\em large deviations rate} for $\b{W}_k$. The theorem shows that $W_k(\theta_1, \theta_2) \to 1$ exponentially fast, and provides an explicit relationship between our choice of $Y$ and $\phi$, and the corresponding exponent. In other words, $1 - W_k = \mathcal{O}(e^{-rk}\text{poly}(k))$.

 Two rating systems can be compared by their respective learning rates: for each design, simply calculate their rates and then compare. The rate function can be calculated numerically given $R (\cdot | Y)$, $\phi$ and $g(\theta)$: in particular, observe that $\sup_z \{ za - \Lambda(z|\theta) \}$ is a concave maximization problem in $z$, and $\inf_{a \in \bbR} \left \{ g(\theta_{i+1}) I(a|\theta_{i+1}) + g(\theta_i) I(a|\theta_i)\right\}$ is a convex minimization problem in $a$.

Our design optimization problem is thus as follows: {\em choose $Y$ and $\phi$ to maximize the large deviations rate $r$ in \eqref{eq:rW}}. We suggest the following approach, supposing the platform has a collection of candidate scales $\{Y_p\}$. First, experiment with each scale $Y_p$ and estimate $R (\theta, y | Y_p)$ and $g(\theta)$. Then use the following  brute force approach to optimization: for each $Y_p$, choose a random, increasing set of scores $\phi(y) \in [0,1], \forall y \in Y_p$ in each iteration, and calculate the learning rate. For each candidate scale $Y_p$, run a large (exponential in $|Y_p|$) number of such iterations. Finally, choose the design $Y_p, \phi$ with the best learning rate.  {While this is a brute force optimization, we envision a platform will not be changing the rating system design very frequently, and thus computation time is not critical.}

Qualitatively,  $\phi(y)$ should be large if higher quality freelancers are much more likely to receive a rating choice of at least $y$ than are lower quality freelancers. If the rating joint distribution $R (\theta, y | Y)$ is such that no such choice $y$ exists, then the scale $Y$ will perform poorly in separating the sellers. In particular, rating scales with inflated responses are uninformative for this reason: every seller independent of quality is likely to receive the most positive rating choice, and so nothing separates low quality from high quality sellers. Similarly, a $\phi$ that does not reward sellers for receiving rare, positive ratings, is suboptimal.

Finally, we note that the two parts of the design -- scale $Y$ and score mapping $\phi$ -- differ in their visibility to raters. Scale $Y$ is presented to raters in order to induce a certain desirable rating behavior $R(\cdot | Y)$. On the other hand, a platform need not share the mapping $\phi$, which is simply a technical tool that maximally leverages aggregate rating behavior to form an internal ranking of sellers. The platform may then choose to share statistics about sellers to buyers, for example whether a seller is in the top $25\%$. The optimal information sharing procedure is a question tackled by other work~\citep{papanastasiou2017crowdsourcing,ifrach_bayesian_2017,acemoglu_fast_2017}.

\makeatletter{}%
\subsection{Application to the online labor market}
\label{sec:laborapplyresults}

\begin{figure}
	\centering
\begin{subfigure}[b]{.47\textwidth}
	\small
\begin{tabular}{l|cc}
	& \multicolumn{2}{c}{\textbf{Learning rates}}                                                                     \\
	\textbf{Condition}        & \textbf{Naive $\phi$}           & \textbf{Optimal $\phi$}       \\ \hline
	Expectations & 0.022 & 0.024 \\
	Adjectives &  0.021 & 0.027 \\
	Average & 0.023 & 0.026 \\
	Average, not affect score & 0.013 & 0.014 \\
	Average, randomized & 0.014 & 0.019 \\
	Numeric & 0.009 & 0.009 \\
\end{tabular}
\vfill
\caption{Large deviation learning rates.}
	\label{fig:learningratesbyscore}
\end{subfigure}\hfill
	\begin{subfigure}[b]{.52\textwidth}
		\includegraphics[width=\linewidth]{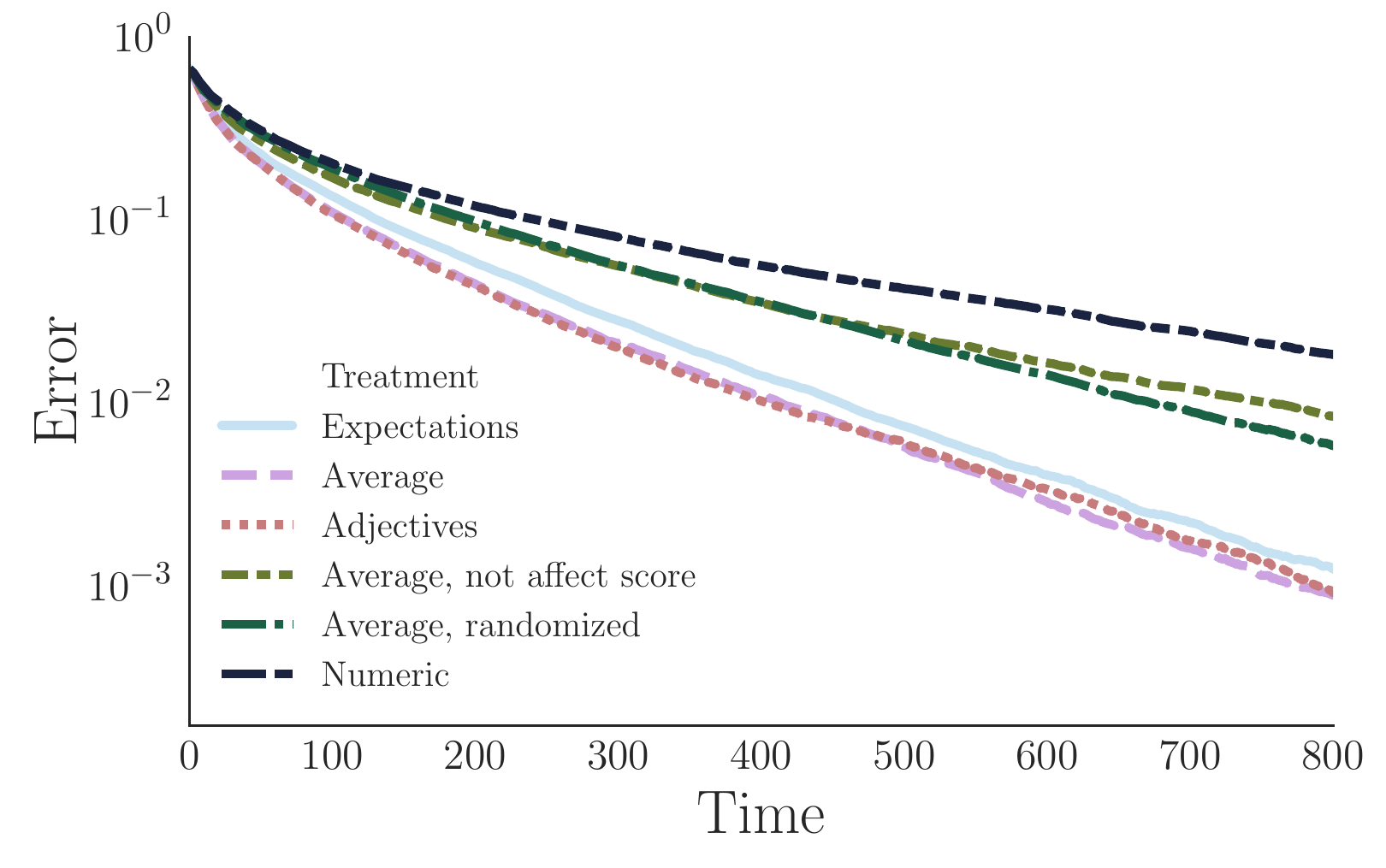}\vfill
	\caption{Simulated performance of each rating scale.}
		\label{fig:labor_simulations}
	\end{subfigure}
	\caption{We apply and test our design approach using experimental data from our online labor market. Large deviation rates are calculated using Equation~\eqref{eq:rW} and the joint distributions generated in Section~\ref{sec:jointdistributions}. Optimal for each treatment corresponds to the highest learning rate among many random scores.}
\end{figure}

We now follow the design approach outlined above using the empirical $R(\cdot | Y_p)$ calculated in Section~\ref{sec:fulljointdist} {for each scale $Y_p$ in the test}. For simplicity, we assume a uniform search rate $g(\theta) = 1$ for all $\theta$; the simulation results are robust to this choice. %

\subsubsection{{Large deviation learning rates for each design}}

First, we calculate the large deviation rates for each treatment scale, assuming {equally spaced} scores $\phi = \{0, 1, 2, 3, 4, 5\}$. All of the verbal treatments have larger learning rates than the \textit{Numeric} treatment, as shown in Figure~\ref{fig:learningratesbyscore}.

Next, we optimize the scores $\phi$ for each of the scales. Figure~\ref{fig:learningratesbyscore} also contains the learning rate achieved by the corresponding optimal score function for each treatment. It suggests that picking the correct labels on the scale is the first order determinant of the rating system's performance, while the optimal scores are second order.

The optimal scores themselves (in Appendix Table~\ref{tab:optscores}) reflect the corresponding joint distributions. For example in the \textit{Numeric} scale, only the frequency of receiving the top rating distinguishes freelancers; thus freelancers receive a lesser score ($3.45/5$, versus $\approx 4/5$) for the second-highest rating in that scale versus in the verbal rating scales. 

Note that perhaps because our estimation procedure on the joint distribution is noisy, the \textit{Average} and \textit{Average, not affect score} treatments differ in their joint distributions and learning rates, even though they have identical marginal distributions and rehire rates.

\subsubsection{{Simulated market performance of each design}}

Finally, we simulate a market for each of the treatment conditions as follows, in order to compare how the scales perform. 

In our simulation, there are 500 sellers with i.i.d.~quality in \{\textit{Low, Medium, High}\}. There are 100 buyers, each of which matches uniformly at random to a \textit{unique} seller per time period. In other words, matching is not independent across sellers, and each seller can only match once per time period; each seller matches approximately once every $5$ time periods. The buyers rate the sellers according to the joint distributions calculated in Section~\ref{sec:fulljointdist}. Ratings are converted to scores according to the optimal score function for each treatment. All sellers enter the market at time $k = 0$ and do not leave. After each time period, the sellers are ranked according to their average scores. The true ranking of sellers (i.e., \textit{Low $<$ Medium $<$ High}) is also constructed. We then calculate the Kendall's $\tau$ rank distance (not counting sellers tied according to true quality) between the two lists.

Figure~\ref{fig:labor_simulations} shows the mean (across many simulations) ranking errors over time for each treatment system as described. The plot and corresponding learning rates for each treatment demonstrate that even though large deviations rates are an asymptotic quantity, they effectively predict the performance of each rating scale even for small horizons. The \textit{Numeric} treatment in particular learns the ranking of sellers at a much slower rate than do the other mechanisms, both in terms of learning rate and simulated performance.

In Appendix Figure~\ref{fig:simulatedperfappendix}, we show other simulations and analyses as robustness checks. First, we show performance over time when each seller independently leaves the market with probability $.01$ at the end of each time $k$, with a new seller with no reputation score taking her place; such entry and exit does not affect the comparative performance of each rating scale. Robustness to such entry and exist further suggests that designed scales will outperform others when only sellers' most recent ratings are used, in order to facilitate and reflect seller improvement, cf.~\citep{aperjis2010optimal}. 

Next, we compare learning with {equally spaced} vs.~optimal $\phi$; as suggested by the learning rates, calculating an optimal $\phi$ has a small but noticeable effect in performance. 

Finally in Appendix~\ref{sec:mturkrepeatanalysis}, we repeat the analysis for a synthetic setting on Mechanical Turk that demonstrates the utility of our methods for survey contexts beyond ratings on online platforms. We find that superficially similar scales may perform dramatically differently in a way that is not a priori knowable before conducting an experiment and calculating learning rates. We evaluate performance of each treatment scale on new data not used for scale optimization and find that performance improvements can transfer to a deployment. 

\makeatletter{}%
\section{Conclusion and discussion}
\label{sec:conclusion}
In this work, we study the the design of informative rating systems. We demonstrate through a field test on a large online labor platform that there can be substantial benefit to changing answer choices and question phrasing in a rating scale.  In particular, we observe that (1) it is possible to choose a design of the verbal descriptions attached to answer choices present in the rating system that lead to deflated ratings, and (2) that these ratings are much more informative than ratings obtained in standard numeric rating systems.  Motivated by this finding, we develop a technical framework to compare and {design} the scales by properly choosing the answer choices available to raters and the mapping of these choices to scores.  We show that applying this framework can lead to designs that appear to substantially outperform {\em ad hoc} choice of the rating scale. We believe this work provides a foundation for a much more systematic approach to the design of rating systems, and that it has direct practical guidance for platforms to build more informative systems.  

\subsection{Challenges, opportunities, and limitations}

\paragraph{Fraud in online reviews and ratings} Our results establish that verbal rating scales can effectively counter \textit{behavioral} norms and implicit pressures to provide maximally positive ratings. However, such scales do not \textit{constrain} rater behavior and thus are ineffective against inflation caused by ratings fraud, in which the seller may fake transactions and rate themselves. There is a large literature on the prevalence of such fraud and techniques to detect it~\citep{hu2011fraud,akoglu2013opinion,zhang2013trust,hooi2016birdnest,luca2016fake}. Our work is complementary to such approaches and is most appropriate for markets where such fraud is not the first order determinant of rating inflation, such as on the labor market in question (as evidenced by the informative verbal scales).

\paragraph{Horizontal vs.~vertical differentiation} In many markets, buyers have heterogeneous preferences over sellers, i.e., there is {\em horizontal} differentiation. Our work assumes that there is an underlying ranking of sellers, i.e., that sellers are {\em vertically} differentiated (at least among the buyers who match with a given seller). If the matching process segments the market, then vertical differentiation may dominate within each segment. For example, price and location may segment the market on AirBnB such that only consumers with similar preferences match with and rate a given host. Then, a single rating scale can be used if rating behavior is similar across segments (recall that we show in Section~\ref{sec:expdiscussion} that scales perform similarly across segments in our labor market). In markets with substantial horizontal differentiation (even given that a buyer and seller have matched), however, the methods in this work can be used either (1) with comparatively objective questions (e.g., rating cleanliness or timeliness), where there may be vertical differentiation; or (2) alongside techniques that detect heterogeneous preferences and create ``virtual'' market segments (when feasible). In particular, our work is applicable wherever rating scales are used under the assumption of some degree of vertical differentiation.

\paragraph{International markets} One potential difficulty in implementing verbal rating scales is that they must be designed in each language, and people in different cultures may interpret the same scale differently. This difficulty is especially acute as modern online platforms often operate globally. We note that verbal scales provide an opportunity as well as a challenge. There is variation across cultures in numeric rating systems, both for response scales in general and for online platforms in particular~\citep{chen_response_1995,hamamura_cultural_2008,koh_online_2010,wang_problem_2015}. In the status quo, the platform is left without a mechanism through which it can equalize the rating distributions. On the other hand, with verbal rating scales, if comparable ratings across regions are important, the platform can choose scales for each region that provide comparable rating distributions.

\paragraph{Using ratings for search and matching}  Another potential concern is that at the moment the answers to these questions are not used on the platform for other functions, such as search or matching.  As illustrated by~\cite{filippas2018reputation}, some inflation for private questions is to be expected once the answers start affecting freelancers, even if freelancers cannot directly identify the client who provided any specific rating (e.g., if freelancers start asking for higher ratings on this question). We cannot completely eliminate this concern and leave the question for future work after a treatment condition is chosen to be implemented permanently on the platform.  However, note that the marginal rating distributions and relation to how often freelancers are rehired by a client are extremely similar for the \textit{Average}, and \textit{Average, not affect score} conditions: either the clients already are aware that the question they are answering is a test question that will not affect freelancers, or this additional information does not substantially influence how clients rate beyond the deflating effects of the answer choices in question. 

\paragraph{Switching to a new rating system} One final practical concern with introducing a new rating system with drastically different behavior is that it may be challenging from a data integrity perspective: how can old, inflated ratings be compared to the new ratings, and how can models throughout the platform be adjusted to handle both types of ratings? In some settings, such as our large online labor market, the new system can simply co-exist with the status quo: multiple questions can be asked in the rating form until enough time has passed with the new system such that older, inflated data is no longer useful. This approach adds friction in the form of additional work for clients, but it may be a price worth temporarily paying for finer resolution information. On other platforms where typically users are only asked one question, the transition may be more challenging. However, such platforms have begun experimenting with their rating systems.  Furthermore, ratings data typically grows stale, as sellers enter and exit the platform or improve over time, and platforms often only use the last few ratings given to each seller~\citep{aperjis2010optimal}. Such factors mitigate the cost of switching to a new system.

\subsection{Future work}

\paragraph{Platform goals} Rating systems should reflect the specific goals and context of a platform. On some platforms, it may be undesirable to attempt to fully recover the ranking of sellers. For example, platforms that provide a commodified experience (e.g., ridesharing or delivery services) may only care about identifying bad actors on the platform. In this setting for example, asking buyers to rate sellers against the ``average'' may place undesirable, excessive pressure on sellers to attempt to distinguish themselves. Rather, the platform should potentially encourage raters to give good ratings unless something truly bad happened. Platforms in practice already do this; for example, when a passenger rates a driver 4 stars out of 5, Lyft describes the choice as ``OK, could have been better.'' 

The methods in this work are most appropriate in settings where true differentiation exists between items or sellers (whether this differentiation is under the control of sellers or not), and it is desirable to identify and encourage comparatively high performers. Future work should closely examine the practical and theoretical relationships between a platform's informational goals and its rating system design. We take a theoretical step in this direction in our work on designing binary rating systems~\citep{garg_binary_2018}.

\paragraph{Dynamic design and combating inflation over time} Even with our non-inflated rating scales, it may be possible that over time norms shift so that again ratings become inflated.  In this event, optimization of comparison points and rating scales may need to be a {\em dynamic process} for a platform. An important direction for future research is to consider a dynamic equilibrium view of rating system design. In particular, online marketplaces and platforms should aim to design systems that are naturally robust to inflation yet provide an good user experience. A complete picture should consider how search, buyer rating behavior, and seller behavior may change in response to changes in the rating system.  Capturing these short- and long-run equilibrium effects remain important challenges.  We believe our work provides an important empirical and theoretical building block in this direction, by suggesting that the meaning raters attach to levels of a scale can substantially influence the quality of information obtained by the platform.
\FloatBarrier
\bibliographystyle{apalike} %
\bibliography{bib}

\pagebreak\newpage

\appendix
\makeatletter{}%
\section{Further analysis of the labor market test}
\label{sec:applabor}
In this section, we report more detail from the test on the online labor market. For much of this section, we analyze a subset of the jobs: some job covariate information is missing in what was given to us by the labor market. We have full covariate data for 100438 jobs (out of 184172).

\subsection{Verifying randomization in allocation of clients}

As noted in Section~\ref{sec:allocation} of the main paper, there was a bug in the allocation code such that $1,086$ clients were assigned to different treatment cells upon submissions of different jobs. Since this could potentially create contamination between our cells, we disregard these clients in our analysis. Here we make sure that neither this bug nor any other affected experimental validity by checking the distribution of client covariates across the treatment cells. We do so as follows.

We have a set of \textit{job} level covariates for a subset of the jobs: \textit{hourly rate of job} (if applicable), \textit{total cost of project if not hourly} (if applicable), \textit{previous number of closed jobs by client at time of job}, \textit{previous spend by client at time of job}, \textit{value of the job} (4 options), \textit{Tier 1 category} (12 options), \textit{Tier 2 category} (88 options), and \textit{expertise level} (3 options). The first four are continuous covariates, and the last 4 are categorical covariates.

For each client, we sample one of that client's jobs and associate the client with that job's covariates. Then we run tests of independence for the samples of each covariate across the treatment cells. Across a variety of tests and all covariates, the results are consistent with the randomization being valid.
\begin{itemize}
	\item For each continuous covariate, using the Kruskal-Wallis H-test for independent samples on all the treatment groups together, the null hypothesis that the population median of all of the groups are equal is not rejected, with $p>.9$.
	\item Similarly, for each continuous covariate, using the one way ANOVA F test,  the null hypothesis that all the treatment groups have the same population mean is not rejected, with $p>.2$.
	\item For each categorical covariate, we run the chi-squared test of independence of variables in a contingency table, which tests whether the observed frequencies of values is independent of the treatment group. The null hypothesis is not rejected with $p>.1$, for each covariate.
\end{itemize}

These tests are consistent with fact that the allocation of \textit{valid} clients we used for analysis across treatment cells was truly random. Note that these tests do not check whether the \textit{invalid} clients (which we threw out) are similar to the \textit{valid} clients. Invalid clients are more likely to be higher volume clients, as those who submitted many jobs during the test period provided more chances for the bug to manifest.

\subsection{Robustness against high volume clients and allocation bug}

Recall that in the main text we further threw out the 7 clients who submitted more than 200 jobs during the test period (``heavy users''). However, the following may still be the case: idiosyncratic rating behavior of medium-volume clients (over 50 or 100 jobs submitted) may be driving the difference in behavior between treatment cells. Here we show that this is not the case, as well as the fact that throwing out the 7 heavy users was not consequential. We further show that including the clients who were thrown out due to the allocation bug does not materially affect results.

\begin{figure}[tbh]
	\centering

	\begin{subfigure}[b]{0.32\textwidth}
		\includegraphics[width=\textwidth]{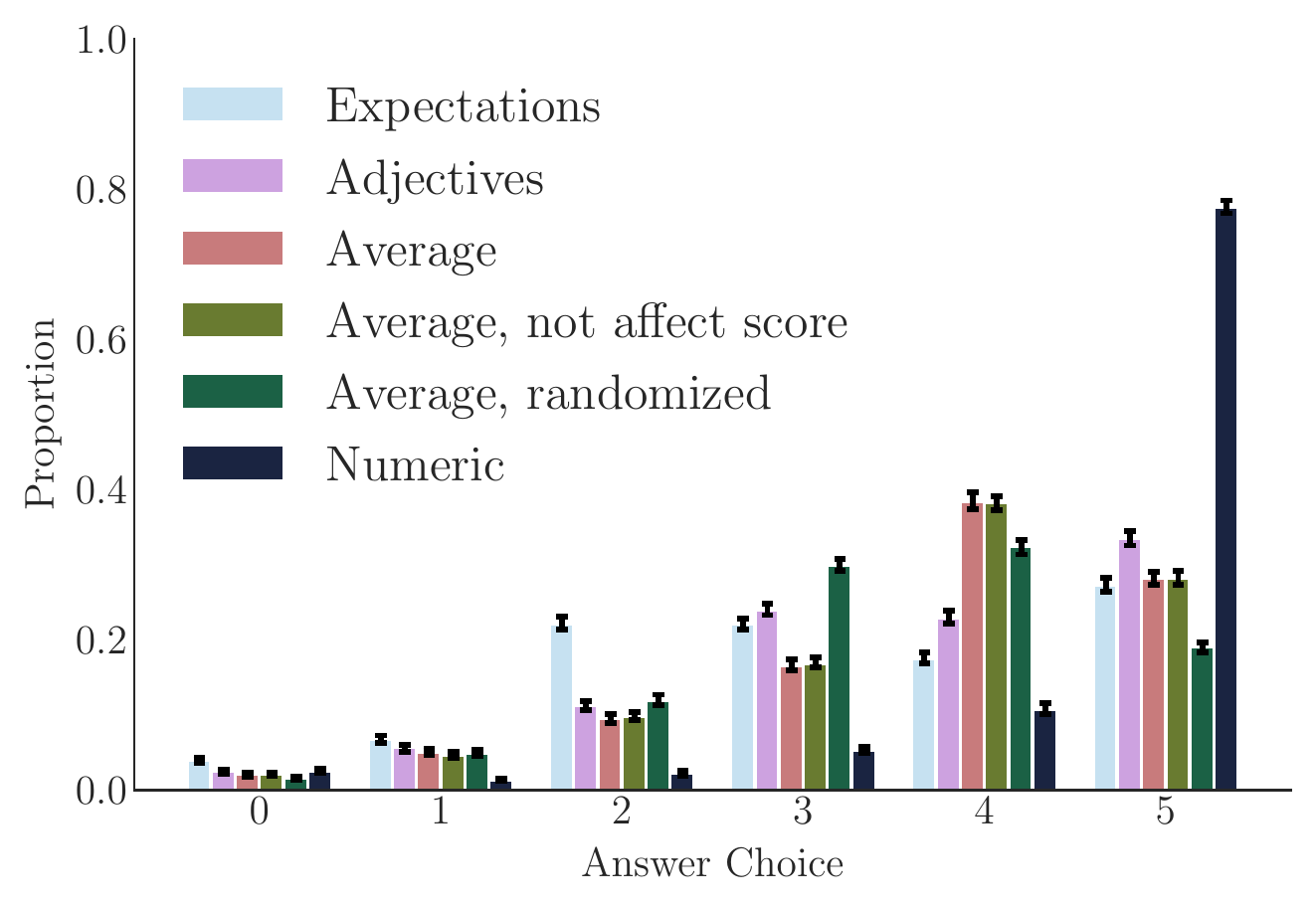}
		\caption{Sampling 1 job per client}
	\end{subfigure}
	\begin{subfigure}[b]{0.32\textwidth}
		\includegraphics[width=\textwidth]{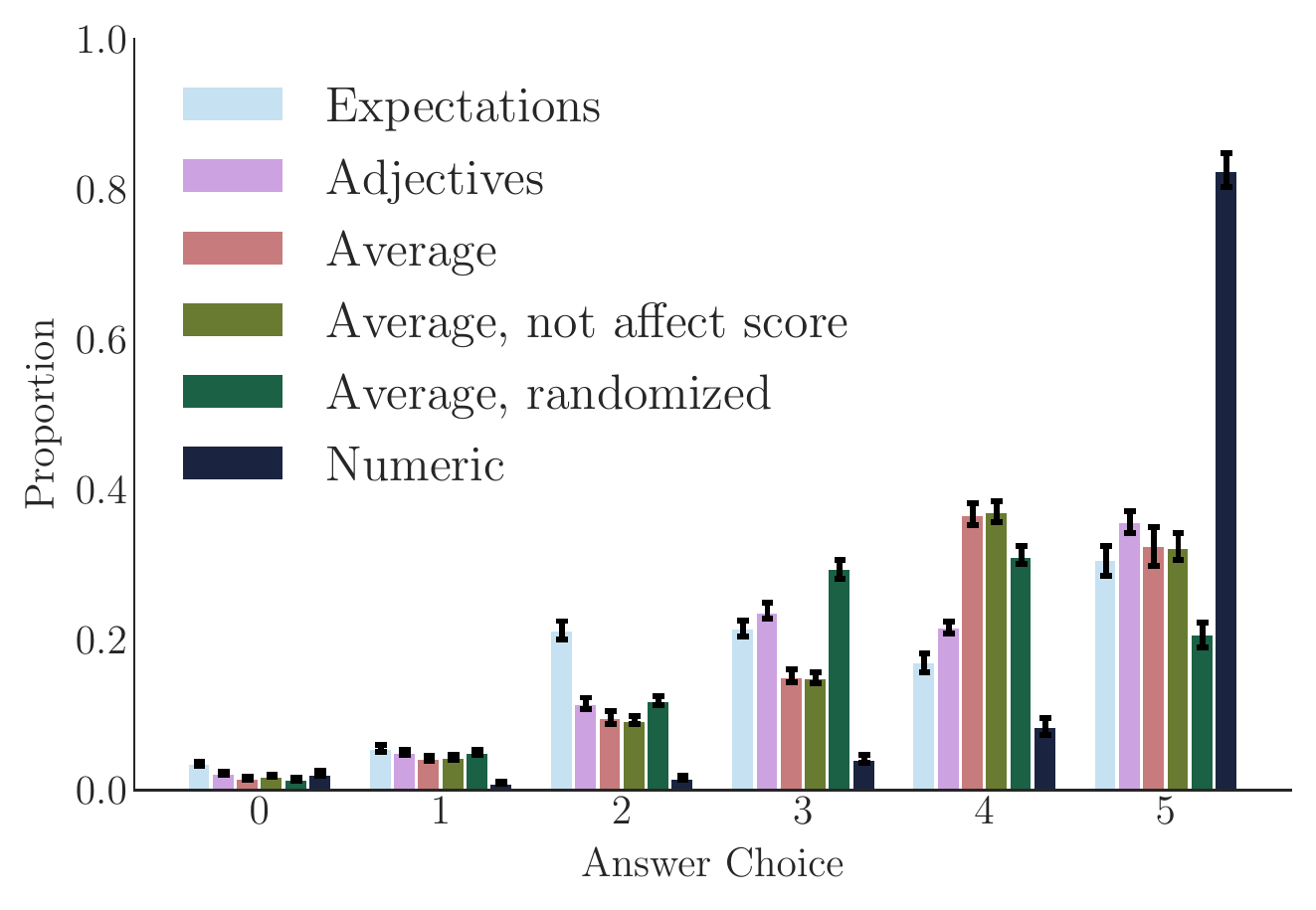}
				\caption{Using all valid clients and jobs}
	\end{subfigure}
	\begin{subfigure}[b]{0.32\textwidth}
	\includegraphics[width=\textwidth]{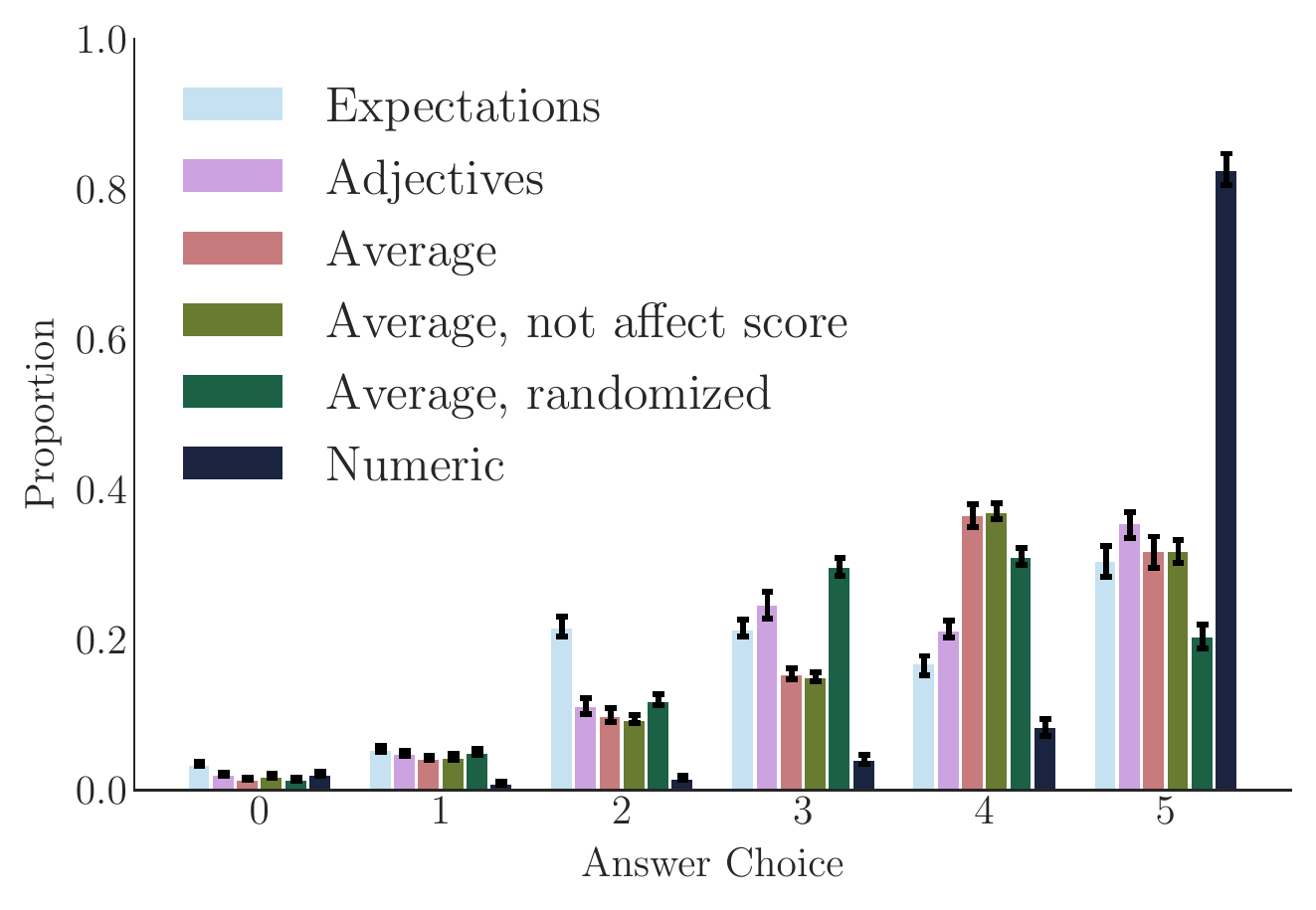}
	\caption{Using all clients and jobs}
\end{subfigure}
\caption{Rating distributions for different client sampling techniques. As in the main text, the confidence intervals are $95\%$ bootstrapped confidence intervals, with bootstrapped sampling at the client level.}
\label{fig:clientsampling}
\end{figure}

In Figure \ref{fig:clientsampling}, we plot the rating distributions when only sampling 1 job/client, including 7 clients excluded for submitting at least 200 jobs during the test period, and using all jobs and clients (even incorrectly allocated clients). The mean treatment responses are also included. Results are similar.

\begin{table}[tbh]
	\hskip-25pt
	\small
	\begin{tabular}{lcccc}
		\textbf{Data sampling policy:}                          & From main text & One job per client & With outlier clients & All clients, even incorrectly allocated \\ \hline
		\multicolumn{1}{l|}{\textit{Expectations}}              & 3.339                 & 3.243              & 3.354                            & 3.350                                   \\
		\multicolumn{1}{l|}{\textit{Adjectives}}                & 3.650                 & 3.597              & 3.650                            & 3.651                                   \\
		\multicolumn{1}{l|}{\textit{Average}}                   & 3.763                 & 3.687              & 3.788                            & 3.774                                   \\
		\multicolumn{1}{l|}{\textit{Average, not affect score}} & 3.777                 & 3.693              & 3.777                            & 3.771                                   \\
		\multicolumn{1}{l|}{\textit{Average, Randomized}}       & 3.465                 & 3.438              & 3.463                            & 3.458                                   \\
		\multicolumn{1}{l|}{\textit{Numeric}}                   & 3.594                 & 4.534              & 4.635                            & 4.639
	\end{tabular}
\caption{Average treatment responses under different data policies}
\end{table}

\subsection{Regressing treatment response with treatment cell and other covariates}
\label{sec:ratingheterogeneinty}
We regress the treatment response with treatment cell and all of our job covariates (except tier 2 category, which had 88 unique values and is a more granular version of tier 1 category). (Note: to maintain full rank, each categorical covariate is encoded such that one of the levels is missing, except for treatment cell, and there is no intercept. As a result, the treatment cell coefficients cannot be interpreted as treatment means -- they are the treatment means conditional on a specific value of each of the categorical covariates and of 0 for the continuous variables). Further note that for simplicity, we only include one set of interaction terms: treatment cell vs. the number of previous treatment responses. Finally, note that the displayed standard errors are cluster-robust standard errors where each client is a cluster, to take into account that ratings given by the same client are correlated. We learn several things from this regression, displayed in Table~\ref{tab:firstregression}:
\begin{itemize}
	\item There is some heterogeneity in ratings across the job covariates, but on the order of .1 points on the average rating. This heterogeneity is dwarfed by the differences between the treatment cells, especially the numeric vs. non-numeric treatments. This relative lack of heterogeneity further supports that the differences between the mean treatment responses are not due to randomness caused by some types of jobs being more present in some treatment groups than others.%

	\item We can directly measure the effect of the number of previous jobs during that testing period a given client has submitted, i.e., estimate the inflation that will result over time as clients submit additional jobs.

	 From the table below, each additional job a client has submitted raises the treatment response for the \textit{Expectations} and the \textit{Averages} treatments, on the order of $.008$ to $.014$ points per previous response. At this rate, these coefficients suggest that only after giving 100 ratings would a client inflate ratings by an average of between .8 and 1.4 points. The \textit{Numeric} treatment cell does not further inflate substantially. %

\end{itemize}

\begin{table}[tbh]
	\tiny
	\hskip-25pt
	\begin{center}
		\begin{tabular}{lclc}
			\toprule
			\textbf{Dep. Variable:}                                 & treatment-response & \textbf{  R-squared:         } &      0.128   \\
			\textbf{Model:}                                         &        OLS         & \textbf{  Adj. R-squared:    } &      0.128   \\
			\textbf{Method:}                                        &   Least Squares    &  \textbf{  Log-Likelihood:    } & -1.6001e+05     \\
			\textbf{No. Observations:}                              &       100438       & \textbf{  AIC:               } &  3.201e+05   \\
			\textbf{Df Residuals:}                                  &       100406       & \textbf{  BIC:               } &  3.204e+05   \\
			\textbf{Df Model:}                                      &           31       & \textbf{                     } &              \\
			\bottomrule
		\end{tabular}
		\begin{tabular}{lcccccc}
			& \textbf{coef} & \textbf{std err} & \textbf{z} & \textbf{P$>$$|$z$|$} & \textbf{[0.025} & \textbf{0.975]}  \\
			\midrule
			\textbf{treatment\_cell[1]}                             &       3.0596  &        0.062     &    49.052  &         0.000        &        2.937    &        3.182     \\
			\textbf{treatment\_cell[2]}                             &       3.3965  &        0.063     &    53.862  &         0.000        &        3.273    &        3.520     \\
						\textbf{treatment\_cell[3]}                             &       3.4516  &        0.062     &    55.353  &         0.000        &        3.329    &        3.574     \\
			\textbf{treatment\_cell[4]}                             &       3.4414  &        0.062     &    55.796  &         0.000        &        3.321    &        3.562     \\
			\textbf{treatment\_cell[5]}                             &       3.1887  &        0.062     &    51.379  &         0.000        &        3.067    &        3.310     \\
			\textbf{treatment\_cell[6]}                             &       4.3745  &        0.062     &    70.044  &         0.000        &        4.252    &        4.497     \\
			\textbf{value\_group[T.lv]}                             &       0.1031  &        0.034     &     2.998  &         0.003        &        0.036    &        0.170     \\
			\textbf{value\_group[T.mv]}                             &       0.0206  &        0.034     &     0.601  &         0.548        &       -0.047    &        0.088     \\
			\textbf{value\_group[T.vlv]}                            &       0.2920  &        0.032     &     9.061  &         0.000        &        0.229    &        0.355     \\
			\textbf{category\_group[T.Admin Support]}               &      -0.0591  &        0.046     &    -1.281  &         0.200        &       -0.150    &        0.031     \\
			\textbf{category\_group[T.Customer Service]}            &      -0.1070  &        0.081     &    -1.320  &         0.187        &       -0.266    &        0.052     \\
			\textbf{category\_group[T.Data Science \& Analytics]}   &       0.1177  &        0.050     &     2.354  &         0.019        &        0.020    &        0.216     \\
			\textbf{category\_group[T.Design \& Creative]}          &       0.1077  &        0.042     &     2.581  &         0.010        &        0.026    &        0.189     \\
			\textbf{category\_group[T.Engineering \& Architecture]} &       0.1235  &        0.058     &     2.122  &         0.034        &        0.009    &        0.238     \\
			\textbf{category\_group[T.IT \& Networking]}            &       0.1277  &        0.049     &     2.595  &         0.009        &        0.031    &        0.224     \\
			\textbf{category\_group[T.Legal]}                       &       0.0643  &        0.061     &     1.047  &         0.295        &       -0.056    &        0.185     \\
			\textbf{category\_group[T.Sales \& Marketing]}          &      -0.0869  &        0.045     &    -1.920  &         0.055        &       -0.176    &        0.002     \\
			\textbf{category\_group[T.Translation]}                 &       0.0405  &        0.060     &     0.676  &         0.499        &       -0.077    &        0.158     \\
			\textbf{category\_group[T.Web, Mobile \& Software Dev]} &       0.0940  &        0.042     &     2.256  &         0.024        &        0.012    &        0.176     \\
			\textbf{category\_group[T.Writing]}                     &      -0.1158  &        0.044     &    -2.638  &         0.008        &       -0.202    &       -0.030     \\
			\textbf{expertise\_tier[T.Expert/Expensive]}            &       0.1465  &        0.020     &     7.276  &         0.000        &        0.107    &        0.186     \\
			\textbf{expertise\_tier[T.Intermediate]}                &       0.0582  &        0.018     &     3.306  &         0.001        &        0.024    &        0.093     \\
			\textbf{hr\_charge}                                     &    1.376e-05  &     2.16e-06     &     6.376  &         0.000        &     9.53e-06    &      1.8e-05     \\
			\textbf{fp\_charge}                                     &     3.64e-05  &     6.73e-06     &     5.409  &         0.000        &     2.32e-05    &     4.96e-05     \\
			\textbf{log(1 +client\_prev\_spend)}                &      -0.0069  &        0.006     &    -1.156  &         0.248        &       -0.018    &        0.005     \\
			\textbf{log(1 +num\_prev\_asg)}                     &      -0.0177  &        0.010     &    -1.769  &         0.077        &       -0.037    &        0.002     \\
			\textbf{treatment\_cell[1]:\# prev. treatment resps. by client }                    &       0.0080  &        0.004     &     2.042  &         0.041        &        0.000    &        0.016     \\
			\textbf{treatment\_cell[2]:\# prev. treatment resps. by client }                    &      -0.0043  &        0.006     &    -0.675  &         0.500        &       -0.017    &        0.008     \\
			\textbf{treatment\_cell[3]:\# prev. treatment resps. by client }                    &       0.0085  &        0.003     &     2.850  &         0.004        &        0.003    &        0.014     \\
			\textbf{treatment\_cell[4]:\# prev. treatment resps. by client }                    &       0.0141  &        0.003     &     5.468  &         0.000        &        0.009    &        0.019     \\
			\textbf{treatment\_cell[5]:\# prev. treatment resps. by client }                    &       0.0024  &        0.005     &     0.485  &         0.628        &       -0.007    &        0.012     \\
			\textbf{treatment\_cell[6]:\# prev. treatment resps. by client }                    &       0.0010  &        0.004     &     0.246  &         0.806        &       -0.007    &        0.009     \\
			\bottomrule
		\end{tabular}
		\begin{tabular}{lclc}
			\textbf{Omnibus:}       & 11064.189 & \textbf{  Durbin-Watson:     } &     1.911  \\
			\textbf{Prob(Omnibus):} &    0.000  & \textbf{  Jarque-Bera (JB):  } & 15421.891  \\
			\textbf{Skew:}          &   -0.876  & \textbf{  Prob(JB):          } &      0.00  \\
			\textbf{Kurtosis:}      &    3.785  & \textbf{  Cond. No.          } &  9.60e+04  \\
			\bottomrule
		\end{tabular}
		\caption{OLS Regression Results with covariate for previous number of treatment responses}
		\label{tab:firstregression}
	\end{center}
\end{table}

\subsection{More on inflation over time}
\label{onlinesuppsec:inflationovertime}
The interpretations above {suffer from selection bias}: the set of clients who submit 10 jobs in the test period are a different cohort than those who submit fewer. This effect is partially captured by the term containing the previous number of client assignments. %
To address this issue, we repeat the regression in Table~\ref{tab:firstregression}, limiting the analysis to those clients who have more than ten treatment responses during the test period (all of which have the job covariates). The table is ommitted; the coefficients for inflation over time are largely the same. %

\begin{figure}[tbh]
	\centering
		\includegraphics[width=.7\linewidth]{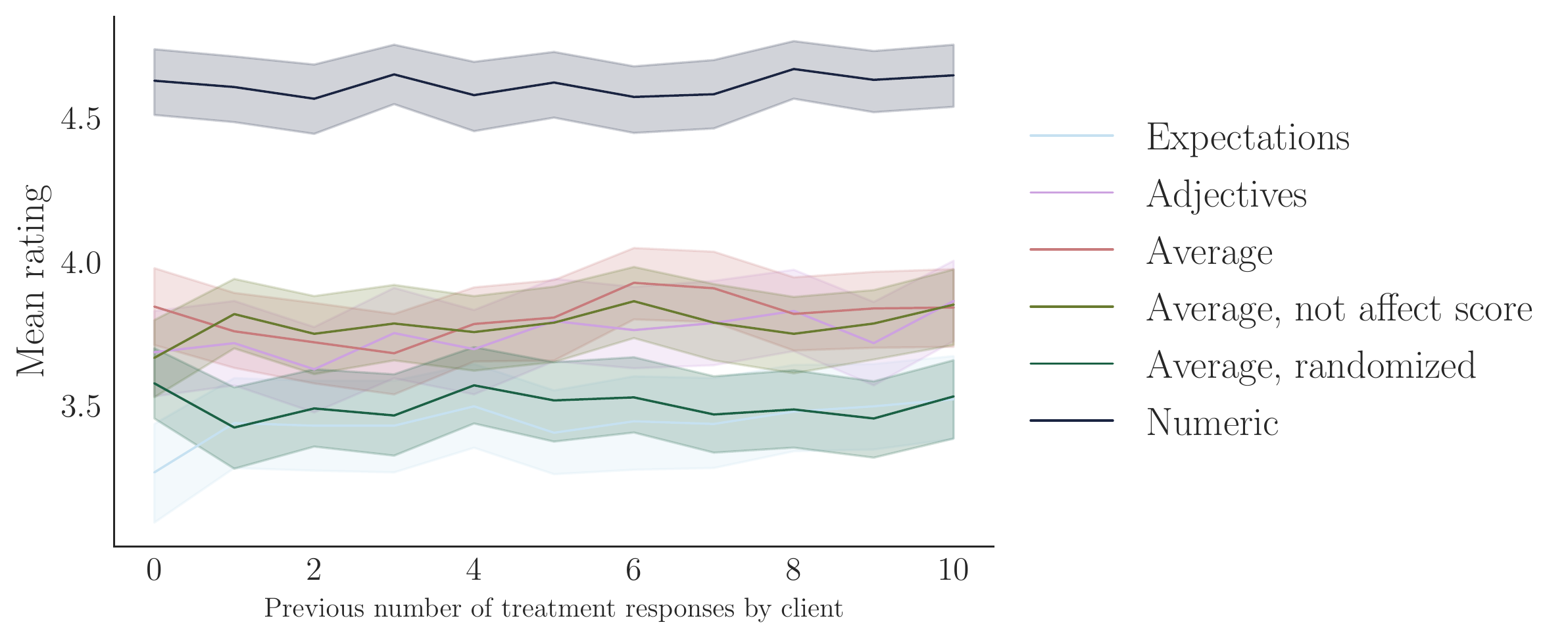}\vfill
				\caption{Mean ratings for each treatment cell by the number of previous treatment responses given during the test period. Error bands are bootstrapped $95\%$ confidence intervals.}
		\label{fig:over_numberratings_plot}
\end{figure}

To further help visualize (the relative lack of) inflation over the number of submitted ratings, Figure~\ref{fig:over_numberratings_plot} shows the mean ratings for each treatment cell by the number of previous treatment responses given during the test period. As the plot has no covariate data, we use the first ten responses for all 2145 clients who submitted at least 10 ratings during the test period. Clients are not substantially more likely to give more positive ratings on their 10th rating during the test than they give on their first rating.

\subsection{Analysis of cell with randomized order of answer choices}

The \textit{Average, Randomized} contained the same question and answer choices as the \textit{Average} condition, but the choices were presented in a random order. If the raters read all the answer choices and pick the most applicable one, then this condition would have returned a rating distribution identical to that of the \textit{Average} condition. However, it does not. Furthermore, the \textit{location} of the chosen choice would be distributed uniformly, i.e., the rater should pick the choice presented first as much as she picks other choices. We find this not to be the case: the first answer choice presented to the rater is picked $6806/26978 = 25.2\%$ of the time. The second through sixth answer choices are picked $17.3\%, 14.7\%, 14.3\%, 13.9\%$, and $14.5\%$ of the time each, respectively.

This phenomenon suggests that (a) a small percentage (up to $10-13\%$) of raters do not read the answer choices at all and simply select the first answer choice, and (b) many raters start reading from the first presented choice and select the first one that approximately describes their experience. Our test design cannot disambiguate between these (or other plausible) explanations.  Nevertheless, this effect is second-order relative to the overall finding that more descriptive scales are substantially more informative than numeric scales, and the \textit{Average, Randomized} treatment results are comparable to those of other verbal scales.

\subsection{Design approach using labor market data}

\begin{figure}[tbh]
	\centering
	\begin{subfigure}[b]{.4\textwidth}

		\includegraphics[width=\linewidth]{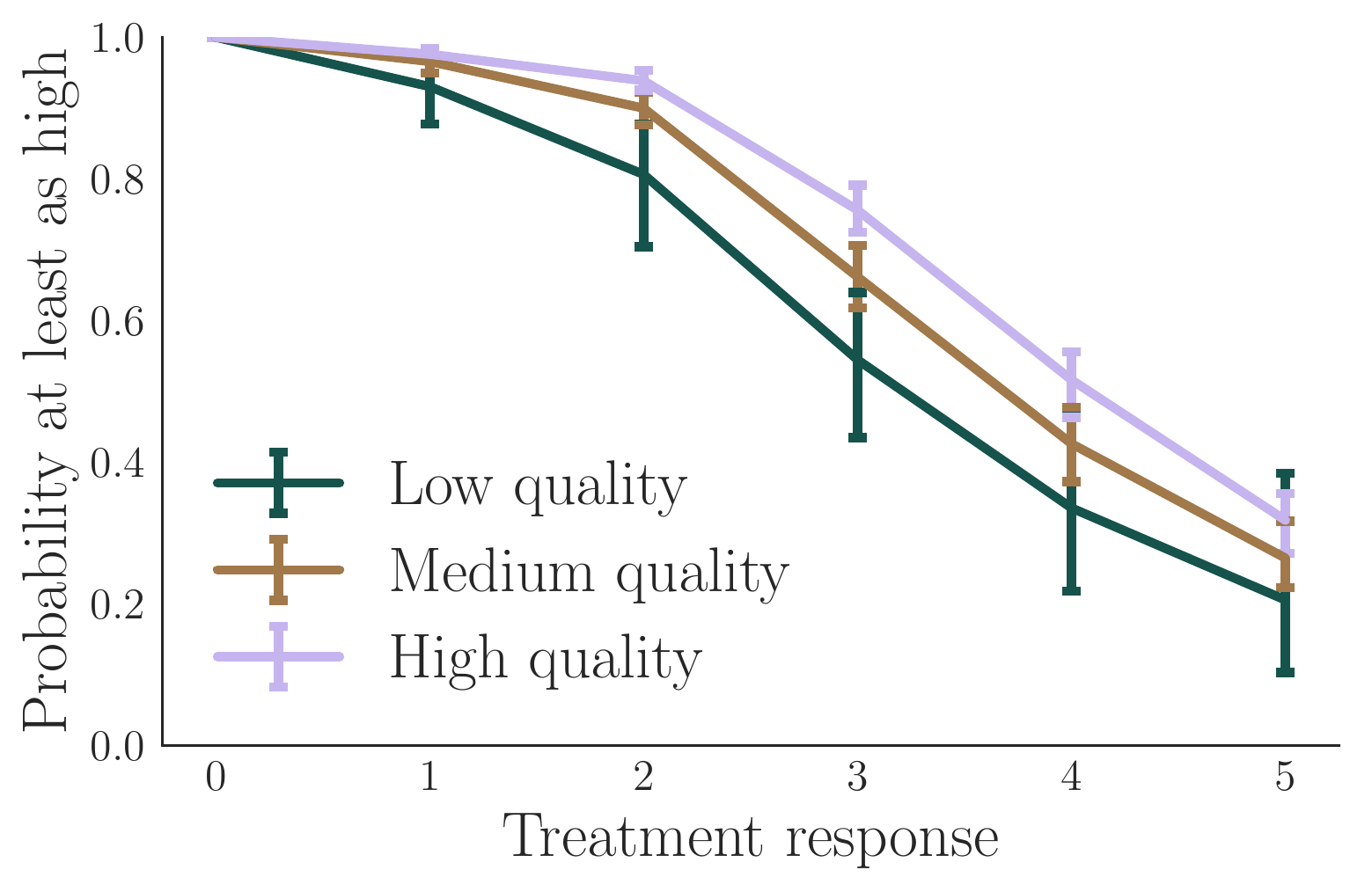}\vfill
		\caption{\textit{Expectations}}
		\label{fig:labor_joint_treatment1}
	\end{subfigure}
	\begin{subfigure}[b]{.4\textwidth}
		\includegraphics[width=\linewidth]{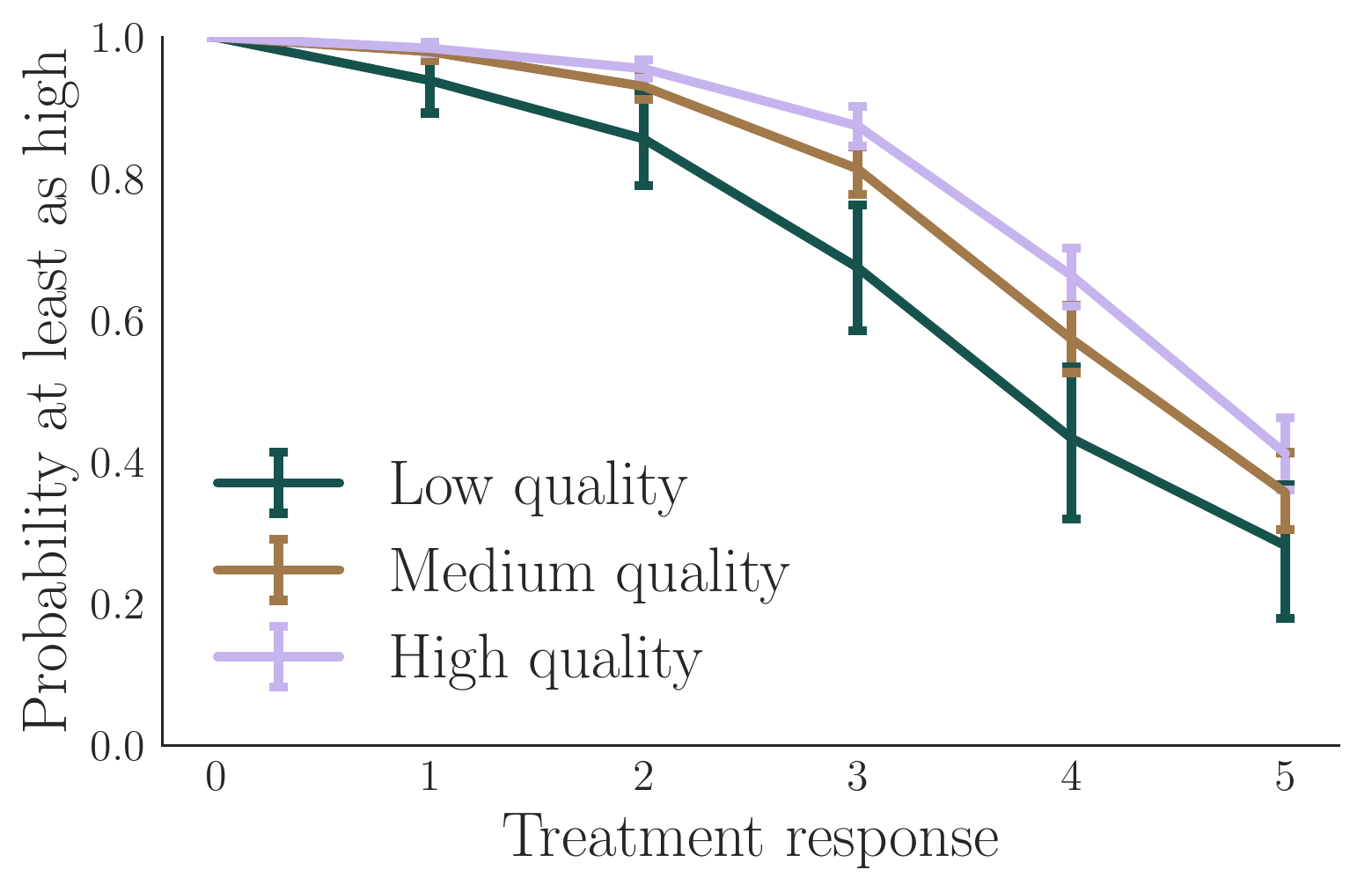}\vfill
		\caption{\textit{Adjectives}}
		\label{fig:labor_joint_treatment2}
	\end{subfigure}\hfill

	\begin{subfigure}[b]{.4\textwidth}
	\includegraphics[width=\linewidth]{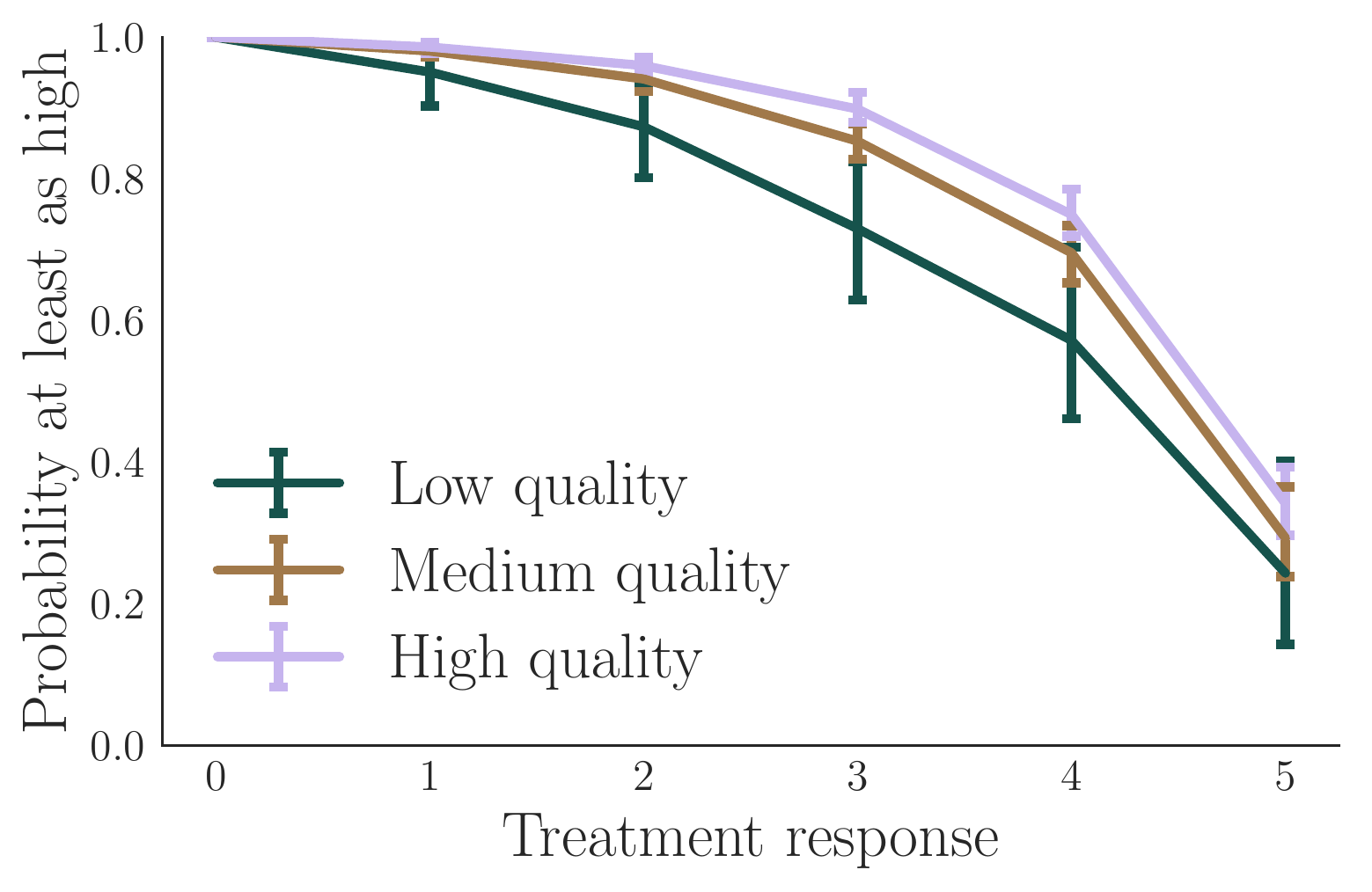}\vfill
	\caption{\textit{Average, not affect score}}
	\label{fig:labor_joint_treatment4}
\end{subfigure}
\begin{subfigure}[b]{.4\textwidth}
	\includegraphics[width=\linewidth]{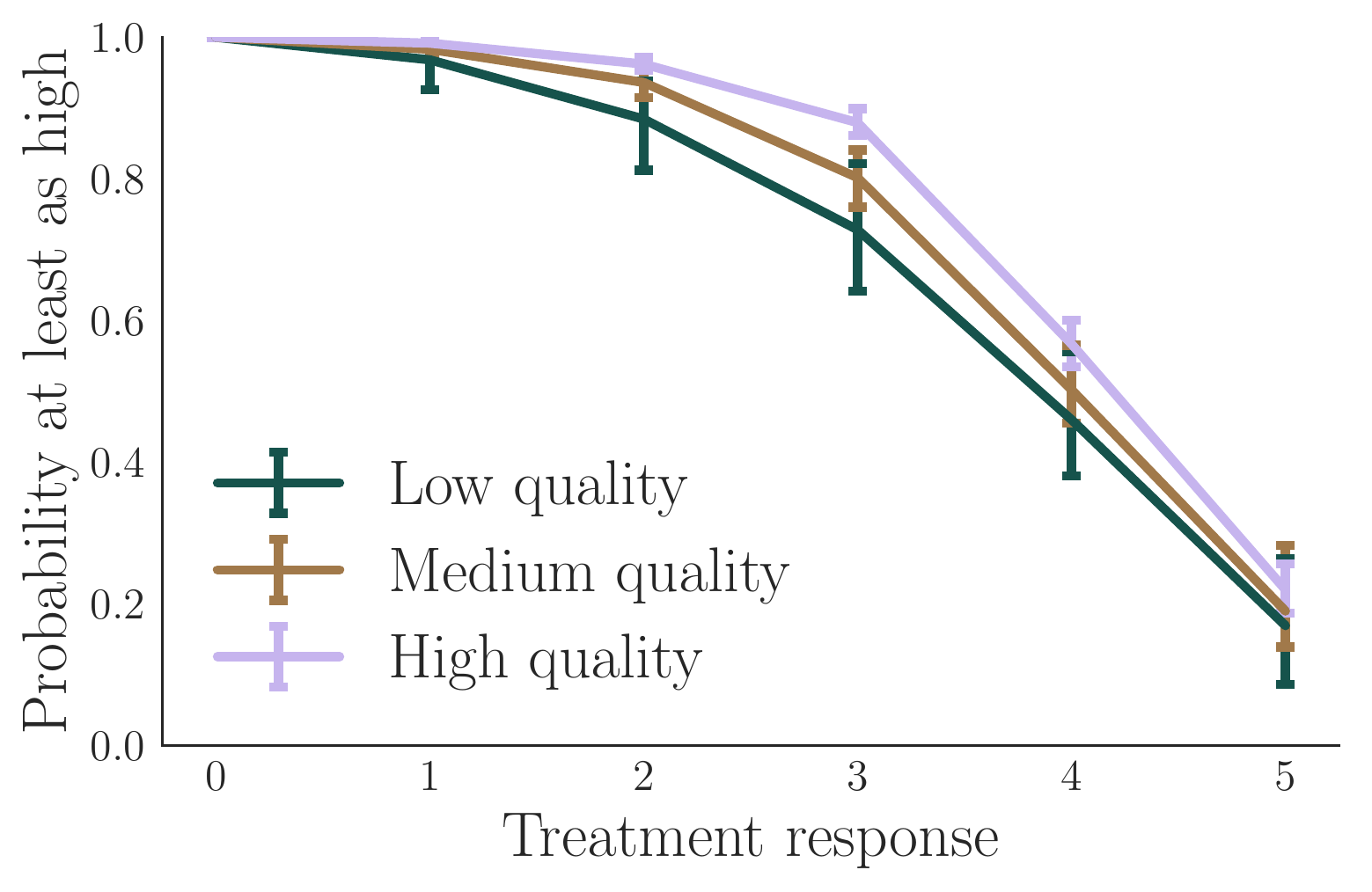}\vfill
	\caption{\textit{Average, randomized}}
	\label{fig:labor_joint_treatment5}
\end{subfigure}\hfill

	\caption{Joint distributions of freelancer quality vs. ratings in the other treatment cells. Low, Medium, and High quality sellers refer to those with other cell average ratings in $[0, 2), [2.5, 3.5)$ and $[4.5, 5]$, respectively. %
	}

	\label{fig:jointdistributions_labor_app}
	\end{figure}

\begin{figure}[tbh]
	\centering
	\begin{subfigure}[b]{.48\textwidth}
	\includegraphics[width=\linewidth]{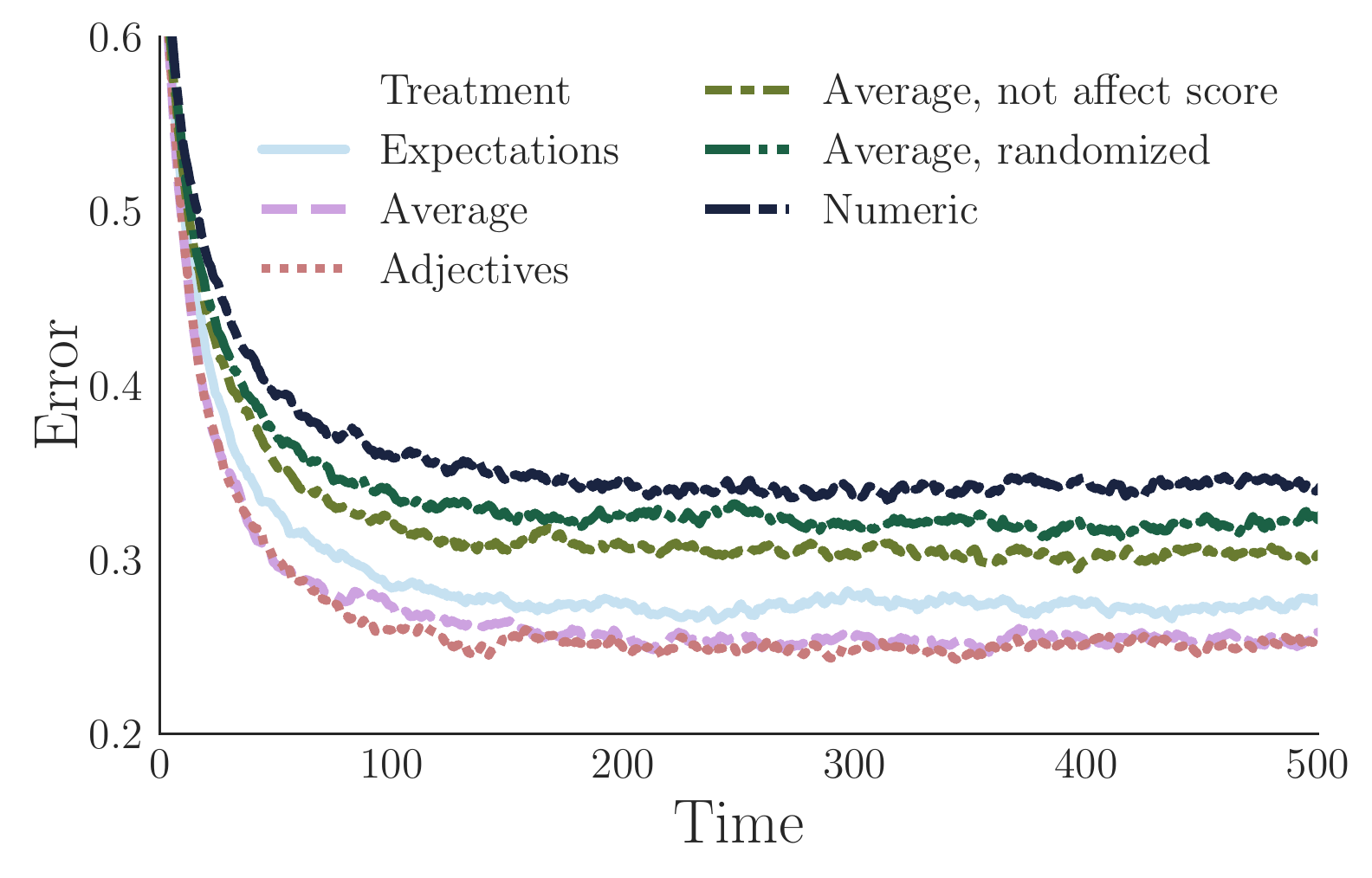}\vfill
	\caption{With optimal $\phi$ and probability of exit of $0.01$.}
	\label{fig:labor_simulationsdeath}
\end{subfigure}
	\begin{subfigure}[b]{.5\textwidth}
		\includegraphics[width=\linewidth]{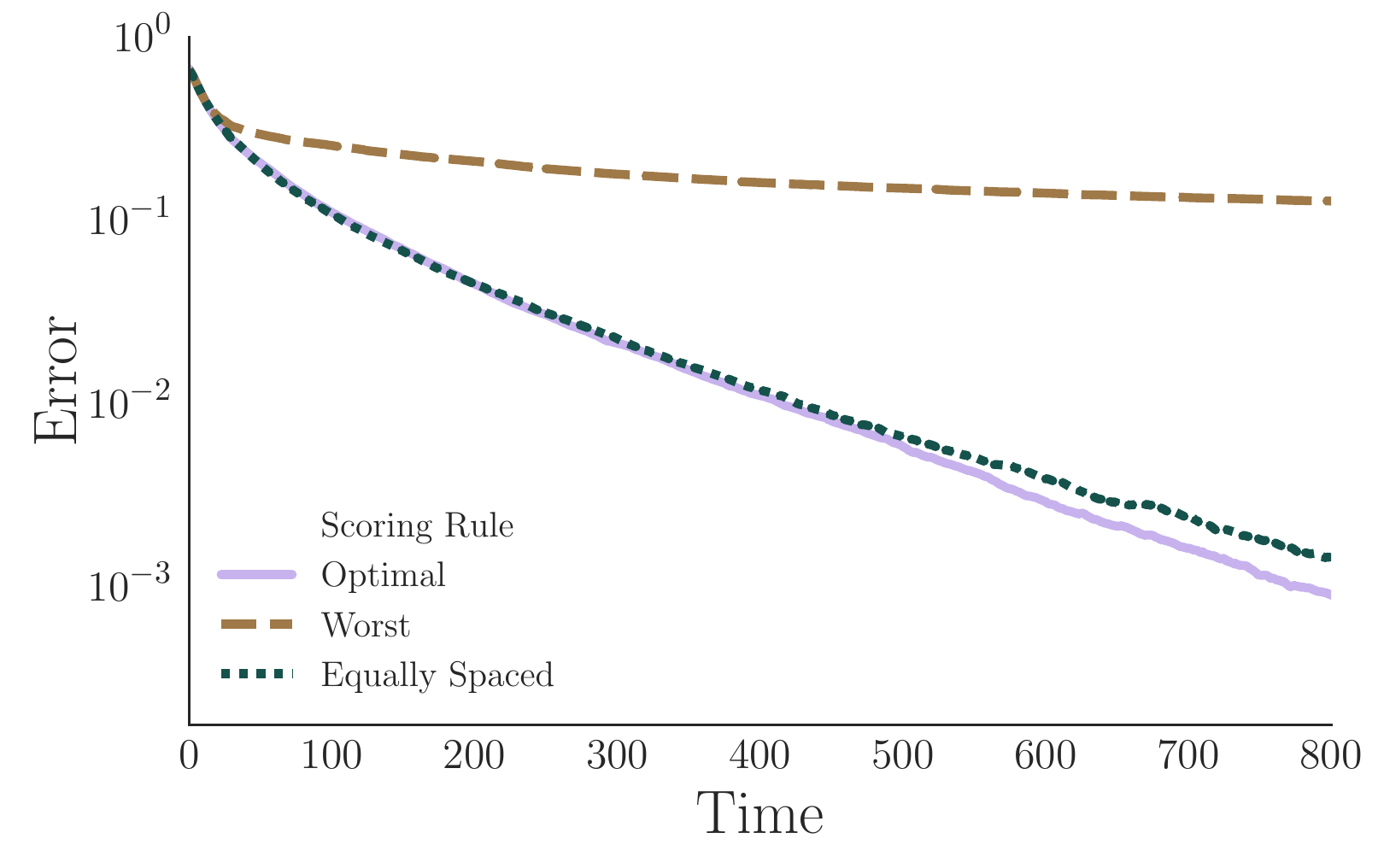}\vfill
		\caption{\textit{Average} treatment with different scoring rules}
		\label{fig:labor_simulationstr2}
	\end{subfigure}\hfill

	\begin{subfigure}[b]{.5\textwidth}

		\includegraphics[width=\linewidth]{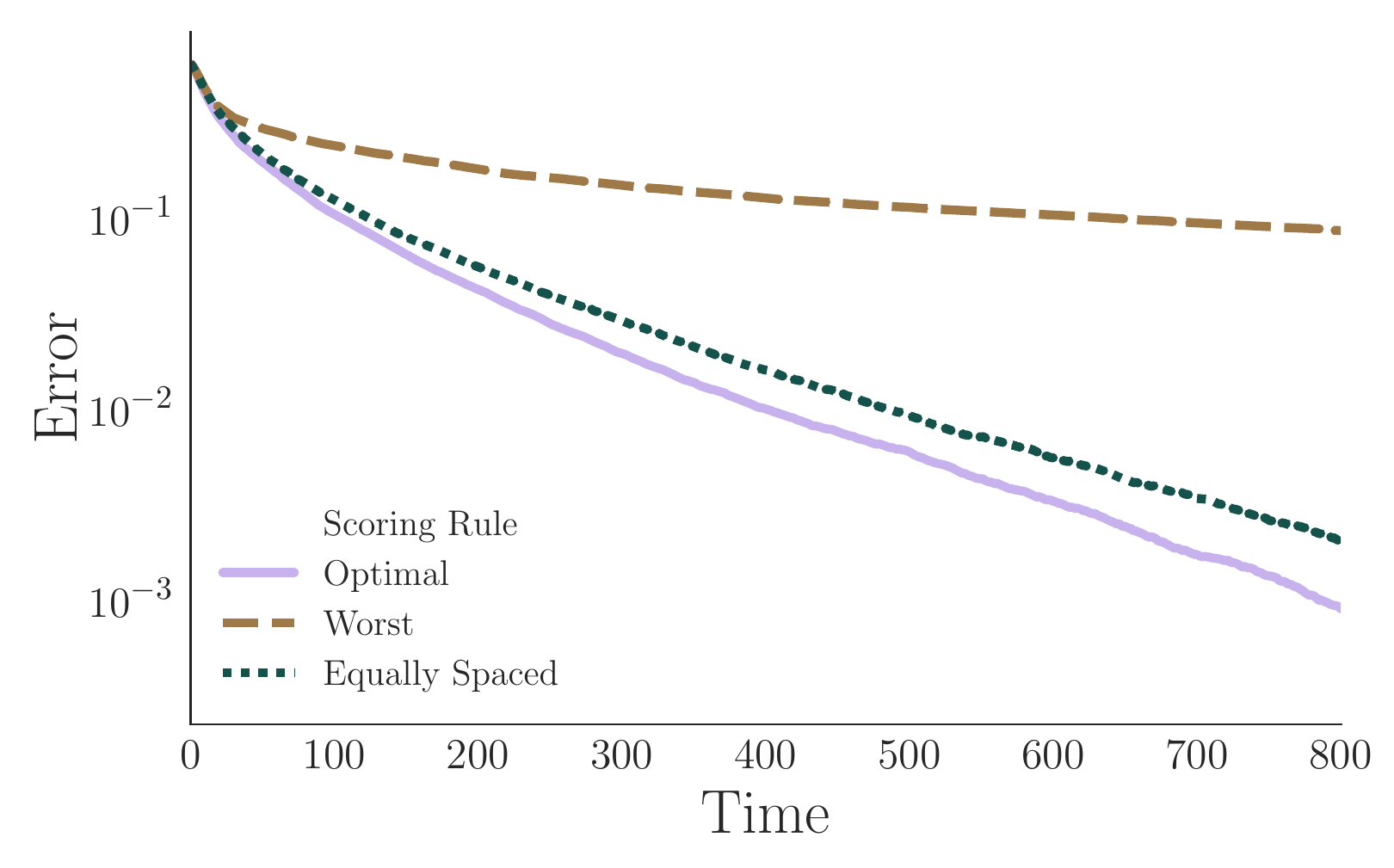}\vfill
		\caption{\textit{Adjectives} treatment with different scoring rules}
		\label{fig:labor_simulationstr3}
	\end{subfigure}\hfill
	\begin{subfigure}[b]{.5\textwidth}
		\includegraphics[width=\linewidth]{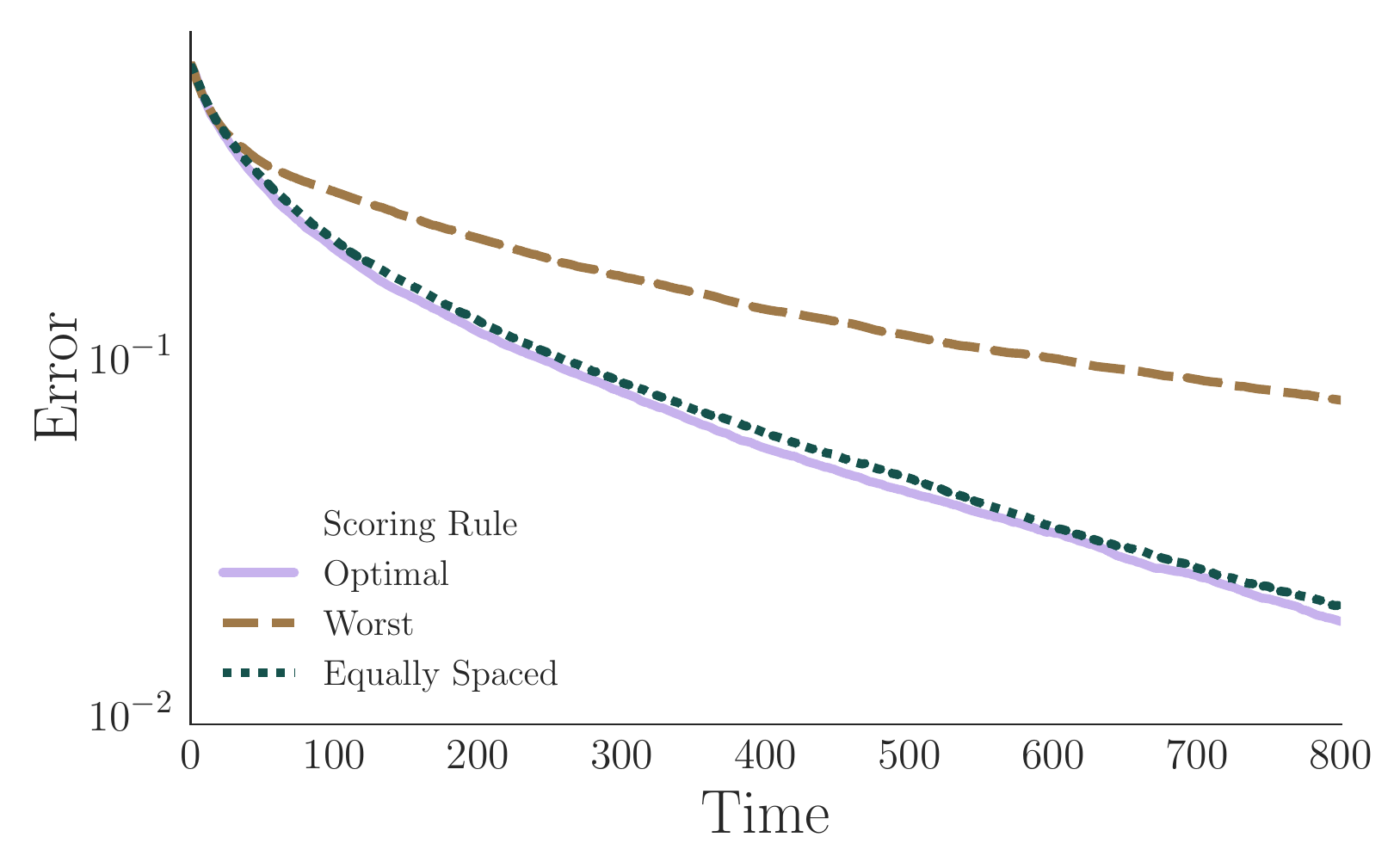}\vfill
		\caption{\textit{Numeric} treatment with different scoring rules}
		\label{fig:labor_simulationstr6}
	\end{subfigure}\hfill
	\caption{Simulated performance over time with various other configurations. The ``Worst'' scoring rule corresponds to the rule $\phi$ with the smallest learning rate found for each treatment.}

\label{fig:simulatedperfappendix}
\end{figure}

Table~\ref{tab:optscores} and Figures~\ref{fig:jointdistributions_labor_app} and \ref{fig:simulatedperfappendix} contain supplementary information regarding our application of the design approach to the labor market data, as described in the main text.

\begin{figure}[tbh]
	\centering
	\begin{subfigure}[b]{.33\textwidth}
		\includegraphics[width=\linewidth]{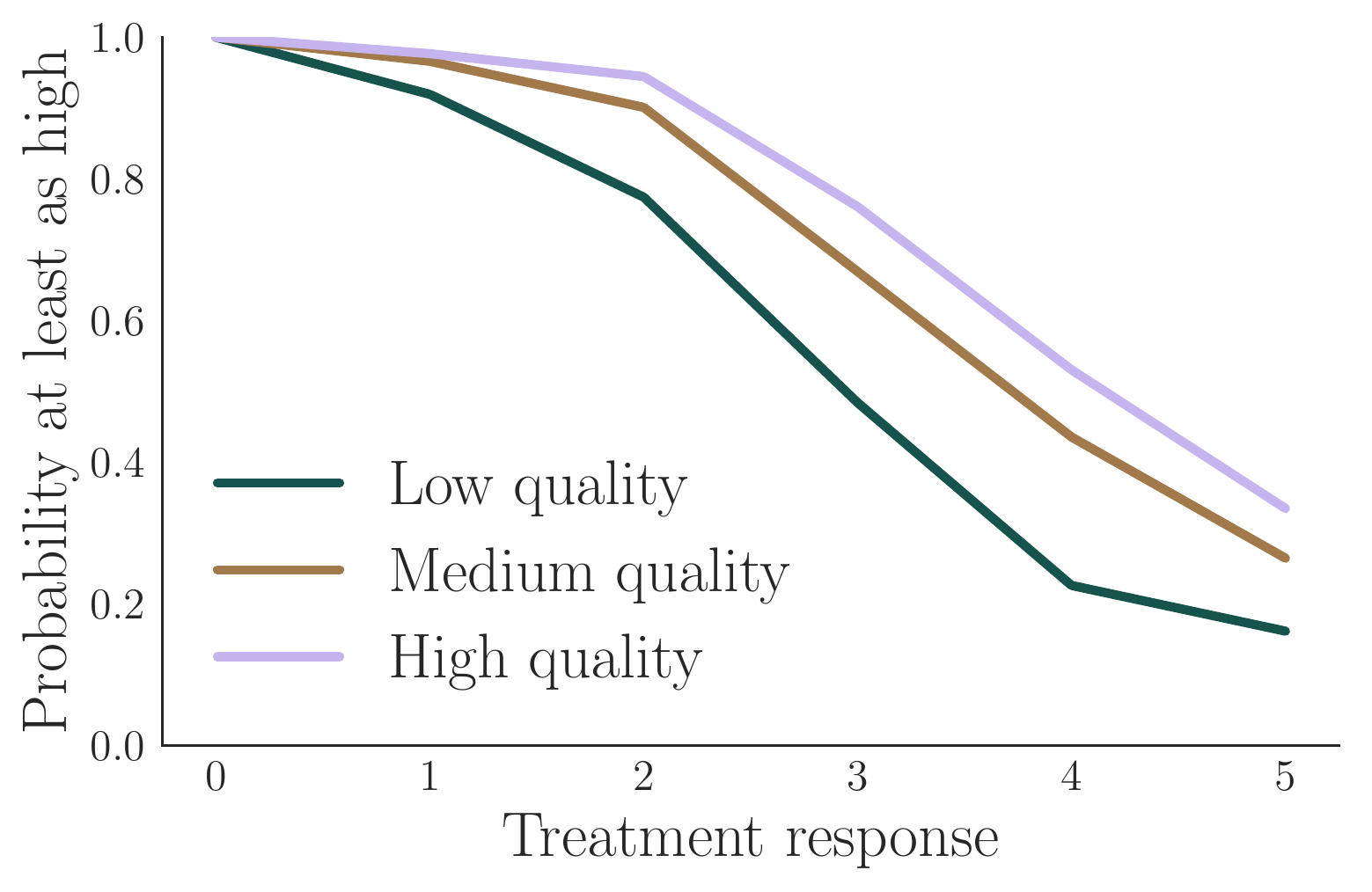}\vfill
		\caption{\textit{Expectations}}
		\label{fig:labor_joint_treatment61}
	\end{subfigure}\hfill
	\begin{subfigure}[b]{.33\textwidth}
		\includegraphics[width=\linewidth]{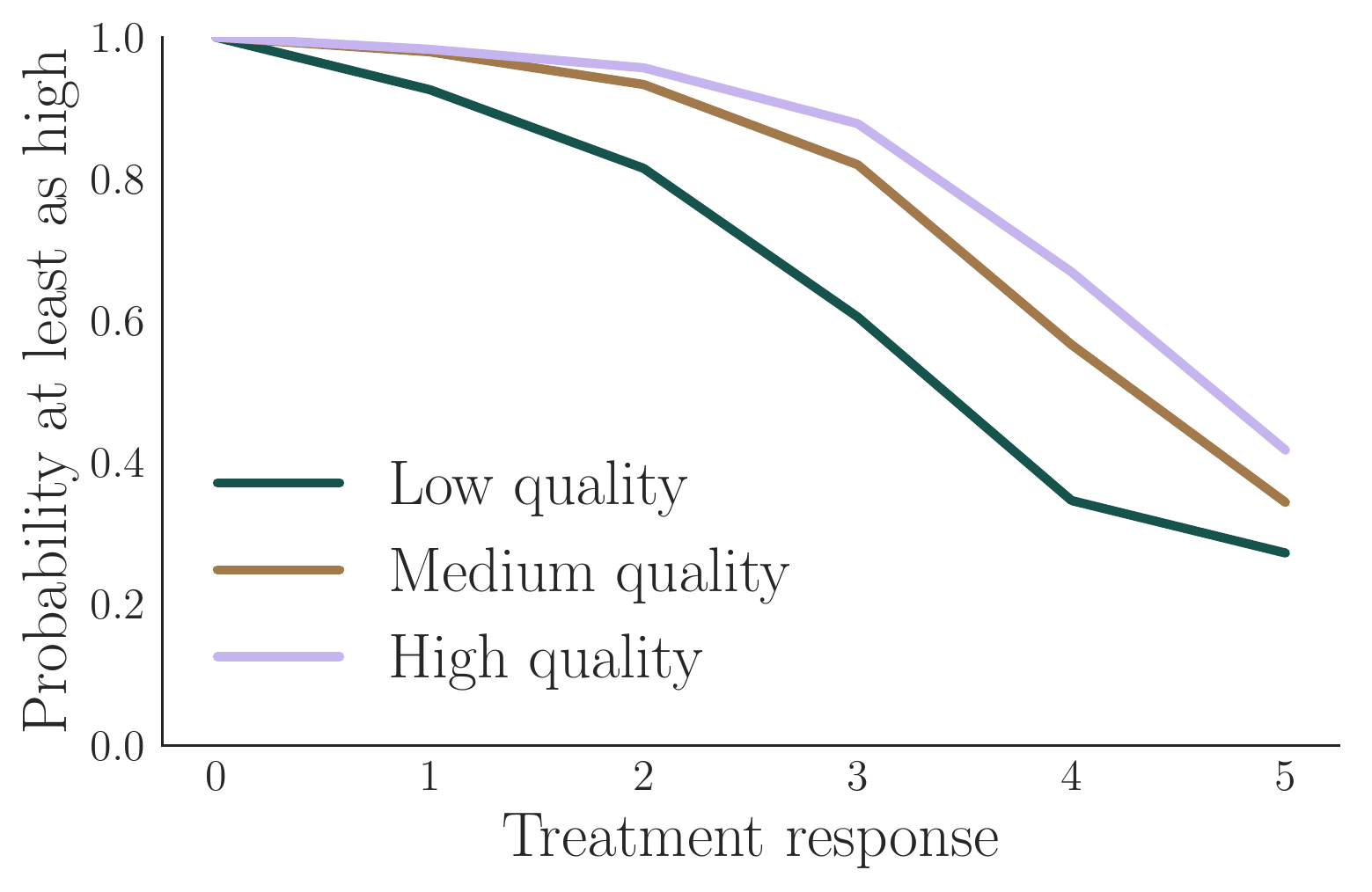}\vfill
		\caption{\textit{Adjectives}}
		\label{fig:labor_joint_treatment63}
	\end{subfigure}\hfill
	\begin{subfigure}[b]{.33\textwidth}
	\includegraphics[width=\linewidth]{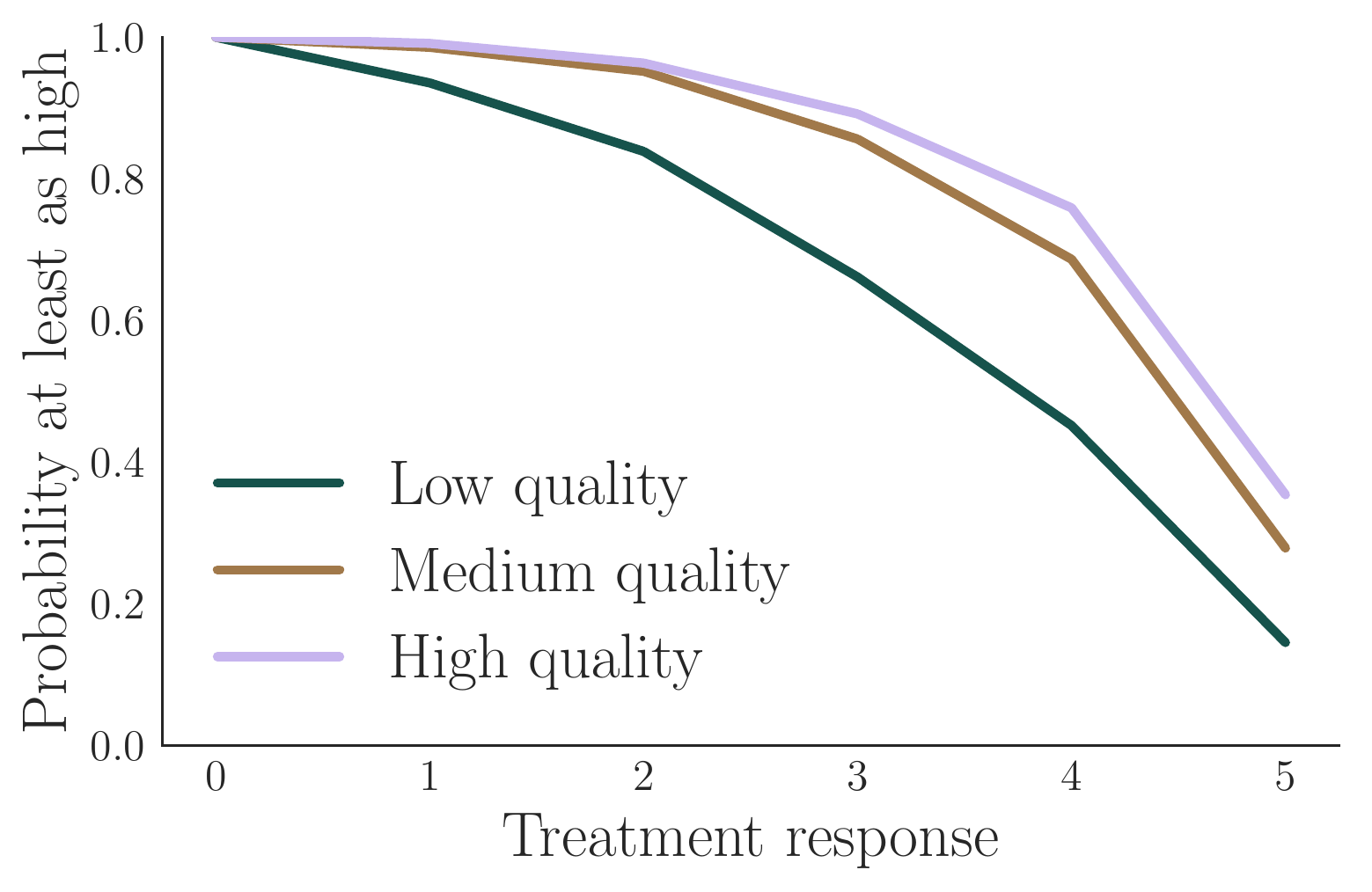}\vfill
	\caption{\textit{Average}}
	\label{fig:labor_joint_treatment62}
\end{subfigure}\hfill
\end{figure}
\begin{figure}[tbh]
\ContinuedFloat
	\begin{subfigure}[b]{.33\textwidth}
	\includegraphics[width=\linewidth]{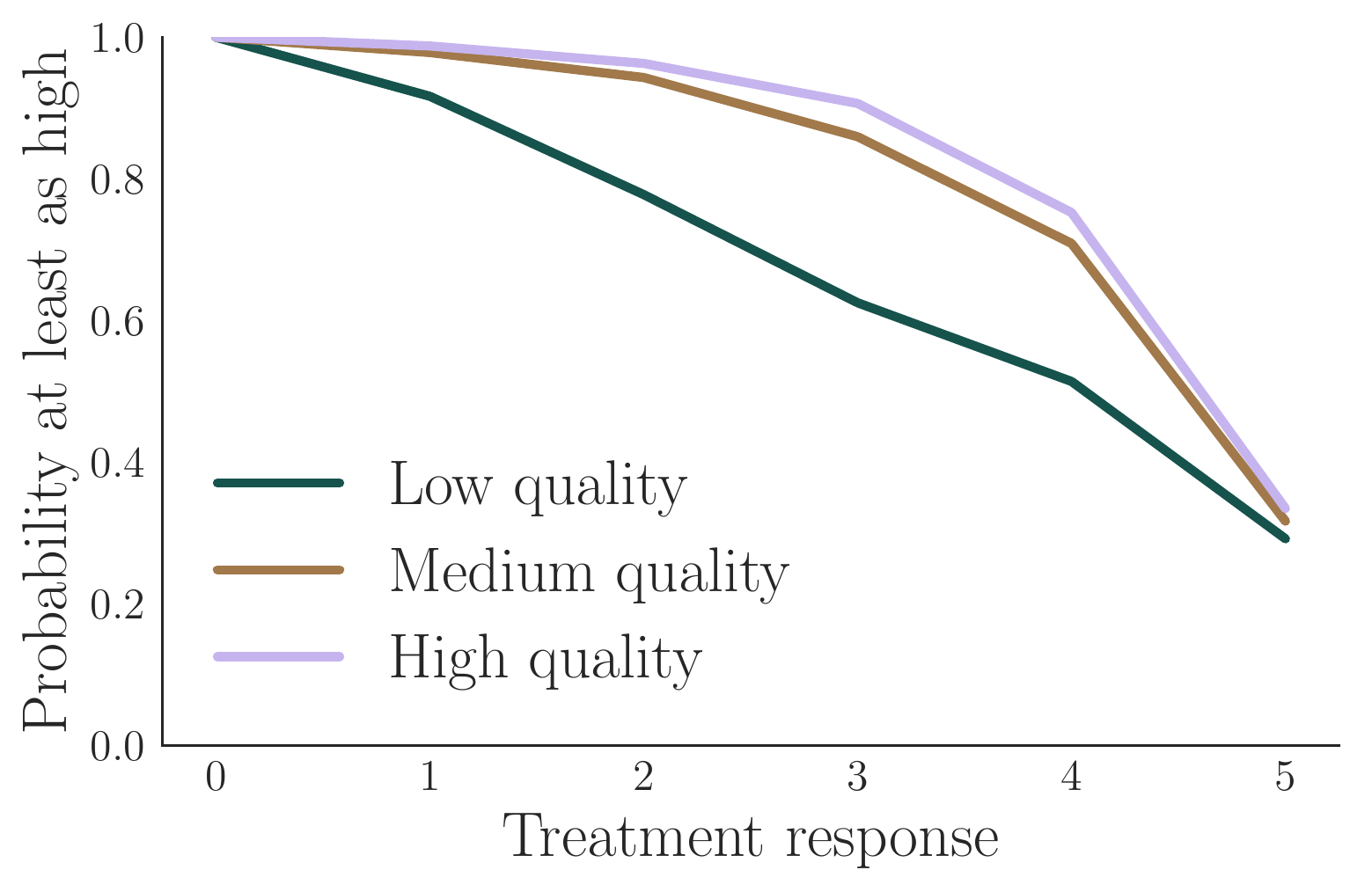}\vfill
	\caption{\textit{Average, not affect score}}
	\label{fig:labor_joint_treatment64}
\end{subfigure}\hfill
\begin{subfigure}[b]{.33\textwidth}
	\includegraphics[width=\linewidth]{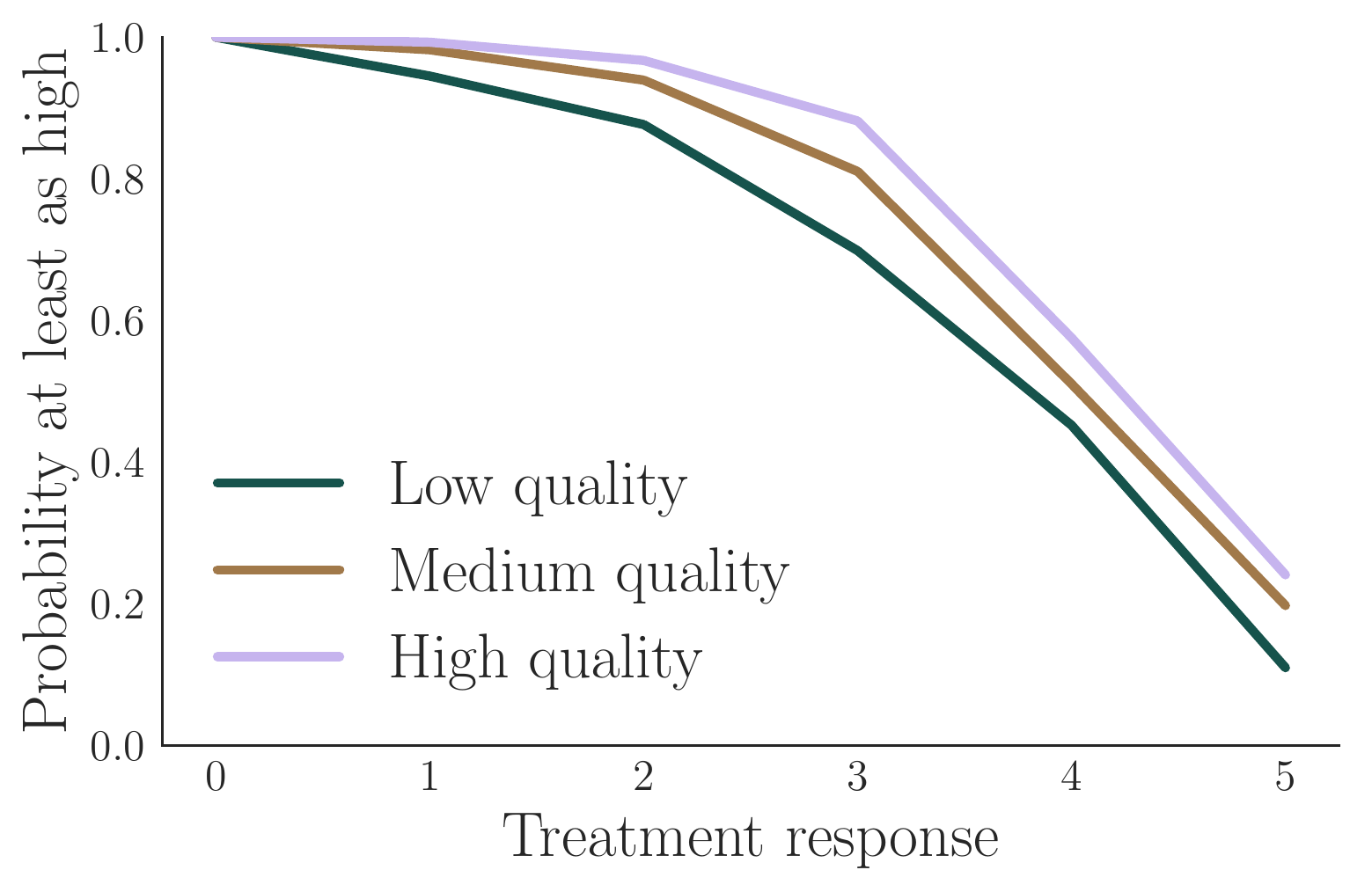}\vfill
	\caption{\textit{Average, randomized}}
	\label{fig:labor_joint_treatment65}
\end{subfigure}\hfill
\begin{subfigure}[b]{.33\textwidth}
	\includegraphics[width=\linewidth]{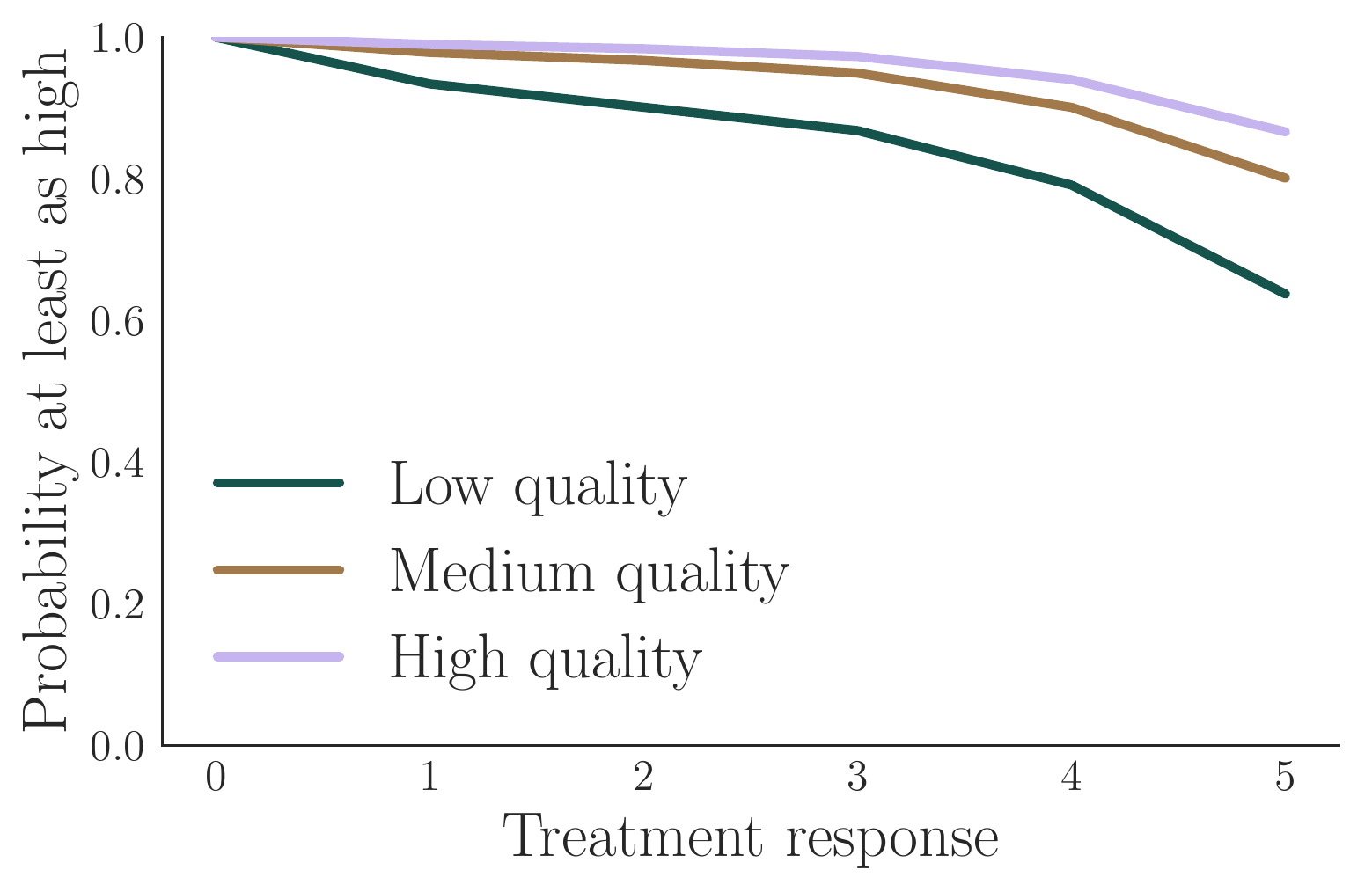}\vfill
	\caption{\textit{Numeric}}
	\label{fig:labor_joint_treatment66}
\end{subfigure}\hfill
	\caption{Joint distributions, where Low, Medium, and High quality sellers refer to those with other cell average ratings in $[0, 2), [2, 4)$ and $[4, 5]$, respectively.}
\end{figure}
\begin{table}[tbh]
	\small
	\centering
	\begin{tabular}{l|cccccc}
		& \multicolumn{6}{c}{\textbf{Response Score}}                                                                     \\
		\textbf{Condition}        & \textbf{0}           & \textbf{1}           & \textbf{2}           & \textbf{3} & \textbf{4} & \textbf{5} \\ \hline
		Expectations & 1.22 & 1.22 & 2.28 & 3.74 & 4.38 & 5.00 \\
		Adjectives & 1.47 & 1.55 & 1.63 & 3.22 & 4.97 & 5.00 \\
		Average & 1.80 & 1.84 & 1.88 & 2.53 & 3.83 & 5.00 \\
		Average, not affect score & 0.89 & 1.57 & 1.59 & 3.32 & 4.04 & 5.00 \\
		Average, randomized & 0.72 & 2.41 & 2.63 & 4.18 & 4.30 & 5.00 \\
		Numeric & 0.50 & 1.20 & 1.98 & 2.88 & 3.45 & 5.00 \\
	\end{tabular}
	\caption{Optimal scores $\phi$ for each treatment, where the score of the top position is normalized to $5$. }
	\label{tab:optscores}
\end{table}

	\pagebreak\newpage\FloatBarrier
\makeatletter{}%

\section{Amazon Mechanical Turk synthetic experiment}
\label{sec:mturkrepeatanalysis}

In this section, we deploy an experiment on Amazon Mechanical Turk (``MTurk'') to repeat and analyze our design approach, in a synthetic setting where we have expert (external) quality information on items. We note that this section is not a replication of the behavioral components of our results, as the MTurk and online labor market settings are too different to meaningfully compare. Furthermore, one should be aware of limitations of using MTurk convenience samples in research~\citep{landers2015inconvenient}; such limitations mean that there will be behavioral biases that differ from those on other platforms. For these reasons, this section should be seen as a synthetic, example application of our overall comparison and design methodology to other domains, and in particular will show how our methods are useful not just to counter rating inflation but also other types of biases.

This appendix section is organized as follows. In~\ref{appsec:mturkexperimentdescription} we describe the task, and in~\ref{appsec:mturkresults} we repeat our analysis from the main text, including: (\ref{appsec:mturkratingdistandR}) showing the resulting marginal and joint distributions of ratings and quality, and (\ref{appsec:mturkratingsimulations}) testing designs on new, unseen data.

\subsection{Experiment description}
\label{appsec:mturkexperimentdescription}

\subsubsection{Task Information}
We asked subjects to rate the English proficiency of 10 paragraphs which are modified TOEFL (Test of English as a Foreign Language) essays with known scores as determined by experts and reported in a TOEFL study guide~\citep{educational_testing_service_toefl_2005}; these are our true quality types for each essay.  Expert scores range from 1 through 5, with two paragraphs with each score. Essays are shortened to a single paragraph of just a few sentences, and the top rated paragraphs are improved and the worst ones are made worse; this is largely to ensure the quality could be sufficiently distinguished between paragraphs despite having shortened them. In other words, for each topic, we improved the language of the best rated paragraph and further degraded the language of the worst one. In principle, our editing of these paragraphs may remove the validity of the expert ratings. However, the estimated $R(\theta,y|Y)$ indicates that this does not substantially occur, suggesting our editing of the paragraphs preserved the quality ordering of the paragraphs per the expert ratings.

Subjects were given one of five possible verbal scales, where the scales were designed using a list of adjectives, \{\textit{Abysmal, Awful, Bad, Poor, Mediocre, Fair, Good, Great, Excellent, Phenomenal}\}, compiled by~\citet{hicks_choosing_2000}. Each scale had five options. The scales are:
\begin{itemize}
\item \textbf{Every Other}: \textit{Awful, Poor, Fair, Great, Phenomenal}
\item \textbf{Close to Every Other}: \textit{Abysmal, Poor, Mediocre, Good, Phenomenal}
\item \textbf{Extremes}: \textit{Abysmal, Awful, Bad, Excellent, Phenomenal}
\item \textbf{Negative-skewed}: \textit{Abysmal, Awful, Bad, Poor, Mediocre}
\item \textbf{Positive-skewed}: \textit{Fair, Good, Great, Excellent, Phenomenal}
\end{itemize}

We note that it is not a priori clear which of these scales will perform well in this setting, or what the optimal scoring mapping should be.

Raters (i.e., mTurk workers) were shown each of the ten paragraphs. The instructions were: ``Please rate on English proficiency (grammar, spelling, sentence structure) and coherence of the argument, but not on whether you agree with the substance of the text.'' The specific question then asked was: ``How does the following rate on English proficiency and argument coherence?'' One paragraph was shown per page; returning to modify a previous answer was not allowed; and paragraphs were presented in a random order. Each rater was shown one of the scales picked at random, and the same scale was used for all paragraphs for that rater.
There were approximately 500 raters overall across the 5 treatment cells, with between 97 and 104 raters in each cell. For each cell, we divide the raters (randomly) into train (75$\%$) and test (25$\%$). We design optimal scoring rules using the training data, and then test performance on the test data.

\subsubsection{Rater logistics}

We did not exclude any data, and all raters were paid $\$1.50$.  Instructions advised raters to spend no more than a minute per question, though this was not enforced. The median rater spent 325 seconds, corresponding to a median wage of \$16.61/hr. About $80\%$ of raters spent 8 minutes or less.

\subsection{Results}
\label{appsec:mturkresults}
\begin{figure}
	\centering
	\begin{subfigure}[b]{.33\textwidth}
		\includegraphics[width=\linewidth]{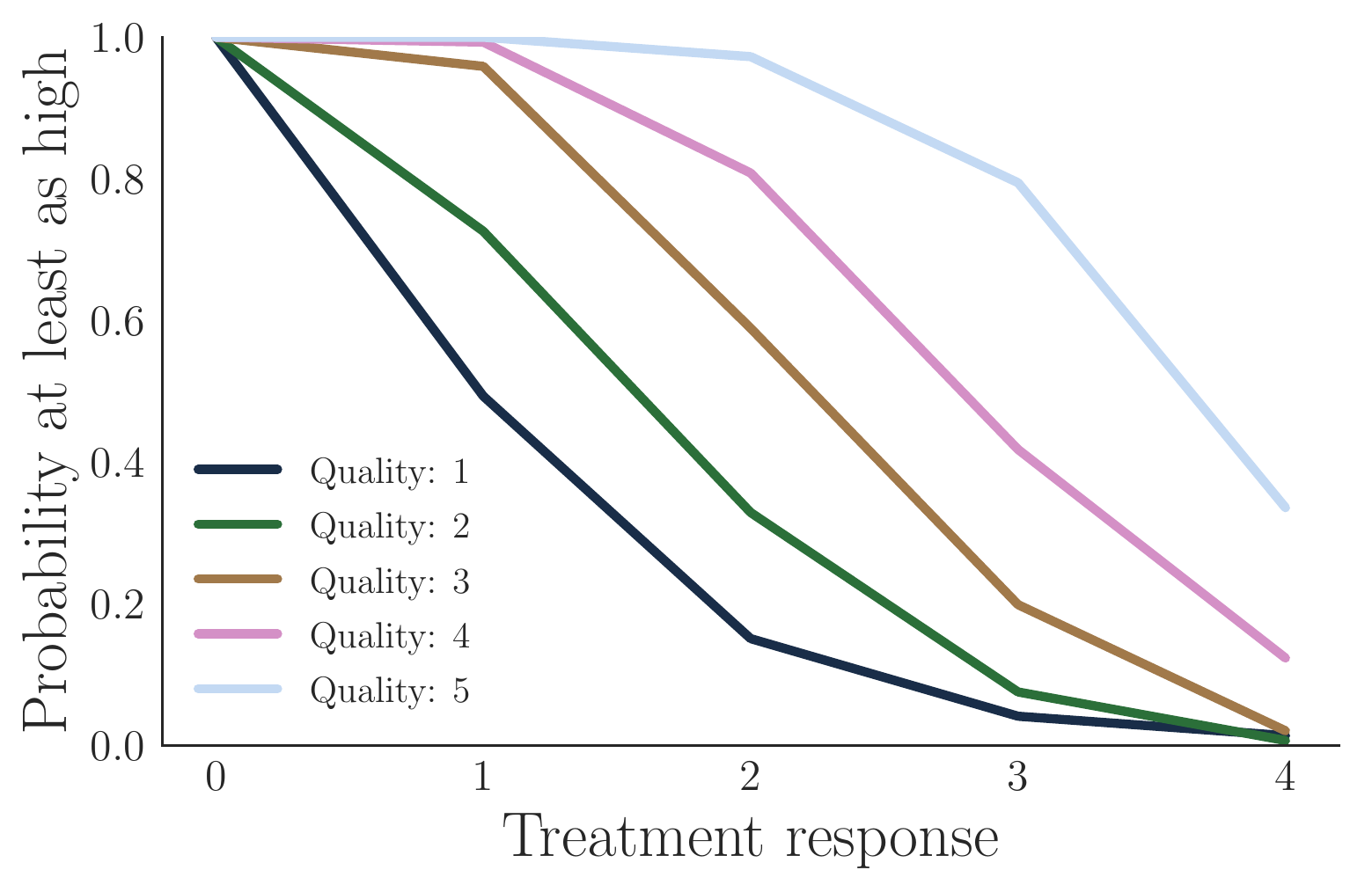}\vfill
		\caption{\textit{Every Other} }
	\end{subfigure}\hfill
	\begin{subfigure}[b]{.33\textwidth}
		\includegraphics[width=\linewidth]{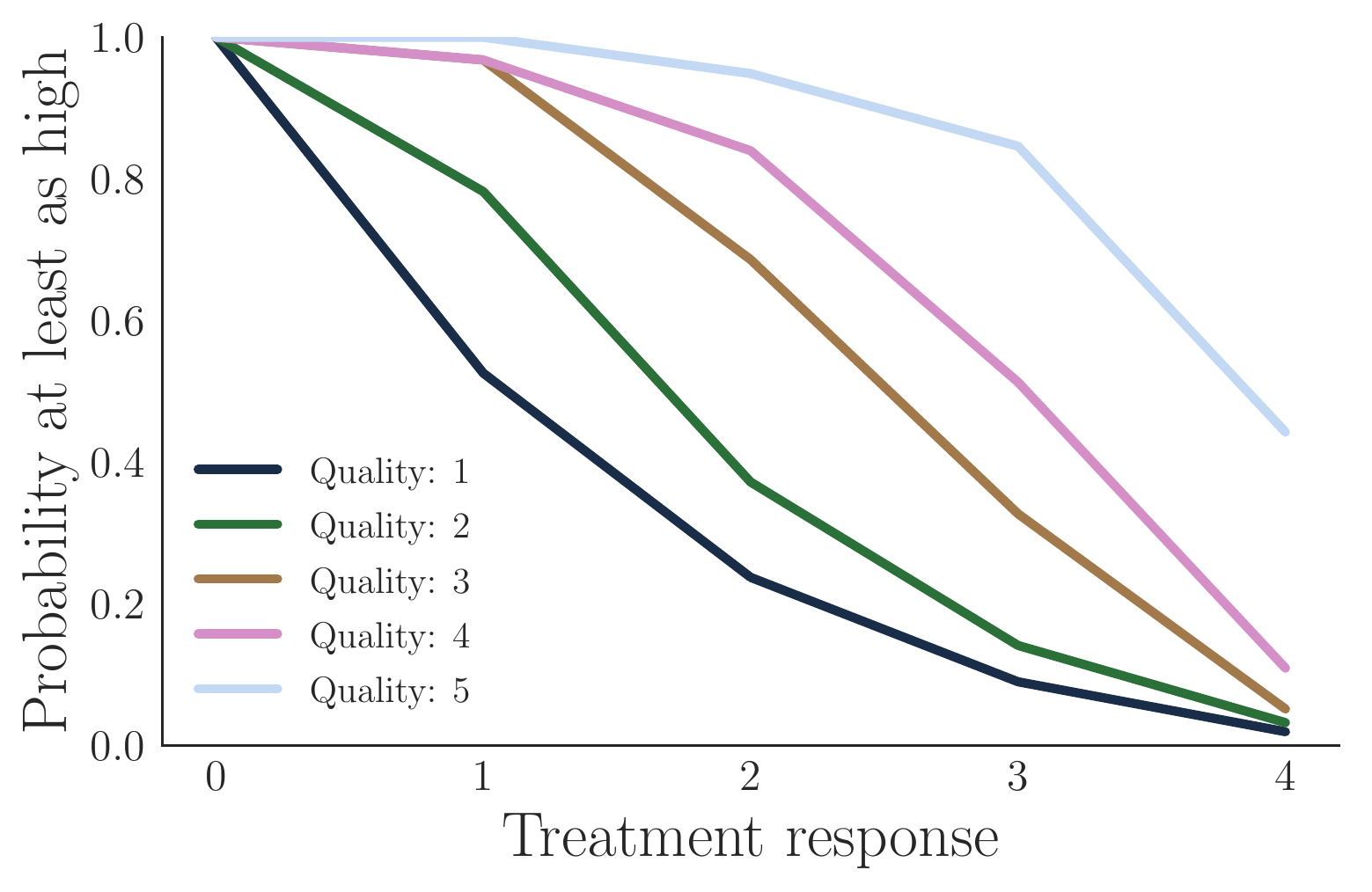}\vfill
		\caption{\textit{Close to Every Other}}
	\end{subfigure}\hfill
	\begin{subfigure}[b]{.33\textwidth}
		\includegraphics[width=\linewidth]{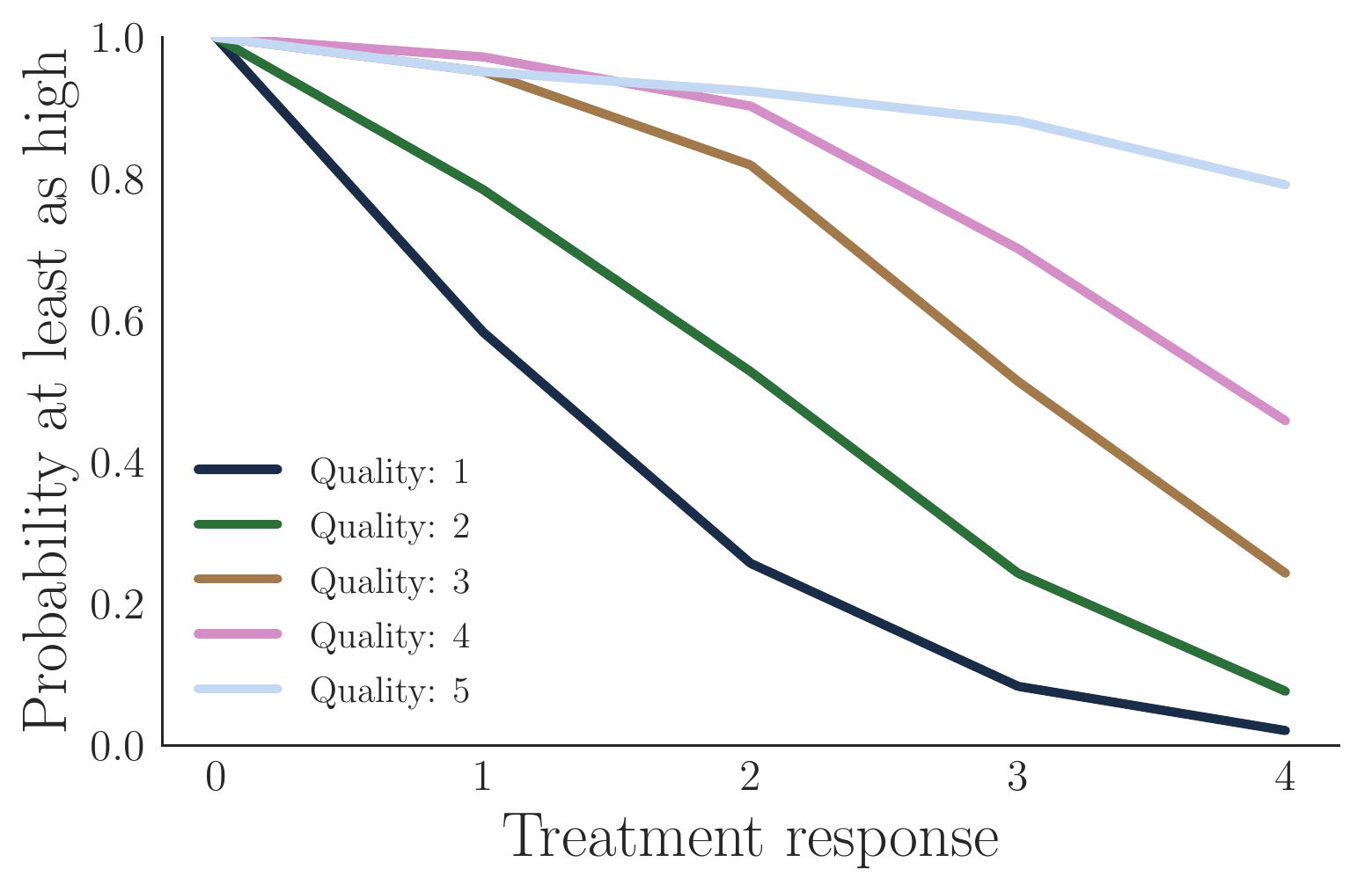}\vfill
		\caption{\textit{Negative-skewed}}
	\end{subfigure}\hfill

	\begin{subfigure}[b]{.33\textwidth}
		\includegraphics[width=\linewidth]{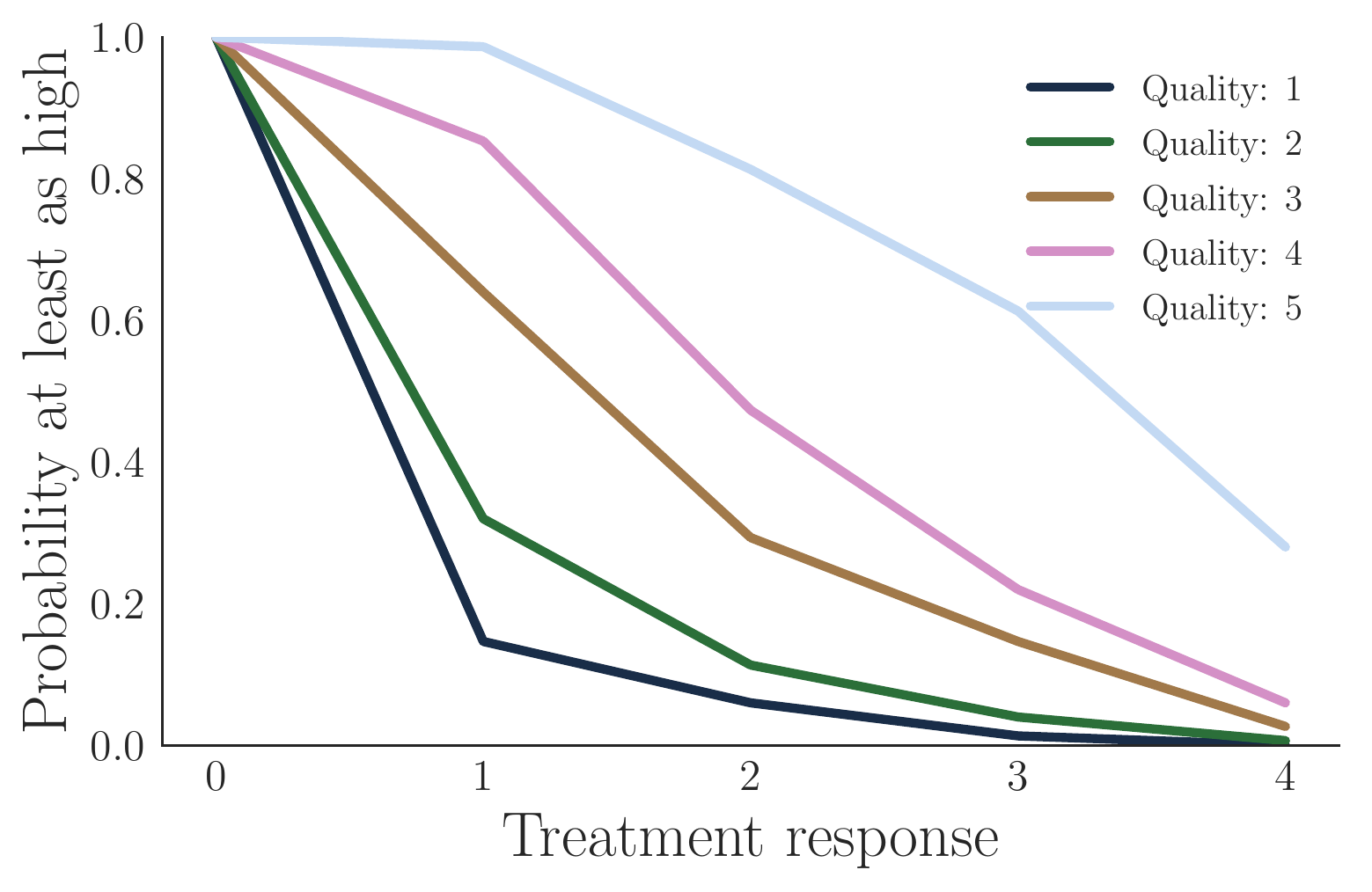}\vfill
		\caption{\textit{Positive-skewed}}
	\end{subfigure}
	\begin{subfigure}[b]{.33\textwidth}
		\includegraphics[width=\linewidth]{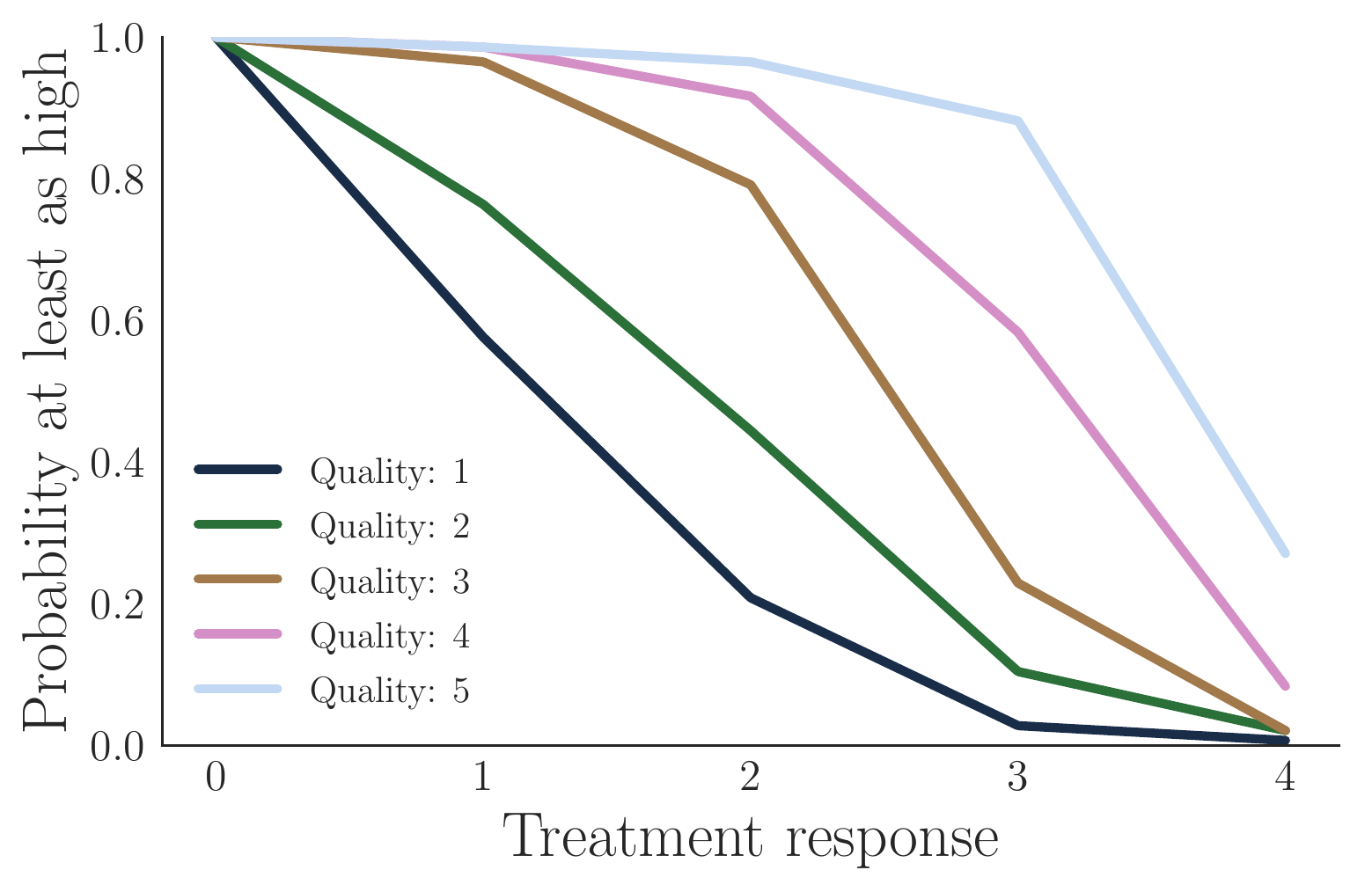}\vfill
		\caption{\textit{Extremes}}
	\end{subfigure}\hfill
	\caption{Joint distributions of rating and expert score on the MTurk training set, by treatment condition.}	
	\label{fig:jointmturktrain}
\end{figure}

\begin{table}
		\small
	\centering
		\begin{tabular}{l|cc}
		& \multicolumn{2}{c}{\textbf{Learning rates}}                                                                     \\
\textbf{Condition}        & \textbf{Equally spaced $\phi$}           & \textbf{Optimal $\phi$}       \\ \hline
		Every Other         & 0.058       & 0.069 \\
		Extremes             & 0.077       & 0.079       \\
		Negative-skewed        & 0.051       & 0.059       \\
		Positive-skewed        & 0.034      & 0.043      \\
		Close to Every Other & 0.043      & 0.044           
	\end{tabular}
	\vfill
	\caption{Large deviation learning rates for each treatment in the Mturk experiment, calculated using Equation~\eqref{eq:rW} and the joint distributions generated using the training data plotted in Figure~\ref{fig:jointmturktrain}. Optimal for each treatment corresponds to the highest learning rate among many random score functions tested.}
	\label{fig:learningratesbyscoremturk}
\end{table}

\begin{figure}
\begin{subfigure}[b]{.5\textwidth}
	\includegraphics[width=\linewidth]{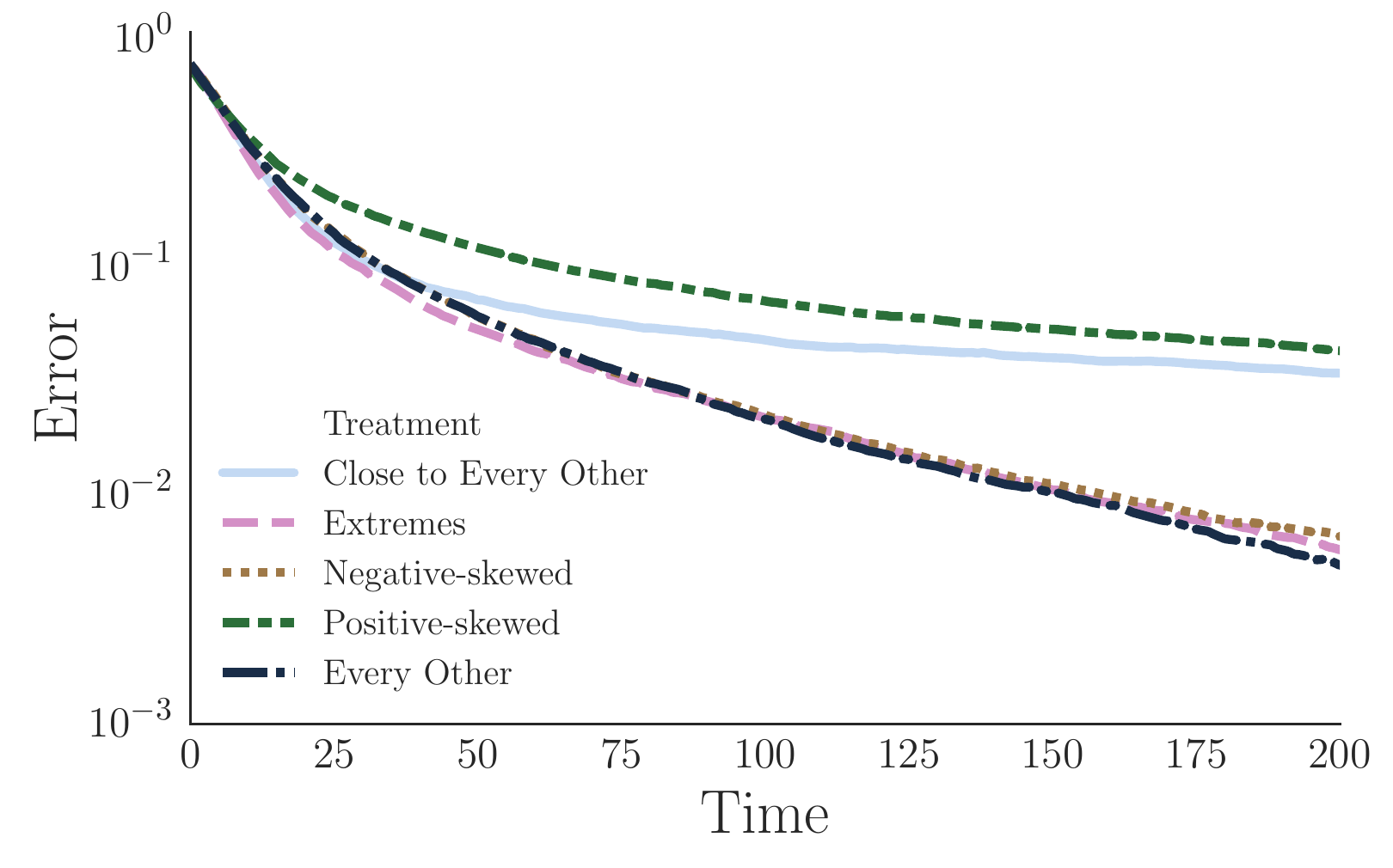}\vfill
	\caption{Training data}
\end{subfigure}
	\begin{subfigure}[b]{.5\textwidth}
	\includegraphics[width=\linewidth]{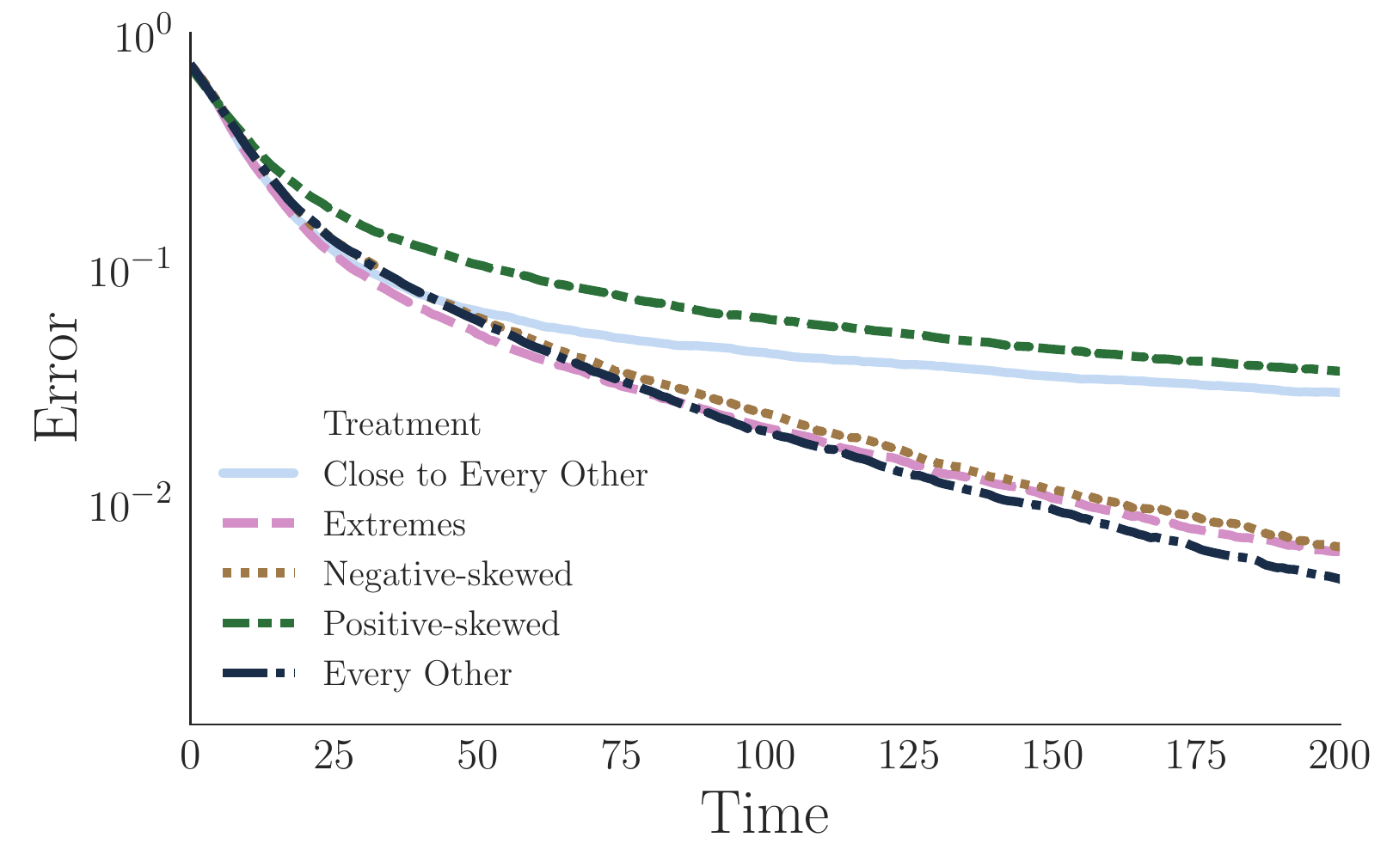}\vfill
	\caption{Test data}
\end{subfigure}
\caption{Simulated performance of each rating scale with Equally Spaced scores.}
	\label{fig:mturk_simulationsoptimalscores}
\end{figure}

We now repeat the design and test procedure from the main text, for this setting. All plots, figures, and scoring rules are generated exactly as in the main text, with the following exceptions: (1) we have true expert scores for the paragraph qualities and so do not use the procedure where we estimate such qualities from the other treatment cells, and (2) we split the rater responses into test and training sets. We show the joint distributions $R(\cdot | Y)$ and optimal scores calculated from the training set, and then we evaluate performance on simulations using the test set.

\subsubsection{Rating distributions and learning rates}
\label{appsec:mturkratingdistandR}

Figure~\ref{fig:jointmturktrain} shows the joint distributions of rating and expert score on the MTurk training set, for each treatment condition. Note that each of the treatment conditions produce different joint distributions, even the two conditions with similar verbal scales, \textit{Every Other} and \textit{Close to Every Other}. Furthermore, note that even with these joint distributions, it is not immediately obvious which one(s) will induce the best learning rates. 

Next, we calculate optimal score functions for each treatment using the training joint distributions. Table~\ref{fig:learningratesbyscoremturk} shows the training set learning rates for each treatment using equally spaced scores, as well as the best performing scores, respectively. The various designs have dramatically different rates, even when the rating scales use similar phrases. 

\subsubsection{Simulations of accuracy over time}
\label{appsec:mturkratingsimulations}

Finally, we simulate the performance of the designs (generated using the training data), following the same simulation technique as outlined in the main text. We also evaluate performance on the test data, in order to demonstrate how a platform would use our design approach. Figure~\ref{fig:mturk_simulationsoptimalscores} shows the resulting errors over time with Equally Spaced scores. Errors with optimal scores are qualitatively similar.

\pagebreak\newpage\FloatBarrier
\makeatletter{}%
\section{Proofs}

\begin{lemma} 
	$$	\lim_{k \to \infty}- \frac{1}{k} \log \left[\mu((x_k(\theta_1) - x_k(\theta_2)) \leq 0 | \theta_1 , \theta_2)\right] = \inf_{a \in \mathcal{R}} \left \{ g(\theta_1) I(a|\theta_1) + g(\theta_2) I(a|\theta_2) \right \}$$ where $I(a|\ell) = \sup_z \{ za - \Lambda(z|\theta) \}$, $\Lambda(z|\theta)$ is the log moment generating function of a single sample from $x(\theta_1)$, and $g(\theta)$ is the sampling rate. \label{lem:problessthan}
\end{lemma}
	\proof
	\noindent$\lim_{k \to \infty}- \frac{1}{k} \log \left[\mu((x_k(\theta_1) - x_k(\theta_2)) \leq 0 | \theta_1 , \theta_2)\right]$
	\begin{align}
	&= \lim_{k \to \infty}- \frac{1}{k} \log \left[\int_{a\in \mathcal{R}} \mu((x_k(\theta_1) = a | \theta_1 ) \mu(x_k(\theta_2) \geq a |\theta_2)da\right]\\
	&= \lim_{k \to \infty}- \frac{1}{k} \log \left[\int_{a\in \mathcal{R}} e^{-kg(\theta_1) I (a | \theta_1)} e^{-kg(\theta_2) I (a | \theta_2)}da\right] \label{eqnpart:applyld}\\
	&= \inf_{a \in \mathcal{R}} \left \{ g(\theta_1) I(a|\theta_1) + g(\theta_2) I(a|\theta_2) \right \} & \text{Laplace principle}
	\end{align}

Where~\eqref{eqnpart:applyld} is a basic result from large deviations, and $kg(\theta_i)$ is the number of samples item of quality $\theta_i$ has received. \Halmos

This lemma also appears in \cite{glynn_large_2004}, which uses the Gartner-Ellis Theorem in the proof. Our proof is conceptually similar but instead uses Laplace's principle. \\
\noindent We can now establish the rate function for $P_k(\theta_1, \theta_2)$. 

Recall $
P_k(\theta_1, \theta_2) = \mu_k(x_k(\theta_1) > x_k(\theta_2)|\theta_1, \theta_2) - \mu_k(x_k(\theta_1) < x_k(\theta_2)|\theta_1, \theta_2)$. Then, we have

\begin{restatable}{lemma}{lemPkld}
	\label{lem:Pk_LD}
	Given $\theta_1, \theta_2$, let $\b{P}_k(\theta_1, \theta_2) = 1 - P_k(\theta_1, \theta_2)$.  Then:
	\begin{equation}
	\label{eq:LDrate}
	- \lim_{k \to \infty} \frac{1}{k} \log \b{P}_k(\theta_1, \theta_2) = \inf_{a \in \mathcal{R}} \left \{ g(\theta_1) I(a|\theta_1) + g(\theta_2) I(a|\theta_2) \right \},
	\end{equation}
	where $I(a|\theta) = \sup_z \{ za - \Lambda(z|\theta) \}$, and $\Lambda(z|\theta)$ is the log moment generating function of a single rating given to seller of type $\theta$:
	\[ \Lambda(z|\theta) = \log \sum_{y \in Y} \rho(\theta, y | Y) \exp( z \phi(y) ). \]
\end{restatable}
\proof
	Follows directly from Lemma \ref{lem:problessthan}. 
	\begin{align*}
	- \lim_{k \to \infty} &\frac{1}{k} \log \b{P}_k(\theta_1, \theta_2 | \beta) \\
	&= \lim_{k \to \infty}- \frac{1}{k} \log \left[1 + \mu_k(x_k(\theta_1) - x_k(\theta_2) < 0 | \theta_1 , \theta_2) - \mu_k(x_k(\theta_1) - x_k(\theta_2) > 0 | \theta_1 , \theta_2)\right] \\
	&= \lim_{k \to \infty}- \frac{1}{k} \log \left[2\mu_k(x_k(\theta_1) - x_k(\theta_2) < 0 | \theta_1 , \theta_2) + \mu_k(x_k(\theta_1) - x_k(\theta_2) = 0 | \theta_1 , \theta_2)\right] \\
	&= \inf_{a \in \mathcal{R}} \left \{ g(\theta_1) I(a|\theta_1) + g(\theta_2) I(a|\theta_2) \right\} &\text{Lemma}~\ref{lem:problessthan}
	\end{align*} \Halmos

Now we show that this rate function transfers to a rate function for $W_k$. 
\\\\\noindent\textbf{Proof of Theorem~1}

\begin{equation}
r \triangleq - \lim_{k \to \infty} \frac{1}{k} \log (1- {W}_k) = \min_{0 \leq i < M} \inf_{a \in \bbR} \left \{ g(\theta_{i+1}) I(a|\theta_{i+1}) + g(\theta_i) I(a|\theta_i)\right\} 
\end{equation}
where 
$I(a|\theta) = \sup_z \{ za - \Lambda(z|\theta) \}$, and $\Lambda(z|\theta) = \log \sum_{y \in Y} \rho(\theta, y | Y) \exp( z \phi(y))$ is the log moment generating function of a single rating given to seller of type $\theta$.

\proof{}
\begin{align}
	- \lim_{k \to \infty} \frac{1}{k} \log (1 - {W}_k) 
	& = - \lim_{k \to \infty} \frac{1}{k} \log \left(1 - \frac{2}{M(M-1)}\sum_{\theta_1 > \theta_2 \in \Theta} P_k(\theta_1, \theta_2)\right)\\
	&= - \lim_{k \to \infty} \frac{1}{k} \log\frac{2}{M(M-1)} \sum_{0  \leq i < j \leq M} \b{P}_k(\theta_j, \theta_i)\\
	&= -\max_{0\leq i < j \leq M}\left( \lim_{k \to \infty} \frac{1}{k} \log \left( \b{P}_k(\theta_j, \theta_i)\right)\right)\label{eqnstep:ldpproperty} 
	= \min_{0\leq i < j \leq M}\left( -\lim_{k \to \infty} \frac{1}{k} \left[\log \left(\b{P}_k(\theta_j, \theta_i)\right)\right]\right)\\
	&= \min_{0 \leq i < j \leq M} \inf_{a \in \mathcal{R}} \left \{ g(\theta_j) I(a|\theta_j) + g(\theta_i) I(a|\theta_i)\right\}\\
	&= \min_{0 \leq i < M} \inf_{a \in \mathcal{R}} \left \{ g(\theta_{i+1}) I(a|\theta_{i+1}) + g(\theta_i) I(a|\theta_i)\right\}
\end{align}
	Where the last line follows from adjacent $\theta_i, \theta_{i+1}$ dominating the rate due to properties of $R$. 	Line~\eqref{eqnstep:ldpproperty} follows from: $\forall a^\epsilon_i \geq 0$, ${\lim \sup}_{\epsilon \to 0} \left[\epsilon\log\left(\sum_i^N a^\epsilon_i\right)\right] = \max^N_i {\lim \sup}_{\epsilon \to 0} \epsilon \log(a^\epsilon_i)$.
	See, e.g., Lemma 1.2.15 in~\cite{dembo_large_2010} for a proof of this property. 

\Halmos 

\FloatBarrier

\end{document}